\documentclass[prd,eqsecnum,twocolumn,amsfonts,amssymb]{revtex4}

\usepackage{graphicx}
\usepackage{amsmath}
\usepackage{bm}
\usepackage{booktabs}
\usepackage{tabularx}

\newcommand{\beq}{\begin{equation}}
\newcommand{\eeq}{\end{equation}}
\newcommand{\beqs}{\begin{eqnarray}}\newcommand{\eeqs}{\end{eqnarray}}
\newcommand{\lsim}{\mathrel{\raisebox{-
.6ex}{$\stackrel{\textstyle<}{\sim}$}}}
\newcommand{\gsim}{\mathrel{\raisebox{-
.6ex}{$\stackrel{\textstyle>}{\sim}$}}}

\begin{document}

\title{Neutrino Masses and Mixing in Models with Large Extra Dimensions
  and Localized Fermions} 

\author{S. Girmohanta$^a$, R. N. Mohapatra$^b$, and R. Shrock$^a$}

\affiliation{a \ C. N. Yang Institute for Theoretical Physics and 
Department of Physics and Astronomy, \\
Stony Brook University, Stony Brook, NY 11794, USA }

\affiliation{b \ Maryland Center for Fundamental Physics and  
Department of Physics, \\
University of Maryland, College Park, MD 20742 }

\begin{abstract}

  Using a low-energy effective field theory approach, we study some
  properties of models with large extra dimensions, in which quarks
  and leptons have localized wave functions in the extra dimensions.
  We consider models with two types of gauge groups: (i) the
  Standard-Model gauge group, and (ii) the left-right symmetric (LRS)
  gauge group.  Our main focus is on the lepton sector of models with
  $n=2$ extra dimensions, in particular, neutrino masses and mixing.
  We analyze the requisite conditions that the models must satisfy to
  be in accord with data and present a solution for lepton wave
  functions in the extra dimensions that fulfills these conditions.  As
  part of our work, we also present a new solution for quark wave
  function centers.  Issues with flavor-changing neutral current
  effects are assessed.  Finally, we remark on baryogenesis and dark
  matter in these models.

\end{abstract}

\maketitle


\section{Introduction}
\label{intro_section}

An interesting idea for physics beyond the Standard Model (BSM) is
that our four-dimensional spacetime is embedded in a
higher-dimensional space with $n$ extra spatial dimensions
compactified on a length scale of $L \sim 10^{-19}$ cm, i.e., $1/L
\sim 100$ TeV, in which SM fermions have strongly localized wave
functions \cite{as,ms}. These are commonly called split-fermion (SF)
models, and we shall follow this terminology. One motivation for
split-fermion models is that they can explain the generational
hierarchy of quark and charged lepton masses by appropriate choices of
locations of the fermion wave function centers in the extra
dimensions, without the necessity of a large hierarchy in the Yukawa
couplings in the higher-dimensional space \cite{as,ms}.

In the present work we shall study some properties
of split-fermion models models. We shall give a number of general
formulas for arbitrary $n$, but for our detailed phenomenological
calculations, we focus on the case of $n=2$ extra dimensions.
Two types of gauge groups are considered: (i) the
Standard-Model gauge group,
$G_{SM} = {\rm SU}(3)_c \otimes {\rm SU}(2)_L \otimes {\rm U}(1)_Y$,
and (ii) the left-right symmetric (LRS) group
\cite{lrs75a}-\cite{lrs81}
\beq
G_{LRS} = {\rm SU}(3)_c \otimes {\rm SU}(2)_L \otimes {\rm SU}(2)_R \otimes 
{\rm U}(1)_{B-L} \ , 
\label{glrs}
\eeq
where $B$ and $L$ denote baryon and (total) lepton number.  We
concentrate on investigating properties of the lepton sector, including,
in particular, neutrino masses and mixing.  An analysis is given of the
necessary conditions that the models with SM and LRS gauge symmetries
must satisfy to be in agreement with constraints from data. We calculate
a solution for lepton wave functions in the extra dimensions that
fulfils these conditions.  Issues pertaining
to flavor-changing neutral current processes and fine tuning in
both the lepton and quark sectors are addressed. As part of our
work, we calculate a new solution for quark wave functions in the extra
dimensions that greatly reduces flavor-changing neutral current effects
to show that there is adequate suppression of
proton decay, the models must also have sufficient separation between
the centers of quark wave functions and lepton wavefunctions, and we
show that this condition is met with our solution for lepton and quark
wave functions.  Finally, we remark on baryogenesis and dark
matter in these models. Some early studies of phenomenological aspects
of split-fermion models after Refs. \cite{as,ms} include
Refs. \cite{ags}-\cite{ghpss}. Among these works, studies of neutrino
masses and mixing focused on the case of $n=1$ extra dimension, and
this was one motivation for our focus on the case $n=2$.

This paper is organized as follows.  In Sec. \ref{extra_dim_section}
we describe the split-fermion models used for our study and the
procedure of integrating over the extra dimensions to derive terms in
the Lagrangian of the (four-dimensional) low-energy effective field
theory (EFT). In Sec.  \ref{gauge_group_section} we review relevant
aspects of the left-right symmetric gauge theory.
Sec. \ref{yukawa_section} describes the Yukawa terms and the resultant
masses and mixing in the quark and lepton sectors.  In Sections
\ref{nu_sm_section} and \ref{nu_lrs_section} we discuss the
determination of lepton wave function centers to fit charged lepton
and neutrino masses and lepton mixing and for the extra-dimensional
models. This section also contains a new solution for quark wave
function centers.  Section \ref{kk_section} is devoted to a discussion
of the contributions of KK modes to various physical processes. In
Section \ref{phenom_section} we study constraints on these models
arising from limits on non-Standard-Model contributions to weak decays
and neutrino reactions, on charged-lepton flavor-violating processes,
electromagnetic properties of (Majorana) neutrinos, and on
neutrinoless double beta decays.  Section \ref{baryogensis_dm_section}
is devoted to a discussion of baryogenesis and dark matter in the
models. Our conclusions are presented in Section
\ref{conclusion_section}.  Some auxiliary formulas and further details
about the calculations are included in several appendices. Our present
work is an extension of previous studies of baryon-number violation in
extra-dimension models, including, in particular, $n-\bar n$
oscillations, in this class of models\cite{nnb02,bvd,nnblrs}.


\section{Extra-Dimensional Framework}
\label{extra_dim_section} 

In this section we describe the extra-dimensional framework
\cite{as,ms} used for our present study. Motivations for hypothesizing
extra (spatial) dimensions go back at least to the effort by Kaluza
and Klein (KK) to unify electromagnetism and gravity and were further
strengthened with the advent of (super)string theories of quantum
gravity. We shall give a number of formulas for a general number $n$ of
extra dimensions and later specialize to the case $n=2$ for our
phenomenological calculations.  Usual spacetime coordinates are
denoted as $x_\nu$, $\nu=0,1,2,3$, and the $n$ extra coordinates as
$y_\nu$. The fermion and boson fields are taken to have a factorized
form; for the fermions, this is
\beq
\Psi(x,y)=\psi(x)\chi(y) \ ,
\label{psiform}
\eeq
and similarly for the bosons.  In each of the extra dimensions these
fields (including right-handed neutrinos) are restricted to a range of
finite length $L$ \cite{lmugen}.  We define an energy corresponding to
the inverse of the compactification scale as $\Lambda_L \equiv 1/L$.
Because of the compactification, the fields have excited KK modes,
which will be discussed further below.

Starting from an effective Lagrangian in the $d=(4+n)$-dimensional
spacetime, one obtains the resultant low-energy effective Lagrangian
in four dimensions by integrating products of operators over the extra
$n$ dimensions.  We use a low-energy effective field theory (EFT)
approach that entails an ultraviolet cutoff, which we denote as $M_*$.
The wave function of a fermion $f$ in the extra dimensions has the
Gaussian form \cite{as,ms}
\beq
\chi_f(y) = A_f \, e^{-\mu^2 \, \| y-y_f \|^2} \ , 
\label{gaussian}
\eeq
where $A_f$ is a normalization factor (see Eq. (\ref{af})), and the
$n$-dimensional vector $y_f$ denotes the position of this fermion in the
extra dimensions, with components $y_f = ((y_f)_1,...,(y_f)_n)$ and
with the usual Euclidean norm of a vector in a compactification of
${\mathbb R}^n$, namely
\beq
\| y_f \| \equiv \Big (\sum_{\lambda=1}^n y_{f,\lambda}^2 \Big )^{1/2} \ .
\label{yfnorm}
\eeq
A measure of the width of the Gaussian fermion wave function is given
by 
\beq
\sigma = \frac{1}{\mu} \ ,
\label{sigma}
\eeq
which is $\sqrt{2}$ times the variance $\sigma_v$ of the Gaussian
(\ref{gaussian}).  We take this width to be the same for all of the
fermions \cite{lmugen}.  (An alternate normalization is $\sigma =
1/(2^{1/2} \mu)$.)  For $n=1$ or $n=2$, this fermion localization can
result from appropriate coupling to scalar localizer field(s) with
kink or vortex solution(s), respectively
\cite{rubakov83}-\cite{volkas2007}.  This may lead to fermion wave
functions in the extra dimensions that are localized with profiles
that are not precisely Gaussian \cite{grossman_perez}, but here, for
technical simplicity, we assume Gaussian fermion wave function
profiles, as in \cite{as,ms}. Some early suggestions for underlying
physics that could provide a deeper explanation for the locations of
the fermion wave function centers were made in \cite{as,qlw}, but
here, in accordance with our low-energy effective field theory
approach, we shall adopt an empirical approach to these locations,
obtaining a solution for them that fits the data of quark and lepton
masses and mixing, and investigating, in particular, the consequences
in the neutrino sector.

We shall use periodic boundary conditions for each of the $n$
compactified dimensions, so that the compact $n$-dimensional space is
the $n$-torus, ${\mathbb T}^n$, i.e., the $n$-fold topological product
of circles. Consequently, the coordinate of a
point $y_f$ along the $\lambda$ axis, namely $y_{f,\lambda}$, is
defined mod $L$, i.e., $y_{f,\lambda} = y_{f,\lambda} \pm L$. Without
loss of generality, we shall define the origin in each compact
dimension to be symmetrically located, so that the range of
$y_\lambda$ is
\beq
-\frac{L}{2} < y_\lambda \le \frac{L}{2} \quad {\rm for} \ \lambda =1,...,n
\ .
\label{yrange}
\eeq
Because of the compactification on ${\mathbb T}^n$, it follows that
along each direction $\lambda$, where $1 \le \lambda \le n$, the
maximal distance between the $\lambda$ components of two points $y_f$
and $y_{f'}$ is $L/2$, i.e., ${\rm
  max}(|y_{f,\lambda}-y_{f',\lambda}|)=L/2$.  Hence, the maximal
distance between two points $y_f$ and $y_{f'}$ on the $n$-torus
${\mathbb T}^n$ is ${\rm max}(\|y_f - y_{f'}\|) = \sqrt{n} \, (L/2)$.
The Euclidean metric is used since $T^2$ is a flat Riemannian manifold.

Although we give a number of formulas abstractly for general $n$, we
focus here on the case of $n=2$ extra dimensions.  The choice of even
$n$ is a necessary (and sufficient) condition for there to exist a
matrix $\gamma_5$ with the property $\{\gamma_5,\gamma_\lambda\}=0
\ \ \forall \ \ \lambda$ and thus for there to exist right- and
left-handed chiral projection operators $P_{R,L} = (1/2)(1 \pm
\gamma_5)$ for fermion fields. A more complicated mechanism to get
chiral fermions is necessary if $n$ is odd; for example, for $n=1$,
one can compactify on the space $S/{\mathbb Z}^2$, which projects out
one chirality of fermions. The normalization factor $A_f$ is determined
by the condition that, after integration over the $n$ higher
dimensions, the four-dimensional fermion kinetic term has its
canonical normalization and correct Maxwellian (free-field) dimension.
This yields the result
\beq
A_f = \bigg ( \frac{2}{\pi} \bigg )^{n/4} \mu^{n/2} \ .
\label{af}
\eeq
Recalling that in $d=4+n$ spacetime dimensions, a fermion field has
dimension $d_f = (d-1)/2 = (3+n)/2$, one sees that the increased mass
dimention of the fermion field $\sim ({\rm mass})^{n/2}$, is
incorporated in this normalization constant, and is set by the inverse
localization length $\mu=1/\sigma$. Because the $A_f$ accounts for
this increased dimension of a fermion field in $d=4+n$ dimensions, the
remaining part of the field operator has its usual Maxwellian
dimension of 3/2 appropriate for four-dimensional spacetime.  The
fermion wave functions are assumed to be strongly localized, with
Gaussian width
\beq
\sigma \equiv \frac{1}{\mu} \ll L
\label{sigma_mu}
\eeq
at various points in the higher-dimensional space. The ratio $\sigma/L$
measures the localization size of the fermions relative to the
compactification size. As in
\cite{as,nnb02,bvd,nnblrs}, we take 
\beq
\frac{\sigma}{L} = \frac{1}{\mu L} = \frac{1}{30} \ .
\label{mul}
\eeq
We define a dimensionless length variable 
\beq
\eta = \mu y \ .
\label{eta}
\eeq
With $\mu L=30$, the range of each component of the $n$-dimensional
vector $\eta$, from Eq. (\ref{yrange}), is
\beq
-15 < \eta_\lambda \le 15 \quad {\rm for} \ \lambda =1,...,n.
\label{etarange}
\eeq
Hence, the maximal distance, in terms of this dimensionless variable $\eta$,
between any two points $\eta_f = \mu y_f$ and $\eta_{f'}=\mu y_{f'}$
on the $n$-torus is
\beq
    {\rm max}(\|\eta_f - \eta_{f'}\|) = \sqrt{n} \, \frac{\mu L}{2} \ . 
    \label{max_dist_eta}
\eeq
For the case $n=2$ on which we focus here, with the value $\mu L=30$
in Eq. (\ref{mul}), this maximal distance is ${\rm max}(\|\eta_f -
\eta_{f'}\|) = 15\sqrt{2} = 21.21$.  We choose 
\beq
\Lambda_L \simeq 100 \ {\rm TeV} \quad i.e., \quad L \simeq 2 \times 10^{-19}
\ {\rm cm} \ .
\label{lambdaL}
\eeq
With $\mu/\Lambda_L=30$, this yields
\beq
\mu \simeq 3 \times 10^3 \ {\rm TeV}, \quad i.e., \ \ 
\sigma \simeq 0.67 \times 10^{-20} \ {\rm cm}.
\label{muvalue}
\eeq
With these values, the particular models that we study are
consistent with bounds on extra dimensions from collider searches
\cite{pdg} and
from flavor-changing neutral current (FCNC) processes and precision
electroweak constraints, as will be discussed further below.  The UV
cutoff $M_*$ is taken to be much larger than any mass scale in the
models to ensure the self-consistency of the low-energy effective
field theory analysis.

Some remarks on baryon number violation are in order here.  In
\cite{mm80} an example was given of a left-right symmetric model in
four dimensions in which proton decay is absent but
neutron-antineutron oscillations can occur at observable levels. For
some additional early works on neutron oscillation, see
\cite{glashow}-\cite{nnb84}.  In \cite{as} it was observed that in
split-fermion models, it is easy to suppress proton and bound neutron
decays well below experimental limits by separating quark and lepton
wave function centers in the extra dimensions.  Ref. \cite{nnb02}
showed that this does not suppress neutron-antineutron oscillations,
which can occur at levels comparable to existing limits.  This was
studied further in \cite{bvd,nnblrs}; recent general reviews include
\cite{nnbwhite,nnb2020}.

 We note that the split-fermion models considered here are quite
different from models in which only the gravitons propagate in these
dimensions (e.g., \cite{a}-\cite{comp}). One may recall that for these
latter models, the fundamental Planck mass in $4+n$ dimensions,
denoted $M_{Pl,4+n}$, is related to the observed Planck mass in
four-dimensional spacetime, $M_{Pl}=(G_N)^{-1/2}=1.2 \times 10^{19}$
GeV, by $M_{Pl}^2 = M_{Pl,4+n}^2(M_{Pl,4+n}r_n)^n$, where $r_n$
denotes the compactification radius.  In the models in
\cite{a}-\cite{comp}, the fundamental Planck mass scale $M_{Pl,4+n}$
of quantum gravity in the higher-dimensional spacetime could be much
less than $M_{Pl}$ if $r_n$ is much larger than the Planck length; for
example, for the illustrative case $n=2$, the value $M_{Pl,4+n} \equiv
M_{Pl,6} = 30$ TeV corresponds to $r_n \equiv r_2 = 2.7 \times
10^{-4}$ cm. This is obviously a much larger compactification size
than the size $L \simeq 2 \times 10^{-19}$ cm in the models used here.
For the models of Refs. \cite{a}-\cite{comp}, a mechanism was
suggested to account for light neutrino masses which hypothesized a
SM-singlet fermion in the ``bulk'', with an exponentially small
overlap integral with the left-handed weak isodoublet neutrinos on the
``brane'', producing small Dirac neutino masses
\cite{ddgnu,addmnu,mnpl,mpl} (see also
\cite{dienes_hossenfelder}).  It should be noted our present framework
is also different from the model considered in \cite{acd}, in which SM
fields propagate in the extra dimensions, but without strong
localization of fermion wave functions.

For  integrals of products of purely fermion fields, although the range of
integration over each of the $n$ coordinates of a vector $y$ is from
$-L/2$ to $L/2$, the strong localization of each fermion field in the Gaussian
form (\ref{gaussian}) with $\sigma \ll L$ means that
the integral is very well approximated
by the result that would be obtained by extending the range of integration
to the interval $(-\infty,\infty)$:
$\int_{-L/2}^{L/2} \cdots \int_{-L/2}^{L/2} d^n y \to
\int_{-\infty}^\infty \cdots \int_{-\infty}^\infty  d^n y$ for
integrands of operator products consisting of fermion fields. As in
earlier work \cite{as,ms,nnb02,bvd,nnblrs}, we shall
use this approximation. In general, we
denote the integration over the extra dimensions with the concise
notation $\int d^n y...$ or $\int d^n \eta...$ in terms of the dimensionless
coordinates $\eta$, where the dots represent the integrands. 
A general integral formula that we use in this case
is (cf. Eq. (A2) in \cite{bvd}) is
\begin{widetext}
\beq
\int d^n \eta \, \exp
\Big [-\sum_{i=1}^m a_i\|\eta-\eta_{f_i}\|^2 \Big ]
= \bigg [ \frac{\pi}{\sum_{i=1}^m a_i} \bigg ]^{n/2} \,
\exp\Bigg [ \frac{-\sum_{j,k=1; \ j < k}^m \, a_j a_k
\|\eta_{f_j}-\eta_{f_k}\|^2}{\sum_{s=1}^m a_s} \Bigg ] \ .
\label{intform}
\eeq
\end{widetext}
As is implicit in Eq. (\ref{intform}), if just one type of field is
involved, so that $m=1$, then the exponential factor is absent.  The
presence of these exponential suppression factors arising from the
integration of various operators over the extra dimensions gives rise
to a number of general properties in the split-fermion models,
including the ability to account for the hierarchy in the spectrum
of SM quarks and charged leptons, the ability to strongly
suppress baryon-number-violating nucleon decays, but also 
an exponential sensitivity to the distances between fermion wave
function centers.

For a given process involving fermions, one part of the analysis involves
terms in an effective Lagrangian in four spacetime dimensions containing
a certain set of $k$-fermion operators, indexed by a subscript $r$,
\beq
{\cal L}_{eff}(x) = \sum_r c_{r,(k)} {\cal O}_{r,(k)}(x) + h.c. \ ,
\label{leff}
\eeq
where the $c_{r,(k)}$ are coefficients. 
The corresponding effective Lagrangian in the
$d=(4+n)$-dimensional space is 
\beq
{\cal L}_{eff,4+n}(x,y) = \sum_r \kappa_{r,(k)} O_{r,(k)}(x,y) + h.c. \ .
\label{leff_higherdim}
\eeq
As a $k$-fold product of fermion fields in $d=4+n$
spacetime dimensions, $O_{r,(k)}(x,y)$ has 
Maxwellian (free-field) mass dimension
$k(d-1)/2 = k(3+n)/2$, and hence, the coefficient
$\kappa_{r,(k)}$ has mass dimension
\beqs
{\rm dim}(\kappa_{r,(k)}) &=& d-k\Big ( \frac{d-1}{2} \Big ) \cr\cr
                    &=& 4+n - k \Big ( \frac{3+n}{2} \Big ) \ .
\label{dim_kappa}
\eeqs
It is useful to write the coefficients
$\kappa_{r,(k)}$ in a form that shows this dimensionality explicitly.
Denoting the mass scale characterizing the physics responsible for this
process in the $d=4+n$ space as $M$, we thus write 
\beq
\kappa_{r,(k)} = \frac{\bar\kappa_{r,(k)}}
{M^{(k(3+n)/2)-4-n}} \ ,
\label{kappagen}
\eeq
where $\bar\kappa_{r,(k)}$ is dimensionless. 
The combination of the normalization factors for a $k$-fold product
of fermion fields and the factor from the integration yields an overall
factor denoted $b_k$ in (Eq.(2.29) of) \cite{bvd},
\beqs
&& b_k = A_f^k \, \mu^{-n}\Big ( \frac{\pi}{k} \Big )^{n/2} \cr\cr
&&=\Big [ 2^{k/4} \, \pi^{-(k-2)/4} \, k^{-1/2} \, \mu^{(k-2)/2} \Big ]^n \
.
\label{bk}
\eeqs
Note that $b_2=1$ to guarantee canonical normalization of a free-field
fermion bilinear operator product in $d=4$ dimensions after the
integration over the extra dimensions.  Thus, the integral of an
operator $O_r$ consisting of a $k$-fold product of fermion fields has
the generic form
\beq
I_{r,(k)} = b_k \, e^{-S_{r,(k)}} \ ,
\label{irgen}
\eeq
where $e^{-S_{r,(k)}}$ is the exponential factor in Eq. (\ref{intform}).
The resultant coefficient in the low-energy
effective four-dimensional Lagrangian was given (as Eq. (2.30)) in
Ref. \cite{bvd} and is 
\begin{widetext}
\beq
c_{r,(k)} = \kappa_{r,(k)} I_{r,(k)}
 =  \frac{\bar\kappa_{r,(k)}}{M^{(3k-8)/2} } 
\Big ( \frac{\mu}{M} \Big ) ^{(k-2)n/2} 
\bigg ( \frac{2^{k/4}}{\pi^{(k-2)/4} \, k^{1/2} } \bigg )^n 
e^{-S_{r,(k)}}
\ .
\label{crgen}
\eeq
\end{widetext}
In previous studies of baryon-number-violating processes including
$\Delta B=-1$ nucleon decay and $|\Delta B|=2$ $n-\bar n$ oscillations 
\cite{nnb02,bvd,nnblrs}, we have denoted $M$ as $M_{Nd}$ or $M_{n \bar n}$.
In many of our calculations here, the mass $M$ will be set by $\Lambda_L$.
In applications where the number $k$ of
fermions in the $k$-fermion operator products is obvious, we shall
sometimes suppress this in the notation.

Concerning normalizations of gauge and Higgs fields in the extra-dimensional
framework, we recall that the Maxwellian mass dimension of a boson field in
$d=4+n$ spacetime dimensions is $d_b = (d-2)/2 = 1+(n/2)$.  Given
that boson fields have support on the compact domain $-L/2$ to $L/2$ in
each of the $n$ extra dimensions, the additional increment of $n/2$ in the
mass dimension of the boson field is incorporated in the normalization factor 
\beq
A_{bos.} = \frac{1}{L^{n/2}} \ ,
\label{ab}
\eeq
This factor guarantees that after the integration of quadratic
free-field products of boson fields over the $n$ higher dimensions,
the resulting terms have their canonical normalization in four
dimensions.  Since a gauge field-strength tensor $F^{\lambda\rho}$ has
dimension $d_F = 1+d_{bos.} = d/2 = 2+(n/2)$, there is a normalization
factor
\beq
A_F=A_{bos.} = \frac{1}{L^{n/2}}
\label{afmunu}
\eeq
accompanying each power of $F^{\lambda\rho}$ in an operator product,
in particular, for the free gauge action
$-(1/4)F_{\lambda\rho}F^{\lambda\rho}$. With the dimensionful normalization
constants $A_{bos.}$ and $A_F$ extracted, the rest of the boson fields and
gauge field strength tensor have the respective mass dimensions that they would
have in four spacetime dimensions. Regarding gauge interactions, 
we also recall that a generic gauge coupling $g$ has mass
dimension ${\rm dim}(g)=(4-d)/2 = -n/2$, and again, this is incorporated
in the normalization constant $L^{n/2}$ that enters in a gauge coupling
appearing in an expression in $d=4+n$ dimensions by writing
\beq
g_{4+n} = g \, L^{n/2} = \frac{g}{(\Lambda_L)^{n/2}} \ ,
\label{gu}
\eeq
where $g$ is dimensionless.


\section{Gauge and Higgs Sectors}
\label{gauge_group_section}

We shall consider two gauge groups and corresponding sets of fields for
our study. In accordance with our low-energy effective field theory
framework, we do not attempt to specify the physics and associated
symmetries at scales much larger than $\mu$. 

The first of these is the Standard-Model gauge group, $G_{SM}$.  We
denote the quark and lepton fields as $Q^{i\alpha}_{a,L}$,
$u^\alpha_{a,R}$, and $d^\alpha_{a,R}$, where $\alpha, \beta,..$ are
SU(3)$_c$ color indices, $i,j...$ are SU(2)$_L$ indices, and $a,..$
are generation indices.  Thus, $Q^\alpha_{1,L} = {u^\alpha \choose
  d^\alpha}_L$, $d^\alpha_{1,R}=d^\alpha_R$,
$d^\alpha_{3,R}=b^\alpha_R$, etc.  The lepton fields are denoted
$L_{a,L}={\nu_{\ell_a} \choose \ell_a}_L$ and $\ell_{a,R}$ with
$\ell_{1,R}=e_R$, $\ell_{2,R}=\mu_R$, etc.  Extending the original SM,
we also include electroweak-singlet neutrinos $\nu_{s,R}$ and take the
range of $s$ to be $s=1,2,3$ to match the number of SM fermion
generations. The Higgs field is denoted $\phi = {\phi^+ \choose
  \phi^0}$, and the vacuum expectation value (VEV) of the lowest KK
mode of this field in the low-energy four-dimensional theory is
denoted $\langle \phi \rangle_0 = {0 \choose v/\sqrt{2}}$, where
$v=246$ GeV, with $G_F/\sqrt{2} = g^2/(8m_W^2) = 1/(2v^2)$.  Here,
$G_F=1.1664 \times 10^{-5}$ GeV$^{-2}$ is the Fermi weak coupling.
This VEV sets the scale of electroweak symmetry breaking (EWSB), i.e.,
the breaking of the ${\rm SU}(2)_L \otimes {\rm U}(1)_Y$ part of
$G_{SM}$ to ${\rm U}(1)_{em}$. An extension of the SM gauge symmetry
with a gauged U(1)$_{B-L}$ symmetry to avoid excessively large
left-handed Majorana neutrino masses will be discussed below. An
additional extension with the addition of a candidate dark matter
fermion will also be discussed. 

A second gauge theory of considerable interest is the left-right
symmetric theory, with gauge group $G_{LRS}$ given in
Eq. (\ref{glrs}). Among of the appeals of this theory is the elegant
relation for the electric charge, $Q_{em}=T_{3L}+T_{3R}+(B-L)/2$
\cite{ps}. The gauge fields for the SU(2)$_L$ and SU(2)$_R$ factor
groups in $G_{LRS}$ are denoted ${\vec A}_{\lambda,L}$ and${\vec
  A}_{\lambda,R}$, respectively, and the gauge field for the
U(1)$_{B-L}$ group is denoted $U_\lambda$.  The quarks and leptons of
each generation transform as
\beq
Q^\alpha_{a,L}: \ (3,2,1)_{1/3} \ , \quad Q^\alpha_{a,R}: \ (3,1,2)_{1/3}
\label{lrs_quarks}
\eeq
and
\beq
L_{\ell_a,L}: \ (1,2,1)_{-1} \ , \quad L_{\ell_a,R}: \ (1,1,2)_{-1} \ ,
\label{lrs_leptons}
\eeq
where the numbers in the parentheses are the dimensionalities of the
representations under the three non-Abelian factor groups in $G_{LRS}$ and the
numbers in subscripts are the values of $B-L$. 
The explicit lepton field are
\beq
L_{\ell_a,L} = {\nu_{\ell_a} \choose \ell_a}_L \ , \quad
L_{\ell_a,R} = {\nu_{\ell_a} \choose \ell_a}_R \ ,
\label{explicitleptons}
\eeq
where $\ell_1=e$, $\ell_2=\mu$, and $\ell_3=\tau$.
We denote SU(2)$_L$ and SU(2)$_R$ gauge
indices as Roman indices $i,j..$ and primed Roman indices $i',j'...$,
respectively, so, e.g., $L^i_{1,L}= \nu_{e,L}$ for $i=1$ and
$L^{i'}_{2,R}= \mu_R$ for $i'=2$.  An extension of the fermion sector of
the LRS model to include a possible dark matter particle will be
discussed below.

The Higgs sector contains a Higgs field $\Phi$ transforming as
$(1,2,2)_0$, which can be written as $\Phi^{i j'}$, or equivalently,
in matrix form, as
\beq
\Phi = \left( \begin{array}{cc}
    \phi_1^0 & \phi_1^+ \\
    \phi_2^- & \phi_2^0 \end{array} \right ) \ .
\label{phi}
\eeq
The Higgs sector also contains two Higgs fields, commonly denoted $\Delta_L$
and $\Delta_R$, which transform as $(1,3,1)_2$ and $(1,1,3)_2$, respectively.
Since the adjoint representation of SU(2) is equivalent to the symmetric rank-2
tensor representation, these may be written as $(\Delta_L)^{ij} =
(\Delta_L)^{ji}$ and $(\Delta_R)^{i'j'} = (\Delta_R)^{j'i'}$ or,
alternatively, as (traceless) matrices:
\beq
\Delta_h= \left( \begin{array}{cc}
    \Delta_h^+/\sqrt{2} & \Delta_h^{++} \\
    \Delta_h^0 & -\Delta_h^+/\sqrt{2} \end{array} \right ) \ ,
\quad h=L, \ R .
\label{Delta}
\eeq

The minimization of the Higgs potential to produce vacuum expectation
values (VEVs) has been analyzed in a number of studies
\cite{lrs81},\cite{ggmko}-\cite{chauhan}.  With
appropriate choices of parameters in the Higgs potential, this
minimization yields the following vacuum expectation values (VEVs) of
the lowest KK modes of the Higgs fields, expressed in terms in the
four-dimensional Lagrangian:
\beq
\langle \Phi \rangle_0 = \frac{1}{\sqrt{2}} \left( \begin{array}{cc}
    \kappa_1 &  0 \\
     0       & \kappa_2 e^{i\theta_\Phi} \end{array} \right ) \ ,
\label{phivev}
\eeq
\beq
\langle \Delta_L \rangle_0 = \frac{1}{\sqrt{2}} \left( \begin{array}{cc}
     0       &  0 \\
     v_L e^{i\theta_{\Delta}} & 0 \end{array} \right ) \
\label{deltaLvev}
\eeq
and
\beq
\langle \Delta_R \rangle_0 = \frac{1}{\sqrt{2}} \left( \begin{array}{cc}
            0   &  0 \\
            v_R &  0  \end{array} \right ) \ .
\label{deltaRvev}
\eeq

The spontaneous symmetry breaking of the $G_{LRS}$ gauge symmetry
occurs in several stages.  At the highest-mass stage, $\Delta_R$ picks
up a VEV, thereby breaking the ${\rm SU}(2)_R \otimes {\rm
  U}(1)_{B-L}$ subgroup of $G_{LRS}$ to U(1)$_Y$, where $Y$ denotes
the weak hypercharge, i.e., ${\rm SU}(2)_R \otimes {\rm U}(1)_{B-L}
\to {\rm U}(1)_Y$.  This gives the $W_R$ a large mass, which, to
leading order, is $m_{W_R} = g_Rv_R/\sqrt{2}$. The second stage of
symmetry breaking,
${\rm SU}(2)_L \otimes {\rm U}(1)_Y \to {\rm U}(1)_{em}$,
occurs at a lower scale and results from the the VEVs
of the $\Phi$ field.  This produces a mass $m_{W_L} = g_L v/2$, where
\beq
v = \sqrt{\kappa_1^2 + \kappa_2^2} = 246 \ {\rm GeV}
\label{v_in_lrs}
\eeq
is the electroweak symmetry breaking (EWSB) scale.  The neutral gauge
fields $A_{3L}$, $A_{3R}$, and $U$ mix to form the photon, the $Z$,
and a much more massive $Z'$.  Since the VEV $v_L$ of the SU(2)$_L$
Higgs triplet $\Delta_L$ would modify the successful tree-level
relation $\rho=1$, where $\rho = m_W^2/(m_Z^2\cos^2\theta_W) = 1$
(where $\theta_W$ is the weak mixing angle), one arranges the
parameters in the Higgs potential so that $v_L \ll \kappa_{1,2}$.  
The non-observation of any $W_R$ from direct searches
at the Large Hadron Collider sets a lower limit of 4.4 TeV on the
$W_R^\pm$ mass from the CMS experiment~\cite{CMS} and 4.7 TeV from the
ATLAS experiment~\cite{ATLAS}. These lower limits are accommodated by
making $v_R \gg v$. There are theoretical lower limits on the extra
neutral Higgs boson from FCNC contributions~\cite{FCNC_higgs} of about
10 to 15 TeV. There are also lower limits in the TeV range for the
singly and doubly charged $\Delta_{R, L}$ from collider data
\cite{dmz,pdg}.  While the masses of the neutral and charged components of
the $\Delta_R$ Higgs field can be comparable to $v_R$,
one requires that $v_L$ must be much smaller than the masses of the
components of $\Delta_L$. A mechanism that could produce this hierarchy
was presented in \cite{cmp}.

In general, there is mixing of the interaction eigenstates
$A^\pm_{\lambda,L}$ with $A^\pm_{\lambda,R}$ to produce mass
eigenstates. For the lowest KK modes, this has the form (suppressing
the Lorentz indices)
\beq
     \left(\begin{array}{c}
       W_L^\pm  \\
       W_R^\pm  \end{array}\right) = 
        \left(\begin{array}{cc}
         \cos\zeta & e^{i \omega} \sin\zeta \\
          -e^{-i\omega} \sin\zeta & \cos\zeta \end{array}\right) \,
      \left(\begin{array}{c}
       A_L^\pm  \\
       A_R^\pm  \end{array}\right) \ , \  
\label{WL_WR_rel}
\eeq
where the angle $\zeta$ is given by
\beq
\tan\zeta = \frac{\kappa_1 \kappa_2}{\kappa_1^2 + \kappa_2^2 + 8v_R^2} \ .
\label{tanzeta}
\eeq
Because $v_R \gg {\rm max}(\kappa_1, \ \kappa_2)$, the mixing angle
$|\zeta| \ll 1$, so this mixing is very small.  This is true in the
four-dimensional LRS theory without any reference to possible BSM
extra-dimensional models.  Indeed, in the LRS split-fermion model
there is an additional constraint; in order for the rate of $n-\bar n$
oscillations to be in agreement with the experimental upper limit, it
is necessary that $v_R \gsim 10^6$ GeV \cite{nnblrs}. Hence
Eq. (\ref{tanzeta}) gives $|\zeta| \lsim 3 \times 10^{-8}$, so this
mixing is negligibly small, and
$W^\pm_L = A^\pm_L$ and $W^\pm_R = A^\pm_R$ to very good accuracy.

Since the $\Delta_R$ has $B-L$ charge of 2, its VEV, $v_R$, breaks
$B-L$ by two units.  As was pointed out in \cite{mm80,lrs81} (in the
usual $d=4$ spacetime context), this provides a natural explanation
for small neutrino masses via the Yukawa interaction 
\beq
-{\cal L}_{\nu_R,Maj} =
 \sum_{a,b} y^{(RR\Delta_R)}_{ab} \, [L^T_{a,R} C L_{b,R}]\, \Delta_R + h.c.
\label{rh_maj_lrs}
\eeq
(where the sum is over the generation indices, $1 \le a,b \le 3$)
which, via the $\Delta_R$ VEV, $v_R$, yields a seesaw mechanism
\cite{other_seesaw}).  The gauge symmetry breaking could also be
dynamical \cite{dynamical_lrs,sml}, or arise because of different
boundary conditions, but here we assume a conventional
Higgs mechanism for this symmetry breaking.

Because of the compactification, the gauge and Higgs fields have KK
mode expansions (equivalent to Fourier expansions). Since the fermions
have localized wave functions, it is necessary that the lowest KK
modes of the gauge fields and Higgs field(s) are constants in the
extra dimensions, in order to agree with experimental data on
universality of the couplings of gauge fields to these fermions and to
guarantee that, after electroweak symmetry breaking, the resultant
vector boson masses are the same throughout the extra dimensions.

Because of the compactification, the gauge and Higgs fields have KK
mode expansions (equivalent to Fourier expansions). Since the fermions
have localized wave functions, it is necessary that the lowest KK
modes of the gauge fields and Higgs field(s) are constants in the
extra dimensions, in order to agree with experimental data on
universality of the couplings of gauge fields to these fermions and to
guarantee that, after electroweak symmetry breaking, the resultant
vector boson masses are the same throughout the extra dimensions.  The
effects of higher-lying KK modes of the gauge and Higgs fields have
been studied in a number of works (e.g.
\cite{dpq2000,kaplan_tait,ng_split1,ng_split2,grossman_perez,abel,hewett_split,rm_fcnc}). These
are discussed further below. The compactification that was commonly
used in previous works with $n=1$ extra dimension was such that the
extra-dimensional space was $S^1/{\mathbb Z}_2$, which, in addition to
removing one chirality of fermions, had the effect of reducing the KK
expansion to a sum of cosine term.  Because we use a simple toroidal
compactification, in our case the KK expansion for a generic Higgs
field, denoted as $\Phi$, has the form
\beq
\Phi(x,y) = \frac{1}{L^{n/2}} \sum_{m \in {\mathbb Z}^n} \Phi^{(m)}(x) \, 
\exp\Big [ \frac{2\pi i (m \cdot y)}{L} \Big ] \ , 
\label{phi_kk}
\eeq
where $m=(m_1,...,m_n)$ is an integer-valued vector in ${\mathbb Z}^n$
and $m \cdot y = \sum_{i=1}^n m_i y_i$ is the Euclidean scalar product
of the vectors $m$ and $y$ in these extra dimensions. As with the use
of complex exponentials in electrodynamic, it is understood that real
parts are taken in calculations involving KK expansions of the form
(\ref{phi_kk}) and (\ref{v_kk}) to obtain results for fields that are
real.  In a similar manner, a generic gauge field, denoted
$V_\lambda$ (suppressing non-Abelian group indices where present) has
the KK expansion
\beq
V_\lambda(x,y) = \frac{1}{L^{n/2}} \sum_{m \in {\mathbb Z}^n}
V^{(m)}_\lambda(x) \, \exp\Big [ \frac{2\pi i (m \cdot y)}{L} \Big ]  \ . 
\label{v_kk}
\eeq
We use these expansions for $n=2$. An $m$'th KK mode of a gauge or
Higgs field has an excitation energy, relative to the lowest KK mode,
of $2\pi \|m\|/L = 2 \pi \| m \| \Lambda_L$.  In contrast, because of
the effective localization of a fermion field to a distance $\sim
\sigma = 1/\mu$, the $m$'th KK mode of a fermion field has an
exitation energy $\propto \| m \|/\sigma$.  Since $\mu \gg
\Lambda_L$, the KK modes for fermions lie much higher in energy than
the KK modes for bosons, and in our low-energy effective field theory
approach, we thus neglect them, as in previous studies (e.g.,
\cite{ng_split1,ng_split2,hewett_split}).


\section{Masses and Mixing for Quarks and Charged Leptons}
\label{yukawa_section}


Although our focus here is on neutrino masses and mixing, it is also
necessary to give some analysis of the quark sector of the models.  We
divide this section into two parts, corresponding to the split-fermion
models with SM and LRS gauge symmetries, respectively.  In the
following, for notational simplicity, we shall often write Lagrangians
with normalization factors implicit in the fields.


\subsection{SM Split-Fermion Model}

The Yukawa terms in the Lagrangian in $4+n$ dimensions for the quarks
in the split-fermion model with a SM gauge group describing the
physics at the scale $\mu$ are
\beqs
-{\cal L}_{Yuk,q}(x,y) &=&
\sum_{a,b} y^{(d)}_{ab}[\bar Q_{a,L}(x,y) d_{b,R}(x,y)]\phi(x,y) \cr\cr
&+&
\sum_{a,b} y^{(u)}_{ab}[\bar Q_{a,L}(x,y) u_{a,R}(x,y)]\tilde\phi(x,y)
\cr\cr
&+& h.c.,  
\label{quark_yukawa_higherdim}
\eeqs
where
\beq
\tilde \phi = i \tau_2 \phi^* \ , 
\label{phi_tilde}
\eeq
$\tau_2$ is the SU(2) Pauli matrix, and, as before, $a,b$ are generation
indices.  

Taking into account that the lowest KK mode of the Higgs field is a
constant as a function of the extra dimensions, extracting the terms
resulting from the Higgs VEV, and performing the integration over
these extra dimensions, one thus obtains the low-energy effective
Lagrangian in $d=4$ dimensions for the quark mass matrices in the
charge $Q=2/3$ ($u$-type) and $Q=-1/3$ ($d$-type) sectors.  The
integration over the extra $n$ dimensions of a given fermion bilinear
operator product $[\bar f_L(x,y) f_R(x,y)]$ in a Yukawa interaction
involves the integral (from the $m=2$ special case of Eq.
(\ref{intform}), including the normalization factor $A_f$ in (\ref{af})):
\beqs
&& A_f^2 \, \int d^n y \, e^{-\|\eta-\eta_{f_L}\|^2- \|\eta-\eta_{f_R}\|^2}
\cr\cr 
&=& \exp \Big [ -\frac{1}{2}\|\eta_{f_L} - \eta_{f_R}\|^2 \Big ] \ . 
\label{mint}
\eeqs
One obtains 
\beqs
-{\cal L}_{q,mass} &=&
\frac{v}{\sqrt{2}}\sum_{a,b}
y^{(u)}_{ab} [\bar u_{a,L} u_{b,R}] \,
e^{-(1/2)\|\eta_{Q_{a,L}}-\eta_{u_{b,R}}\|^2} \cr\cr
&+&
\frac{v}{\sqrt{2}}\sum_{a,b}
y^{(d)}_{ab} [\bar d_{a,L} d_{a,R}] \,
e^{-(1/2)\|\eta_{Q_{a,L}}-\eta_{d_{b,R}}\|^2}  \cr\cr
&+& h.c. \crcr
&=& \sum_{a,b} \sum_{q=d,u} [\bar q_{a,L} M^{(q)}_{ab} q_{b,R}] + h.c.,
\label{quark_yukawa}
\eeqs
where 
\beq
M^{(q)}_{ab} = \frac{v}{\sqrt{2}} y^{(q)}_{ab} 
e^{-(1/2)\|\eta_{Q_{a,L}}-\eta_{q_{b,R}}\|^2}  \ , \quad q=u,d \ .
\label{quark_mass_matrices}
\eeq
The corresponding Yukawa couplings and integration over extra dimensions
for the charged leptons yields the mass matrices terms
\beq
M^{(\ell)}_{ab} = \frac{v}{\sqrt{2}} y^{(\ell)}_{ab} 
e^{-(1/2)\|\eta_{L_{a,L}}-\eta_{\ell_{b,R}}\|^2}  \ . 
\label{lepton_mass_matrices}
\eeq

The Cabibbo-Kobayashi-Maskawa (CKM) quark mixing matrix $V$
has a hierarchical form, with off-diagonal entries that are
smaller in magnitude than diagonal entries and become smaller as one
moves further away from the diagonal. Hence, one may begin by
neglecting these off-diagonal entries and solving for the relevant
separation distances in the absence of quark mixing, and then take
into account this mixing.  In this approximation, for the generation
$a$ quark in the $Q=2/3$ $(u)$ and $Q=-1/3$ $(d)$ quark sectors (with
$u_1 \equiv u$, $u_2 \equiv c$, $u_3 \equiv t$; $d_1 \equiv d$,
$d_2=s$, $d_3=b$), one obtains $m_{q_a} = M^{(q)}_{aa}$, where
$M^{(q)}_{ab}$ was given in Eq. (\ref{quark_mass_matrices}).
Equivalently, one has, for the separation distance $\|\eta_{Q_{a,L}} -
\eta_{q_{a,R}} \|$, the relation
\beq
\|\eta_{Q_{a,L}} - \eta_{q_{a,R}}\| = \bigg [ 2\ln\bigg ( 
\frac{y^{(q)}_{aa} v}{\sqrt{2} \, m_{q_a}} \bigg ) \bigg ]^{1/2} \ . 
\label{mqa_distance_constraint}
\eeq
Analogously, for the charged leptons,
\beq
\|\eta_{L_{a,L}} - \eta_{\ell_{a,R}}\| = \bigg [ 2\ln\bigg ( 
\frac{y^{(\ell)}_{aa} v}{\sqrt{2} \, m_{\ell_a}} \bigg ) \bigg ]^{1/2} \ . 
\label{mella_distance_constraint}
\eeq
Since the generation of the quark and charged lepton masses occurs at
the electroweak symmetry breaking, one uses the running masses
evaluated at this scale in these equations.  As noted, a major
achievement of these split-fermion models was that they could explain
the large hierarchy in the values of quark and charged lepton masses
with roughly equal dimensionless Yukawa couplings for different
generations, by the choices of the locations of respective centers of
wave functions of the chiral components of these fields in the extra
dimensions \cite{as,ms}.

As in \cite{ms}, we shall choose the locations of lepton wave function
centers so that the charged lepton mass matrix is diagonal up to small
corrections.  While Ref. \cite{ms} also chose the locations of the
$Q=2/3$ quark wave function centers so as to render the up-type quark
mass matrix diagonal, up to small corrections, here we shall carry out
this procedure for the down-quark, instead of up-quark, wave function
centers, so as to make the down-quark mass matrix diagonal, up to
small corrections.  This greatly suppresses FCNC effects due to higher
KK modes of gauge fields
\cite{dpq2000,kaplan_tait,ng_split1,ng_split2,grossman_perez,abel,hewett_split,rm_fcnc},
as discussed in Appendix \ref{kk_appendix}.  Our choice of arranging
down-type quark wave function centers so as to render the $Q=-1/3$
mass matrix nearly diagonal is made to satisfy the particularly
stringent constraints on FCNC effects in $K^0 - \bar K^0$ and $B^0 -
\bar B^0$ mixing. Since we use a low-energy effective field theory
approach, we may leave a deeper explanation of these choices of wave
function centers of charged leptons and down-type quarks to future
work on an ultraviolet completion of the theory; however, the
necessity of this strategem of engineering the charged-lepton and
down-quark mass matrices to be nearly diagonal may be regarded ia a
weakness in these split-fermion models.

Using as inputs the charged lepton masses evaluated at $m_Z$ from
\cite{koide} in Eq.  (\ref{mella_distance_constraint}), we obtain the
distances
\beq
\|\eta_{L_{1,L}}-\eta_{\ell_{1,R}}\|=5.06 
\label{distance_L1left_eright}
\eeq
\beq
\|\eta_{L_{2,L}}-\eta_{\ell_{2,R}}\|=3.86
\label{distance_L2left_muright}
\eeq
and
\beq
\|\eta_{L_{3,L}}-\eta_{\ell_{3,R}}\|=3.03 \ .
\label{distance_L3left_tauright}
\eeq
%


\subsection{LRS Split-Fermion Model}

Here we discuss the Yukawa terms and resultant mass terms for quarks and
charged leptons in the extra-dimensional LRS model. The neutrino sector
will be analyzed in the next section. The quark terms are
\beqs
 -{\cal L}_{Yuk,q,LRS} &=& \sum_{a,b} [ \bar Q_{a,L} (y^{(q)}_{ab} \Phi +
   h^{(q)}_{ab} \tilde \Phi) Q_{b,R}]  + h.c. \ , \cr\cr
 &&
\label{yukterm_lrs}
\eeqs
where $\tilde \Phi = \sigma_2 \Phi^* \sigma_2$, and here
$y^{(q)}_{ab}$ and $h^{(q)}_{ab}$ are matrices of Yukawa couplings.
Inserting the VEV of (the lowest KK mode of) $\Phi$ from
Eq. (\ref{phivev}) and performing the integration, over the extra
dimensions, of the quark bilinears gives the mass terms
\beqs
&& \frac{1}{\sqrt{2}}\sum_{a,b}
[\bar u_{a,L}( y^{(q)}_{ab} \kappa_1 +
h^{(q)}_{ab} \kappa_2 e^{i \theta_\Phi}) u_{b,R}]\, e^{-S_{yQ,ab}}
+ \cr\cr
&&
\frac{1}{\sqrt{2}}\sum_{a,b}
[\bar d_{a,L}( y^{(q)}_{ab} \kappa_2 e^{-i\theta_\Phi}
  + h^{(q)}_{ab} \kappa_1)d_{b,R}] \, e^{-S_{yQ,ab}} + h.c., \cr\cr
&&
\label{quarkmasses_lrs}
\eeqs
where
\beq
S_{yQ,ab} =
\frac{1}{2}\|\eta_{Q_{a,L}} - \eta_{Q_{b,R}} \|^2 \ .
\label{sq}
\eeq
Note that even if one imposes left-right symmetry at a high scale in the UV,
this symmetry is broken at the scale $v_R \gg \kappa_1, \kappa_2$, so that
at the electroweak scale, $\eta_{Q_{a,L}}$ and $\eta_{Q_{a,R}}$ are 
different from each other. For illustrative purposes, let us neglect
the small off-diagonal terms in these mass
matrices.  We obtain two relations for the relevant
separation distances, namely 
\beqs
&& \|\eta_{Q_{a,L}} - \eta_{Q_{a,R}}\| =
\Bigg [ 2 \ln \Bigg ( \frac{| y^{(q)}_{aa} \kappa_1 + h^{(q)}_{aa}
    \kappa_2 e^{i\theta_\Phi}|}{\sqrt{2} \, m_{u_a}} \Bigg ) \Bigg ]^{1/2}
\cr\cr
&&
\label{qlqrdistance_mu}
\eeqs
and
\beqs
&&\|\eta_{Q_{a,L}} - \eta_{Q_{a,R}}\| =
\Bigg [ 2 \ln \Bigg ( \frac{| y^{(q)}_{aa} \kappa_2 e^{-i\theta_\Phi}
 + h^{(q)}_{aa}\kappa_1|}
  {\sqrt{2} \, m_{d_a}} \Bigg ) \Bigg ]^{1/2} \ .\cr\cr
&&
\label{qlqrdistance_md}
\eeqs
For given values of $\kappa_1$ and $\kappa_2$, the Yukawa
couplings $y^{(q)}_{aa}$ and $h^{(q)}_{aa}$, and the phase factor
$e^{i\theta_\Phi}$ can be chosen to satisfy these relations. Taking
$y_{11}^{(q)} \sim O(1)$ and $h_{11}^{(q)} \sim O(1)$ as above, and
using the values of the running quark masses $m_u$ and $m_d$ at the
EWSB scale from Ref. \cite{koide}, one can then compute a value of
$\|\eta_{Q_L}-\eta_{Q_R}\|$ that satisfies Eqs.
(\ref{qlqrdistance_mu}) and (\ref{qlqrdistance_md}).  For example,
this yields the following value for this separation distance
for the first generation:
\beq
\|\eta_{Q_{1,L}}-\eta_{Q_{1,R}}\| \simeq 4.7 \ .
\label{distance_QL_Qr}
\eeq
We use the same model-building strategy for the fermions in this LRS
model as we did for the SM split-fermion model, namely to obtain
solutions for the wave function centers of the charged leptons and
down-type quarks so as to make $M^{(d)}$ and $M^{(\ell)}$ diagonal, up
to small corrections. The reason is the same as in the SM version, namely
to avoid excessive FCNC processes due to higher KK modes of gauge and Higgs
fields. 


\section{Neutrinos in the SM Split-Fermion Model}
\label{nu_sm_section}

In this section we analyze neutrino masses and mixing in the SM split-fermion
model with $n=2$ extra dimensions.  Here, one 
expands the original lepton content with 
the addition of a number $n_s$ of electroweak-singlet
neutrinos, $\nu_{s,R}$, $s=1,...,n_s$. We shall take
$n_s=3$.  To avoid confusion with left-handed neutrinos after charge
conjugation, we set $\nu_{s,R} \equiv \omega_{s,R}$. Restricting to
renormalizable terms in the four-dimensional Lagrangian, the resultant
neutrino mass terms have the form
\beqs
-{\cal L}_{\nu, m} &=& \sum_{a,b} \bigg [
  [\bar\nu_{a,L} M^{(D)}_{ab} \omega_{b,R}] +
  [\omega_{a,R}^T C M^{(R)}_{ab} \omega_{b,R}] \bigg ] \cr\cr
&+& h.c. \ ,
\label{numass_sm}
\eeqs
where $C$ is the conjugation Dirac matrix.
Here, $M^{(D)}$ is, in general, a complex matrix and $M^{(R)}$ is, in
general, a complex symmetric matrix: $[M^{(R)}]^T = M^{(R)}$.  The
right-hand side of Eq. (\ref{numass_sm}) 
can be written compactly by defining the six-dimensional vector
$\Omega_R = (\nu^c_R,\omega_R)^T$. Then, taking into account of the fact
that $\bar \Omega^c_L = (\bar\nu_L,\bar\omega^c_L)^T$, one has
\beq
-{\cal L}_{\nu, m} =  \frac{1}{2} \,
\overline{\Omega^c}_L {\cal M} \Omega_R + h.c.
\label{numass_sm_compact}
\eeq
where
\beq
    {\cal M} = \left(\begin{array}{cc}
M^{(L)}     & M^{(D)}  \\
M^{(D) \ T} & M^{(R)} \end{array}\right) \ .
\label{compact_matrix_SM}
\eeq
Here, the $M^{(L)}$ submatrix arises from
the dimension-5 operator yielding Majorana masses for the active
neutrinos,
\beq
\sum_{a,b} \frac{c^{(LL\phi\phi)}}{\Lambda_{ab}} \, 
  (\epsilon_{ik}\epsilon_{jm} + \epsilon_{im}\epsilon_{jk})
  [L_{a,L}^{i \, T} C L^j_{b,L}] \phi^k \phi^m + h.c. \ ,
\label{lh_majorana_sm}
\eeq
where $i,j,k,m$ are SU(2)$_L$ group indices.  

In order to avoid fine tuning, one would like to have an operative seesaw
mechanism in this model, so that the neutrino mass eigenvalues split
into a heavy set with masses of order $\Lambda_L$ and a light set with
sub-eV masses.  A problem that one encounters was noted in the original
work on the model \cite{as} and can be seen in the
low-energy effective theory, even before considering the embedding in
higher dimensions.  From the VEVs of $\phi$, the dimension-5 operators
in Eq. (\ref{lh_majorana_sm}) yield Majorana mass terms of the left-handed
neutrinos 
\beq
\sum_{a,b} \frac{c^{(LL\phi\phi)}}{\Lambda_{ab}} \, (v/\sqrt{2})^2 \,
  [\nu_{a,L}^T C \nu_{b,L}] + h.c. \ ,
\label{lh_majorana_sm2}
\eeq
The natural size for $\Lambda_{ab}$ is $\Lambda_L$. For the terms that are
diagonal in generation, i.e., with $a=b$, the integration over the extra
dimensions does not yield any exponential suppression factor, so in the
low-energy effective field theory in four dimensions, these give
left-handed Majorana mass eigenvalues 
\beq
\frac{c^{(LL\phi\phi)}_{aa} \, (v/\sqrt{2})^2}{\Lambda_{aa}} \ .
\label{lh_majorana_mass3}
\eeq
In order not to spoil the seesaw, these must be smaller than the
eigenvalues arising from the diagonalization of $M_\nu$ in
Eq. (\ref{mnu_matrix}) below, the largest of which is $\simeq 0.05$ eV
(see Eq. (\ref{mminus}) in Appendix \ref{mixing_appendix}). But with
$\Lambda_L=100$ TeV, the masses in Eq. (\ref{lh_majorana_mass3}) have
magnitudes $(0.3 \ {\rm GeV})|c^{(LL\phi\phi)}_{aa}|$. Without an
artificial fine tuning $|c^{(LL\phi\phi)}_{aa}| \ll 1$, this is
much too large.  One modification of the model to deal with this
problem was suggested in Ref. \cite{as}, namely to extend the SM gauge
group $G_{SM}$ to include a gauged U(1)$_{B-L}$. A number of studies
of such U(1)$_{B-L}$ extensions of $G_{SM}$, in addition to works on
LRS models, have been carried out and bounds set on the mass of the
resulting $Z'$ (e.g.  \cite{buch,langacker,cdmz},\cite{pdg} and
references therein). The U(1)$_{B-L}$ gauge symmetry might play a role
in explaining the overall separation between the wave function centers
of the quarks and leptons in the extra dimensions \cite{as}.  We note
that at mass scales above the breaking scale for this U(1)$_{B-L}$
symmetry, it would also forbid $n-\bar n$ oscillations.

The LRS version of the split-fermion model has the advantage of being
able to suppress the left-handed Majorana mass terms for neutrinos
without requiring any extension, provided that the VEV $v_L$ of the
$\Delta_L$ Higgs is sufficiently small, namely $v_L \lsim 0.05$ eV for
Yukawa couplings of O(1). Although the masses of the components of
$\Delta_L$ are must be larger than O(TeV), this can be arranged
\cite{cmp}. Since the $G_{LRS}$ gauge symmetry must be broken to the
SM gauge symmetry at $v_R \sim 10^3$ TeV in the LRS split-fermion
model to adequately suppress $n-\bar n$ oscillations \cite{nnblrs}
(see Eqs.  \ref{vr_min})-(\ref{vr_value}) below), in the mass range
from $v_r$ down to the electroweak symmetry breaking scale $v \simeq
250$ GeV, one may analyze the physics in terms of SM fermion fields and
the relevant gauge and Higgs fields.

Thus, we proceed with our analysis of the lepton sector in the split-fermion
model.  The light neutrino masses are eigenvalues of the matrix
\beq
M_\nu = - M^{(D)} [M^{(R)}]^{-1} M^{(D) T} \ . 
\label{mnu_matrix}
\eeq
Thus, $M_\nu^T = M_\nu$, i.e., $M_\nu$ is a (complex) symmetric matrix.
We take this to be diagonalized by a unitary transformation
$U_\nu$ thus \cite{hornjohnson}:
\beq
U^{(\nu) T} M_\nu U^{(\nu)} = M_{\nu,diag.} \ . 
\label{mnu_diagonalization}
\eeq
The unitary transformation $U_\nu$ is determined by the relation
\beq
U_\nu^\dagger (M_\nu M_\nu^\dagger) U_\nu = M_{\nu,diag.}^2 \ .
\label{unu_calc}
\eeq
Note that if $M_\nu$ is transformed to $M_\nu' = M_\nu {\cal U}$, where
${\cal U}$ is unitary, then $M_\nu'$ is diagonalized by the same
$U_\nu$, since $M_\nu' [M_\nu']^\dagger = M_\nu M_\nu^\dagger$ in
Eq. (\ref{unu_calc}).

A general charged lepton mass matrix $M^{(\ell)}$, is diagonalized by
a bi-unitary transformation analogous to
Eq. (\ref{ur_m_uradjoint}) for the quarks, as follows:
\beq
U^{(\ell) \dagger}_L M^{(\ell)} U^{(\ell)}_R = M^{(\ell)}_{diag.} \ .
\label{umulep}
\eeq
The Pontecorvo-Maki-Nakagawa-Sakata (PMNS) lepton mixing matrix $U$
that enters in the charged weak current is then
given by
\beq
J_\lambda = \bar\ell_L \gamma_\lambda \nu_L =
\bar\ell_{L} U \gamma_\lambda \nu_{L} \ ,
\label{ju}
\eeq
where here $\ell_L$ and $\nu_L$ denote vectors of mass eigenstates and 
\beq
U = U^{(\ell) \dagger} U^{(\nu)} \ .
\label{uleptonic}
\eeq
With our assumption that $M^{(\ell)}$ is diagonal, it follows that
\beq
U^{(\ell)}_L = U^{(\ell)}_R = {\mathbb I} \ .
\label{uell_identity}
\eeq
The distances between left- and right-handed chiral components of charged
leptons are then fixed, with the values given in Eqs.
(\ref{distance_L1left_eright})-(\ref{distance_L3left_tauright}).  In
standard notation, $\Delta m^2_{ij} = m_{\nu_i}^2-m_{\nu_j}^2$.  The
lepton mixing matrix is given by Eq. (\ref{ugen}) in Appendix
\ref{mixing_appendix}, in terms of the angles $\theta_{12}$,
$\theta_{23}$, and $\theta_{13}$ and the CP phase $\delta$
\cite{alpha12_maj}.  The neutrino oscillation data determine values of
these angles that depend on whether the neutrino masses exhibit the
normal ordering, $m_{\nu_3} > m_{\nu_2} > m_{\nu_1}$, or the inverted
ordering, $m_{\nu_2} > m_{\nu_1} > m_{\nu_3}$ (where we have
incorporated the fact that solar neutrino data imply that $m_{\nu_2} >
m_{\nu_1}$). However, for our present purposes, the differences in the
resultant values are not large enough to be important.  A fit to
current data \cite{cggfit} yields the values
\beq
|\Delta m^2_{32}| = (2.517^{+0.026}_{-0.028}) \times 10^{-3} \ {\rm eV}^2
\label{dmsq32}
\eeq
and
\beq
\Delta m^2_{21} =  (0.742^{+0.021}_{-0.020}) \times 10^{-4} \ {\rm eV}^2 \ ,
\label{dmsq21}
\eeq
and, for the case of normal ordering, the three rotation angles and CP phase
angle (in degrees, in the standard parametrization (\ref{ugen}) )  
\beq
\theta_{23}/^\circ = 49.2^{+0.8}_{-1.2}
\label{theta23}
\eeq
\beq
\theta_{12}/^\circ = 33.44^{0.77}_{-0.74}
\label{theta12}
\eeq
\beq
\theta_{13}/^\circ = 8.57 \pm 0.12
\label{theta13}
\eeq
and
\beq
\delta/^\circ = 197^{+27}_{-24} \ . 
\label{delta_cp}
\eeq
Another recent fit yielded similar results \cite{vallefit}; a recent
review is \cite{mocioiu}.  Substituting the central values of these
angles in the leptonic mixing matrix (\ref{ugen}), one obtains
\begin{widetext}
\beqs
U &=& \left(\begin{array}{ccc}
  0.825                  & 0.545                  & -0.149 e^{(-17^\circ)i} \\
 -0.2715 e^{(-5.8^\circ)i} & 0.605 e^{(1.7^\circ)i} & 0.7485 \\
 0.495 e^{(2.8^\circ)i} & -0.581 e^{-(1.6^\circ)i} & 0.646 
\end{array}\right) \ . \cr\cr
&&
 \label{U_exp}
 \eeqs
 \end{widetext}

Although we shall use this experimentally determined lepton mixing
matrix for our analysis, a parenthetical historical remark is useful
concerning a simple approximate form for the matrix.  The fact that
$\sin^2(2\theta_{23})$ is close to 1 (maximal 2-3 mixing), i.e.,
$\theta_{23}$ is close to $\pi/4$, was evident in the first
atmospheric data analysis by the Super-Kamiokande experiment in
1998. By the early 2000s, it was also known from solar neutrino data
from the Davis, SAGE, GALLEX, Super-Kamiokande, and SNO experiments,
that $\sin^2\theta_{12} \simeq 1/3$. The data from atmospheric, solar,
and terrestrial neutrino oscillation experiments also showed that
$\theta_{13}$ was substantially smaller than $\theta_{23}$ and
$\theta_{12}$ by this time.  This motivated the suggestion \cite{hps}
that these mixing angles have a so-called tribimaximal (TBM) values
\beqs
TBM: \ && \theta_{23}=45^\circ, \quad
\theta_{12} ={\rm arcsin}\Big (\frac{1}{\sqrt{3}} \Big ) = 35.26^\circ, \cr\cr
&& \theta_{13}=0 \ .
\label{theta_tbm}
\eeqs
Substituting these into the lepton mixing matrix (\ref{uleptonic})
(with $U^{(\ell)}={\mathbb I}1$) yields the tribimaximal form 
\beqs
U &=& U_\nu = U_{TBM} = \left(\begin{array}{ccc}
    \sqrt{\frac{2}{3}}  &  \frac{1}{\sqrt{3}}  &         0 \\
   -\frac{1}{\sqrt{6}}  &  \frac{1}{\sqrt{3}}  & \frac{1}{\sqrt{2}} \\
    \frac{1}{\sqrt{6}}  & -\frac{1}{\sqrt{3}}  & \frac{1}{\sqrt{2}}
\end{array}\right) \cr\cr
&=&                     \left(\begin{array}{ccc}
    0.816  & 0.577  & 0     \\
   -0.408  & 0.577  & 0.707 \\
    0.408  &-0.577  & 0.707  \end{array}\right) \ . 
 \label{U_tbm}
\eeqs
As one can see by comparing $U$ in Eq. (\ref{U_exp}) and
(\ref{U_tbm}), the form of the lepton mixing matrix determined by
experimental measurements is moderately close to $U_{TBM}$, with the
exception of the $U_{e3} \equiv U_{13}$ element and the fact that the
$U_{TBM}$ is real. Thus, one can express a realistic lepton matrix
as a perturbation of the TBM form \cite{hycheng}. 

We proceed with our analysis, using the lepton mixing
matrix determined by the (central values of the)
experimentally measured rotation angles
and CP-violating phase in Eq. (\ref{U_exp}).  Since we take the 
charged lepton mass matrix to be diagonal, it follows that
$U^{(\ell)}={\mathbb I}$ and so $U = U_\nu$. The 
Eq. (\ref{mnu_diagonalization}) is equivalent to the relation 
\beq
U_\nu M_{\nu,diag.} U_\nu^T = M_\nu \ . 
\label{umut}
\eeq
We shall assume a hierarchical neutrino mass spectrum, i.e.,
$m_{\nu_3}^2 \gg m_{\nu_2}^2 \gg m_{\nu_1}^2$, so that, to a good
approximation,
\beq
m_{\nu_3} = \sqrt{\Delta m^2_{32}} = 5.0 \times 10^{-2} \ {\rm eV}
\label{mnu3}
\eeq
and
\beq
m_{\nu_2} = \sqrt{\Delta m^2_{21}} = 0.86 \times 10^{-2} \ {\rm eV} \ . 
\label{mnu2}
\eeq
The mass $m_{\nu_1}$ is undetermined by this procedure; for
definiteness, we shall use the illustrative value $m_{\nu_1} = 1.0
\times 10^{-3}$ eV.  We take the elements of $M_R$ to be set by the
overall mass scale inherent in the compactification, namely
$\Lambda_L$, and, for simplicity, we further assume that it is
proportional to the identity:
\beq
M_R = -r \times {\mathbb I} \ , \quad r=\Lambda_L \ .
\label{mr}
\eeq
In general, combining Eq. (\ref{mnu_matrix}) with
(\ref{mnu_diagonalization}), we can write
\beqs
M_{\nu,diag.} &=& U_\nu^T M_\nu U_\nu
= U_\nu^T (-M^{(D)} [M^{(R)}]^{-1} M^{(D) T}) U_\nu \cr\cr
&=& [ r^{-1/2} U^T_\nu M^{(D)}] [ r^{-1/2} M^{(D) T} U_\nu] \ , 
\label{mmrel}
\eeqs
so that 
\beq
M^{(D)} = r^{1/2} U_\nu [M_{\nu,diag.}]^{1/2} \ .
\label{mdgeneral}
\eeq
Evaluating this, we find the following numerical results for $M^{(D)}$,
where the entries are in units of MeV:
\begin{widetext}
\beq
M^{(D)}=  \left(\begin{array}{ccc}
        0.261                      &  0.506                     & -0.334 e^{-(17.0^\circ)i}     \\
       -0.0858 e^{-(5.82^\circ)i}  &  0.561 e^{(1.72^\circ)i}   &  1.68 \\
        0.157 e^{2.75^\circ)i}     & -0.539 e^{-(1.55^\circ)i}  &  1.45 \end{array}\right) \ .
\label{md}
\eeq
\end{widetext}
The minus signs and complex phases can be accommodated by the
requisite complex entries in the Yukawa coupling matrices. The
corresponding distances $\| \eta_{L_{a,L}} - \eta_{\nu_{b,R}} \|$
between the 
$L_{a,L}$ and $\nu_{b,R}$ wave function centers with
$1 \le a, \, b \le 3$ are then
determined from the magnitudes of these entries in $M^{(D)}$. These
distances are listed in Table \ref{L_nu_distances_table} \cite{sigfigs}.
We focus on these distances henceforth. 


%
\begin{table}
  \caption{\footnotesize{Distances $\| \eta_{L_{a,L}}-\eta_{\nu_{b,R}}
      \|$,  determined from the Dirac neutrino mass
      matrix $M^{(D)}$ in Eq. (\ref{md}).  As defined in the
      text, the numerical subscript on each fermion field is the
      generation index of the weak eigenstate, with $1 \le a,b \le 3$.}}
\begin{center}
\begin{tabular}{|c|c|c|} \hline\hline
$a$ & $b$ & $\| \eta_{L_{a,L}}-\eta_{\nu_{b,R}} \|$ \\ \hline
1 & 1 & 5.179   \\
1 & 2 & 5.050  \\
1 & 3 & 5.131  \\ \hline
2 & 1 & 5.389  \\
2 & 2 & 5.029   \\
2 & 3 & 4.807   \\ \hline
3 & 1 & 5.276   \\
3 & 2 & 5.037   \\
3 & 3 & 4.837  \\
\hline\hline
\end{tabular}
\end{center}
\label{L_nu_distances_table}
\end{table}
%


The next step in our analysis is to find a set of wave function
centers of the lepton fields that satisfies these distance
constraints. Recall that we use periodic boundary conditions for the
compactification, and with $\mu L=30$, the range of each coordinate
$\eta_\lambda$ is $-15 < \eta_\lambda \le 15$ for $1 \le \lambda \le
n$.  The full problem to solve requires one to (a) specify a set of
wave function centers for the $Q=2/3$ and $Q=-1/3$ quarks so as to
yield acceptable quark masses and the CKM quark mixing matrix; (b)
specify a set of wave function centers for the lepton fields that
yield the required form for the Dirac neutrino mass matrix $M^{(D)}$
and charged lepton mass matrix $M^{(\ell)}$ (with the Majorana mass
matrix $M^{(R)}$ in Eq. (\ref{mr})); and (c) arrange so that the wave
function centers of the quarks are sufficiently distant from those of
the leptons that baryon-number-violating nucleon decays are suppressed
enough to satisfy current experimental limits.

For our determination of lepton wave function centers, it will be
convenient to choose a coordinate system, denoted $\eta^{(\ell)}$,
whose origin is approximately in the middle of the set of these lepton
wave function centers.  Then we will carry out an analogous
calculation of quark wave function centers using a coordinate
system $\eta^{(q)}$.  For our overall assignment of locations for
centers of wave functions for the full set of quarks and leptons,
we determine translation vectors and rotation angles
of the $\eta^{(\ell)}$ and $\eta^{(q)}$ coordinate systems relative to
the $\eta$ system.  With no loss of generality, we
pick an intermediate point between the quark and lepton wave function
centers and denote this as the origin of the $\eta$ coordinate system.
Furthermore, we take both of the rotation angles to be zero, so that
the horizontal directions in the $\eta^{(\ell)}$,
$\eta^{(q)}$, and $\eta$ coordinate systems are all the same, and
similarly with the vertical directions. Anticipating our results to
be presented below, we choose these translation vectors
to be such that a quark field with coordinates
$(\eta^{(q)}_1,\eta^{(q)}_2)$ has the coordinates $(\eta_1,\eta_2)_q$
given by
\beq
(\eta_1,\eta_2) = (\eta^{(q)}_1,
                   \eta^{(q)}_2) - (8,8) \quad (\rm{for \ quarks})
\label{eta_etaq_relation}
\eeq
and a lepton field with coordinates $(\eta^{(\ell)}_1,\eta^{(\ell)}_2)$ 
has coordinates $(\eta_1,\eta_2)$ given by
\beq
(\eta_1,\eta_2) = (\eta^{(\ell)}_1,
                   \eta^{(\ell)}_2) + (5,3) \quad (\rm{for \ leptons}) \ .
\label{eta_etaell_relation}
\eeq
The overall translation between the wave function centers of the quarks
and leptons is thus in a roughly diagonal direction. The choices
of the translation vectors in Eqs. (\ref{eta_etaq_relation}) and
(\ref{eta_etaell_relation}) is made on the basis of the last step of
our analysis, namely step (c), ensuring that the distances between
quark and lepton wave function centers are large enough to produce
adequate suppression of baryon-number violating nucleon decays.

We now carry out steps (b) and (c) of the analysis.  For step (b), the
abstract mathematical problem can be stated as follows (denoting the
number of SM fermion generations as $n_{gen.}$): Let ${\mathbb T}^n$
denote an $n$-torus in which each circle $S^1_j$, $j=1,...,n$ has
circumference $c$.  Specify a set of $n_{gen.}^2$ Euclidean distances
$\|\eta_{L_{a,L}}-\eta_{\nu_{b,R}}\|$, where
$1 \le a,b \le n_{gen.}$ between the positions of the wave function
centers of the SU(2)$_L$-doublet left-handed lepton fields
$L_{a,L}$ and the SU(2)$_L$-singlet right-handed neutrino fields
$\nu_{b,R}$.  Find an actual set of points
$\eta_{L_{a,L}}$ and $\eta_{\nu_{b,R}}$, $1 \le a,b \le n_{gen.}$ in
the $n$-torus ${\mathbb T}^n$ satisfying these distance constraints.
Then, for the given set of Euclidean distances 
$\|\eta_{L_{a,L}}-\eta_{\ell_{a}}\|$ between the positions of the
SU(2)$_L$-doublet lepton wave function centers and the SU(2)$_L$-singlet
charged lepton wave function centers, with $1 \le a \le n_{gen.}$ and
with $\eta_{L_{a,L}}$ fixed from the previous calculation, find a set
of wave function centers for the right-handed charged leptons
$\ell_{a,R}$. If the embedding space were ${\mathbb R}^n$ rather than
${\mathbb T}^n$, then each one of the $n_{gen.}^2$ 
distance constraints involving $\eta_{L_{a,L}}$ and $\eta_{\nu_{b,R}}$
implies two geometric conditions, namely that (i) the point
$\eta_{\nu_{b,R}}$ must lie on the $(n-1)$-sphere $S^{n-1}$ centered at
$\eta_{L_{a,L}}$ with radius $r_{ab}=\|\eta_{L_{a,L}}-\eta_{\nu_{b,R}}\|$
and (ii) the point $\eta_{L_{a,L}}$ must lie on the $(n-1)$-sphere
centered at $\eta_{\nu_{b,R}}$ with radius $r_{ab}$.  With the positions
$\eta_{L_{a,L}}$ fixed, the second distance constraint implies the condition
that the point $\eta_{\ell_{a,R}}$ must lie an $(n-1)$ sphere centered at
$\eta_{L_{a,L}}$ with radius $\|\eta_{L_{a,L}}-\eta_{\ell_{a,R}}\|$.
Since the embedding space is ${\mathbb T}^n$ rather than ${\mathbb R}^n$,
these distances and positions are understood to be defined for this $n$-torus.
For the case of $n=2$ extra dimensions that we consider here, the
$(n-1)$-spheres are circles, $S^1$.  Depending on
$n$, $n_{gen.}$, and the specified distances, this
mathematical problem may have no solution, a unique solution, or
multiple solutions.

Let us give some pedagogical examples concerning this general problem
of choosing lepton wave function centers that satisfy the distance
constraints to reproduce the Dirac mass matrix $M^{(D)}$.  To make
these as simple as possible, we take $n=1$ for these examples, so that
the compactified space is a circle (of circumference $c$). The
simplest sub-case is $n_{gen.}=1$. Then there are three possibilities
for the distance $\|\eta_{L_{1,L}}-\eta_{\nu_{1,R}}\|$ required to fit
the Dirac matrix $M^{(D)}$ (which reduces to a scalar for
$n_{gen.}=1$): (i) if $0 < \|\eta_{L_{1,L}}-\eta_{\nu_{1,R}}\| < c/2$,
then there are two solutions, depending on whether one proceeds in a
clockwise or counterclockwise manner along the circle to get from the
point $\eta_{L_{1,L}}$ to the point $\eta_{\nu_{1}}$; (ii) in the
special case where $\|\eta_{L_{1,L}}-\eta_{\nu_{1,R}}\|=c/2$, there is
a unique solution, in which $\eta_{L_{1,L}}$ and $\eta_{\nu_{1,R}}$
are located at opposite points on the circle; and (iii) if
$\|\eta_{L_{1,L}}-\eta_{\nu_{1,R}}\| > c/2$, then there is no
solution. 

\begin{figure}
  \begin{center}
    \includegraphics[height=0.9\linewidth]{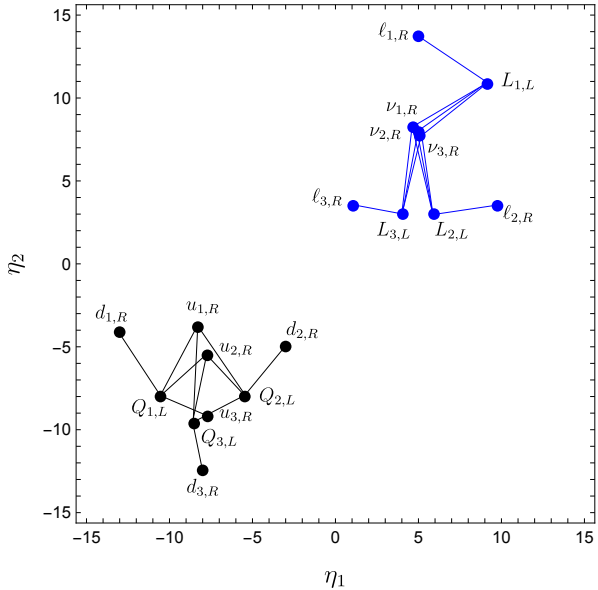}
  \end{center}
  \caption{Plot showing locations of fermion wave function centers in
    the split-fermion model with $n=2$. As defined in the text, the
    numerical subscript on each fermion field is the generation index
    of the weak eigenstate. Toroidal compactification is used, so that
    $\eta_\lambda$ is equivalent to $\eta_\lambda \pm \mu L = \eta_L
    \pm 30$. In the online figure, the lepton wave functions are
    colored blue.}
\label{fermion_centers_figure}
\end{figure}

Returning to the realistic value $n_{gen.}=3$ and the case $n=2$
considered here, we discuss the method that we use to solve for a set
of lepton wave function centers satisfying the distance constraints.
Further details on this are given in Appendix
\ref{lepton_centers_appendix}.  The $L_{2,L}$ and $L_{3,L}$ wave
functions centers are taken to lie along the horizontal
$\eta^{(\ell)}$, axis, equidistant from the vertical $\eta^{(\ell)}$
axis; that is, we set $\eta^{(\ell)}_{L_{2,L}}=(d,0)$ and
$\eta^{(\ell)}_{L_{3,L}}=(-d,0)$, where the parameter $d$ is allowed
to have either sign. From the nine
$\|\eta_{L_{a,L}}-\eta_{\nu_{b,R}}\|$ distance constraints in Table
\ref{L_nu_distances_table} we solve for the nine points
$\eta_{L_{a,L}}$, $\eta_{\nu_{b,R}}$, and $\eta_{\ell_{c,R}}$ for the
lepton wave function centers. Since the distance constraints are
nonlinear equations, they yield several solutions, all of which
produce identically the same $M^{(D)}$ and lepton mixing matrix $U$
(with $M^{(R)}$ as in Eq. (\ref{mr})). We focus on one of these
solutions for our analysis. Although this solution is not unique, it
demonstrates the ability of this model to fit observed data on
neutrino masses and mixing and also to satisfy other phenomenological
constraints.  We note that the fact that a set of solutions for lepton
wave function centers that yield the form of $M^{(D)}$ in
Eq. (\ref{md}) does not, in and of itself, guarantee that this set
also yields predictions in accord with all electroweak data, so the
fact that we find solutions that are in accord with this data is a
further achievement.  We list the results for one of our solutions to
these constraints in Table \ref{lepton_centers_table}, expressed in
the $\eta^{(\ell)}$ and $\eta$ coordinates.  In
Fig. \ref{fermion_centers_figure} we show the locations of the lepton
wave function centers graphically. With the toroidal boundary
conditions, the left edge of the figure is identified with the right
edge and the lower edge is identified with the upper edge, i.e.,
$\eta_\lambda$ is equivalent to $\eta_\lambda \pm \mu L = \eta_\lambda
\pm 30$. As discussed above, the $L_{2,L}$ and $L_{3,L}$ wave function
centers lie along the horizontal axis of the $\eta^{(\ell)}$
coordinate system defined by $\eta^{(\ell)}_2=0$, i.e., $\eta_2=3$,
spaced equidistant from the vertical axis of the $\eta^{(\ell)}$
coordinate system, defined by $\eta^{(\ell)}_1=0$, i.e.,
$\eta_1=5$. In Table \ref{lepton_distances_table} we list the
distances between the different wave function centers of the lepton
fields given in Table \ref{lepton_centers_table}.  The minimal
distance for this set of lepton wave function centers occurs between
the $L_{2,L}$ and $L_{3,L}$ fields, with
$\|\eta_{L_{2,L}}-\eta_{L_{3,L}}\| = 1.878$. With our procedure for
determining locations for lepton wave function centers, we find that
property that one pair of SU(2)$_L$-doublet leptons have a relatively
small separation distance is rather general, but we do not exclude the
possibility that a viable set of lepton wave function centers exists
in which the separation distances between all pairs of lepton fields,
including in particular, these SU(2)$_L$-doublets, are larger than
this value.  Further discussion of our procedure for determining these
wave function centers is given in Appendix
\ref{lepton_centers_appendix}.


\begin{table}
  \caption{\footnotesize{Locations of lepton wave function centers,
      expressed in the $\eta^{(\ell)}$ and $\eta$ coordinate systems,
      related by the translation (\ref{eta_etaell_relation}).  As
      defined in the text, the numerical subscript on each fermion
      field is the generation index of the weak eigenstate. Toroidal
      compactification is used, so that $\eta_\lambda$ is equivalent
      to $\eta_\lambda \pm \mu L = \eta_L \pm 30$.}}
\begin{center}
\begin{tabular}{|c|c|c|} \hline\hline
field      & $(\eta^{(\ell)}_1,\eta^{(\ell)}_2)$ & $(\eta_1,\eta_2)$  \\ \hline
$L_{1,L}$    & $(4.157,7.843)$    & $(9.157,10.843)$    \\
$L_{2,L}$    & $(0.939,0.000)$    & $(5.939,3.000)$     \\
$L_{3,L}$    & $(-0.939,0.000)$   & $(4.061,3.000)$     \\ \hline
$\nu_{1,R}$  & $(-0.320,5.240)$   & $(4.680,8.240)$    \\
$\nu_{2,R}$  & $(0.0219,4.944)$   & $(5.022,7.944)$    \\
$\nu_{3,R}$  & $(0.0783,4.729)$   & $(5.078,7.729)$    \\ \hline
$\ell_{1,R}$ & $(0.000,10.723)$   & $(5.000,13.723)$   \\
$\ell_{2,R}$ & $(4.763,0.500)$    & $(9.763,3.500)$  \\ 
$\ell_{3,R}$ & $(-3.931,0.500)$   & $(1.069,3.500)$   \\ 
\hline\hline
\end{tabular}
\end{center}
\label{lepton_centers_table}
\end{table}


\begin{table*}
  \caption{\footnotesize{Distances between wave function centers of
      lepton fields, as determined from the lepton wave function
      centers listed in Table \ref{lepton_centers_table}. As defined
      in the text, the numerical subscript on each fermion field is
      the generation index of the weak eigenstate. The horizontal
      entries at the top of the table and the vertical entries on the
      left-hand side of the table list the fields. Thus, for example,
      the (1,4) entry in the table is the distance $\|
      \eta_{L_{1,L}}-\eta_{\nu_{1,R}} \|$, and the (1,7) entry is the
      distance $\| \eta_{L_{1,L}}-\eta_{\ell_{1,R}} \|$.}}
\begin{center}
\begin{ruledtabular}
\begin{tabular}{|c|c|c|c|c|c|c|c|c|c|}
$f_1$  & $L_{1,L}$    & $L_{2,L}$   & $L_{3,L}$ &
         $\nu_{1,R}$  & $\nu_{2,R}$  & $\nu_{3,R}$ &
         $\ell_{1,R}$ & $\ell_{2,R}$ & $\ell_{3,R}$ \\ \hline
$L_{1,L}$  & 0     & 8.477 & 9.353  & 5.179 & 5.050 & 5.131 & 5.057 & 7.368 & 10.924  \\
$L_{2,L}$  & 8.477 & 0     & 1.878  & 5.389 & 5.029 & 4.807 & 10.764 & 3.856 & 4.896  \\
$L_{3,L}$  & 9.353 & 1.878 & 0      & 5.276 & 5.037 & 4.837 & 10.764 & 5.724 & 3.034  \\
\hline
$\nu_{1,R}$ & 5.179 & 5.389 & 5.276 & 0      & 0.4522  & 0.6482  & 5.492  & 6.950 & 5.959 \\
$\nu_{2,R}$ & 5.050 & 5.029 & 5.037 & 0.4522 & 0       & 0.2229  & 5.778  & 6.498 & 5.948 \\
$\nu_{3,R}$ & 5.131 & 4.807 & 4.837 & 0.6482 & 0.2229 & 0        & 5.994  & 6.311 & 5.828 \\
  \hline
$\ell_{1,R}$ & 5.057 & 10.764 & 10.764 & 5.492 & 5.778 & 5.994  & 0      & 11.278  & 10.9525 \\
$\ell_{2,R}$ & 7.368 & 3.856  & 5.724 & 6.950  & 6.498 & 6.311  & 11.278 & 0       & 8.694 \\
$\ell_{3,R}$ & 10.924 & 4.896 & 3.034 & 5.959  & 5.948 & 5.828  & 10.9525 & 8.694 & 0 \\
\end{tabular}
\end{ruledtabular}
\end{center}
\label{lepton_distances_table}
\end{table*}
%


Using methods similar to those for our determination of lepton wave
function centers, we have obtained a new solution for quark
wave function centers in the split fermion model. An earlier
solution was given in \cite{ms}.  In view of later studies on FCNC
effects due to higher KK modes of gluons and other gauge fields
\cite{dpq2000,kaplan_tait,ng_split1,ng_split2,grossman_perez,abel,hewett_split,rm_fcnc}),
we have carried out an analysis designed to greatly reduce these FCNC
effects.  The method that we use for this purpose is similar to the
method that we used above for the leptonic sector; there we chose
lepton locations so as to render the charged lepton mass matrix
diagonal, and here we calculate a new solution for quark wave function
centers that renders the $Q=-1/3$ quark mass matrix diagonal, up to
small corrections. This diagonality of the $M^{(d)}$ mass matrix
removes what would otherwise be excessive FCNC contributions to
processes such as $K^0 - \bar K^0$ and $B^0 - \bar B^0$ mixing.  We
have also checked that FCNC contributions to processes such as
$D^0 - \bar D^0$ mixing are sufficiently small. (Recall that the dominant
contributions to $D^0 - \bar D^0$ mixing actually arise from
long-distance contributions \cite{pdg}.) We list our new solution for
these quark wave function centers in Table \ref{quark_centers_table}
and the resultant distances between quark and lepton wave function
centers in Table \ref{quark_lepton_distances_table} \cite{sigfigs}.
\begin{table} 
  \caption{\footnotesize{Locations of quark wave function centers,
      expressed in the $\eta^{(q)}$ and $\eta$ coordinate systems,
      related by the diagonal translation (\ref{eta_etaq_relation}).
      As defined in the text, the numerical subscript on each fermion
      field is the generation index of the weak eigenstate. Toroidal
      compactification is used, so that $\eta_\lambda$ is equivalent
      to $\eta_\lambda \pm \mu L = \eta_L \pm 30$.}}
\begin{center}
\begin{tabular}{|c|c|c|} \hline\hline
field      & $(\eta^{(q)}_1,\eta^{(q)}_2)$ & $(\eta_1,\eta_2)$  \\ \hline
$Q_{1,L}$  & $(-2.539,0.000)$   &  $(-10.539,-8.000)$     \\
$Q_{2,L}$  & $( 2.539,0.000)$   &  $(-5.461, -8.000)$     \\
$Q_{3,L}$  & $(-0.511,-1.628)$  &  $(-8.511, -9.628)$     \\ \hline
$u_{1,R}$  & $(-0.288,4.185)$   &  $(-8.288,-3.815)$      \\
$u_{2,R}$  & $( 0.288,2.486)$   &  $(-7.712,-5.514)$      \\
$u_{3,R}$  & $(0.303,-1.198)$   &  $(-7.698,-9.918)$      \\ \hline
$d_{1,R}$  & $(-5.000,3.883)$   &  $(-13.000,-4.117)$     \\
$d_{2,R}$  & $( 5.000,3.016)$   &  $(-3.000,-4.984)$     \\
$d_{3,R}$  & $(0.000, -4.450)$  &  $(-8.000,-12.450)$     \\
\hline\hline
\end{tabular}
\end{center}
\label{quark_centers_table}
\end{table}


\begin{table}
  \caption{\footnotesize{Distances between quark and lepton wave
      function centers for our assignments of locations of quark and
      lepton wave function centers in Tables
      \ref{lepton_centers_table} and \ref{quark_centers_table}. As
      defined in the text, the numerical subscript on each fermion
      field is the generation index of the weak eigenstate. Toroidal
      compactification is used.}}
\begin{center}
\begin{tabular}{|c|c|c|c|c|c|c|c|c|c|} \hline\hline
  quark & $L_{1,L}$ & $L_{2,L}$ & $L_{3,L}$ &
  $\nu_{1,R}$ & $\nu_{2,R}$ & $\nu_{3,R}$ &
  $\ell_{1,R}$ & $\ell_{2,R}$ & $\ell_{3,R}$ \\ \hline 
  $Q_{1,L}$ &
  15.2 & 17.4 & 18.3 &
  20.2 & 20.2 & 20.3 &
  16.7 & 15.0 & 16.3 \\
  $Q_{2,L}$ &
  18.4 & 15.8 & 14.5 &
  17.1 & 17.5 & 17.7 &
  13.3 & 18.7 & 13.2 \\
  $Q_{3,L}$ &
  15.6 & 19.2 & 17.8 &
  17.9 & 18.4 & 18.6 &
  15.1 & 17.6 & 16.3 \\
  \hline
  $u_{1,R}$ &
  19.3 & 15.8 & 14.1 &
  17.7 & 17.8 & 17.7 &
  18.2 & 14.0 & 11.9 \\
  $u_{2,R}$ &
  18.9 & 16.1 & 14.5 &
  18.5 & 18.5 & 18.4 &
  16.7 & 15.4 & 12.6 \\
  $u_{3,R}$ &
  16.5 & 18.3 & 16.9 &
  17.6 & 18.1 & 18.3 &
  14.5 & 17.8 & 15.4 \\
  \hline
  $d_{1,R}$ &
  16.9 & 13.2 & 14.8 &
  17.4 & 17.0 & 16.8 &
  17.1 & 10.5 & 16.0 \\
  $d_{2,R}$ &
  18.7 & 12.0 & 10.7 &
  15.3 & 15.2 & 15.1 &
  13.8 & 15.3 & 9.41 \\
  $d_{3,R}$ &
  14.5 & 20.1 & 18.9 &
  15.7 & 16.2 & 16.4 &
  13.6 & 18.6 & 16.7 \\
\hline\hline
\end{tabular}
\end{center}
\label{quark_lepton_distances_table}
\end{table}
%


The last step, namely step (c), is to relate the $\eta^{(q)}$ and
$\eta^{(\ell)}$ coordinate systems to each other.  We choose the
separation vector between the quarks and leptons to be approximately
in the diagonal direction, with the separation distances between
quarks and leptons chosen so as to achieve sufficient suppression of
baryon-number-violating nucleon decays.  For this purpose, we recall
some results from Ref. \cite{bvd,nnblrs}.  Let us denote the sum of
squares of wave function separation distances that occur in the
integration over the extra dimensions of an operator $O_r$
contributing to nucleon decay ($Nd$) as $S^{(Nd)}_r$.  The current
limits on nucleon decay \cite{pdg} imply \cite{bvd}
\beq
S_r > (S^{(Nd)}_r)_{\rm min} \ ,
\label{srmin2}
\eeq
where
\beqs
(S^{(Nd)}_r)_{\rm min} &=& 48 -\frac{n}{2} \ln \pi
- 2 \ln \Big ( \frac{M_{BNV}}{100 \ {\rm TeV}} \Big ) \cr\cr
&-& n \, \ln \Big ( \frac{M_{BNV}}{\mu} \Big ) \ ,
\label{srminvalue}
\eeqs
where $M_{BNV}$ denotes the mass scale characterizing the physics responsible
for baryon-number-violating (BNV) nucleon decay. 
In our model with $n=2$ extra dimensions 
(and value $\mu=3 \times 10^3$ TeV, as given in (\ref{muvalue})), with the
illustrative value $M_{BNV} = 100$ TeV, this is the inequality
$\| \eta_{Q_L} - \eta_{L_{\ell,L}}\| > 8.4$, while for $M_{BNV}=\mu$,
this is the inequality $\| \eta_{Q_L} - \eta_{L_{\ell,L}}\| > 7.3$.
Since $S^{(Nd)}_{\rm min}$ depends only logarithmically on the mass
scale $M_{BNV}$, it follows that the lower bounds on the fermion
separation distances also depend only logarithmically on $M_{BNV}$,
i.e., only rather weakly on this scale.  A very conservative solution
to the coupled quadratic inequalities would require that each of the
relevant distances $\|\eta_{f_i}-\eta_{f_j}\|$ that occur from the
integrals over the extra dimensions of the various four-fermion
operators giving the leading contrbutions to nucleon decay should be
larger than the square root of the right-hand side of
Eq. (\ref{srminvalue}).  As is evident from Table
\ref{quark_lepton_distances_table}, the inequality (\ref{srmin2}) is
satisfied by our solutions for quark and lepton wave function centers.

We also recall a constraint from searches for neutron-antineutron ($n-\bar n$)
oscillations, namely that \cite{nnb02,bvd}
\beqs
M_{n \bar n} &>& (44 \ {\rm TeV}) \Big (
\frac{\tau_{n \bar n}}{2.7 \times 10^8 \ {\rm sec}} \Big )^{1/9} \cr\cr
&\times& 
\Big ( \frac{\mu}{3 \times 10^3 \ {\rm TeV} } \Big )^{4/9} 
\bigg ( \frac{|\langle \bar n | {\cal O}^{(n \bar n)}_4 | n\rangle | }
      {\Lambda_{QCD}^6} \bigg )^{1/9}  \ , \cr\cr
      &&
\label{M_nnbar_min}
\eeqs
where $\tau_{n \bar n}$ is the free $n-\bar n$ oscillation time and
$\Lambda_{QCD}=0.25$ GeV, and ${\cal O}^{(n \bar n)}_4$ was the
six-quark operator that gives the dominant contribution to $n-\bar n$
oscillations in this model \cite{nnb02,bvd}. This bound is not very
sensitive to the precise size of $\langle \bar n |{\cal O}^{(n \bar
  n)}_4 | n\rangle$ because of the 1/9 power in the exponent. The
operator ${\cal O}^{(n \bar n)}_4 = - Q_3$ in the notation of a
lattice calculation of these matrix elements in \cite{nnblgt}, which
obtains $|\langle \bar n | Q_3 | n \rangle | = 5 \times 10^{-4}$
GeV$^6$ $= 2\Lambda_{QCD}^6$; substituting the resultant factor of
$2^{1/9} = 1.08$ in Eq. (\ref{M_nnbar_min}) yields the lower bound
$M_{n \bar n} > 48$ TeV.  The current best published lower limit on
$\tau_{n \bar n}$ is $\tau_{n \bar n} > 2.7 \times 10^8$ sec from the
SuperKamiokande (SK) experiment \cite{sknnb}, and hence this is used
for normalization in Eq. (\ref{M_nnbar_min}).  The SK experiment has
reported a preliminary result that would raise this lower limit by
approximately a factor of 2 \cite{sknew}; the resultant factor of
$2^{1/9}$ would increase the lower bound on $M_{n \bar n}$ to 51 TeV.
In this SM split-fermion model, one thus requires that $M_{n \bar n}$ must
satisfy this lower bound.


\section{Neutrinos in the LRS Split-Fermion Model}
\label{nu_lrs_section}

The LRS version of the split-fermion model is considerably better than
the SM version in accounting for light neutrinos.  In this section we
explain this difference. First, we discuss a relevant constraint on
the scale at which the LRS gauge symmetry is broken to the SM gauge
symmetry.

The analysis of proton decay and $n-\bar n$ oscillations in the LRS
split-fermion model in Ref. \cite{nnblrs} showed that, although it is
easy to suppress baryon-number-violating nucleon decays well below
experimental bounds (by appropriate separation of quark and lepton
wave functions), this does not suppress $n-\bar n$ transitions, which
may occur at levels comparable to current limits.  Furthermore, it was
shown that in the LRS split-fermion model, the integration of certain
six-quark operators mediating $n-\bar n$ oscillations over the extra
dimensions does not yield any exponential factors, in contrast to the
situation in the SM split-fermion model.  As a consequence, the
experimental limit on $n-\bar n$ oscillations implied a lower limit on
the mass scale $M_{n \bar n}$ characterizing the physics responsible
for $n-\bar n$ oscillations in the LRS split-fermion model that is
significantly higher than in the SM split-fermion model, namely (for
$n=2$ extra dimensions) \cite{nnblrs}
\beqs
 M_{n\bar n} &>& (1 \times 10^3 \ {\rm TeV}) \Big (
 \frac{\tau_{n \bar n}}{2.7 \times 10^8 \ {\rm sec}} \Big )^{1/9} \cr\cr
 &\times&
\Big ( \frac{\mu}{3 \times 10^3 \ {\rm TeV} } \Big )^{4/9}
\bigg ( \frac{|\langle \bar n | {\cal O}^{(n \bar n)}_4 | n\rangle | }
      {\Lambda_{QCD}^6} \bigg )^{1/9}  \ . \cr\cr
      &&
\label{vr_min}
\eeqs
Since the vacuum expectation value, $v_R$, of the $\Delta_R$ Higgs field in
the LRS model breaks $(B-L)$ by two units and this is the largest mass scale
associated with $n-\bar n$ oscillations in this model, it follows that
\beq
M_{n \bar n} = v_R \ ,
\label{mnnb_vr}
\eeq
so
\beq
v_R \gsim 10^3 \ {\rm TeV}
\label{vr_value}
\eeq
in the LRS split-fermion model.

In contrast to the SM, where a right-handed Majorana mass term can
occur as a gauge-singlet operator, neither an $[L_{a,L}^T C L_{b,L}]$
nor a $[L_{a,R}^T C L_{b,R}]$ term can occur in a theory with
$G_{LRS}$ gauge symmetry, since they violate the U(1)$_{B-L}$ and,
respectively, the SU(2)$_L$ and SU(2)$_R$ gauge symmetries. Similarly
$B-L$ conservation also forbids the term $L\Phi L\Phi$ ($\Phi$ being
the bi-doublet field) which in LRS would be the analog of the $LHLH$
operator in the SM.

  The LRS model features a profound relation between the breaking of
  total lepton number and the breaking of baryon number and also
  features a natural basis for a seesaw mechanism that explains light
  neutrino masses \cite{mm80,lrs81}.  For the $v_R$ scale of about
  1000 TeV, the observed neutrino masses would require leptonic Yukawa
  couplings of order $10^{-4}$ which is of the same order as the
  leptonic and quark Yukawa couplings in the SM (\ref{yukterm_lrs})
  and (\ref{quarkmasses_lrs}).  There is also a direct type II seesaw
  contribution coming from the left triplet Yukawa coupling given by
\beq
-{\cal L}_{\nu_L,Maj} =
\sum_{a,b}[y^{(LL\Delta_L)}_{ab}[L^{T}_{a,L} C L_{b,L}]\, \Delta_L
  + h.c. \ , 
\label{lh_maj_lrs}
\eeq
(where the SU(2)$_L$ and SU(2)$_R$ group indices are left implicit).
The seesaw mechanism proceeds naturally since $v_R$ is much larger
than the VEVs $\kappa_1$ and $\kappa_2$ of the $\Phi$ field in Eq.
(\ref{phivev}). As noted above, in order for the left-handed Majorana
mass terms arising from (\ref{lh_maj_lrs}) not to spoil the seesaw, it
is necessary that the left-handed Majorana mass terms
$y^{(LL\Delta_L)}_{ab}[\nu^T_{a,L} C \nu_{b,L}] \, v_L$ arising from
the Yukawa interaction in Eq. (\ref{lh_maj_lrs}) must be small
compared with the respective seesaw terms in
Eq. (\ref{numass_sm}). Again, we focus on the terms that are diagonal
in generation indices, i.e., have $a=b$, since for these, the
integration over the extra dimensions does not yield any exponential
suppression factor. Since the maximum physical neutrino mass is $\sim
0.05$ eV, it is necessary that $v_L$ should not be much larger than
the eV scale, unless one uses a small Yukawa coupling
$y^{(LL\Delta_L)}$.  Although the masses of the components of
$\Delta_L$ must be larger than O(TeV), the necessary condtion that
$v_L$ is much less than these masses can be arranged \cite{cmp}. The
mechanism in Ref \cite{cmp} involves the breaking of parity separately
at a high scale leaving the $SU(2)_R\times U(1)_{B-L}$ breaking to TeV
scale.

For mass scales below $v_R$, the gauge symmetry is reduced to the SM
gauge group, $G_{SM}$, and, following the usual application of
low-energy effective field theory, one analyzes the physics in terms
of the fields of the SM model.  This is true, in particular, in the
mass range from $v_R \sim 10^6$ GeV down to the electroweak symmetry
breaking scale of $v \simeq 250$ GeV where the matrix $M^{(D)}$ is
generated by the vacuum expectation values $\kappa_1$ and $\kappa_2$
in $\langle \Phi \rangle_0$.  Hence, the analysis in Section
\ref{nu_sm_section} applies, and we reach the same conclusion, that
this model is able to fit the constraints from limits on proton decay
and $n-\bar n$ oscillations. 


%
\section{Contributions of KK Modes to Physical Processes}
\label{kk_section} 

It is evident from Eq. (\ref{v_kk}) that the KK modes of the SM gauge
bosons have non-flat profiles in the extra dimensions. The higher KK
modes of gauge fields (and Higgs fields) lead, in general, to
tree-level flavor-changing neutral currents, as has been discussed in
a number of works (e.g.,
\cite{dpq2000,kaplan_tait,ng_split1,ng_split2,grossman_perez,abel,hewett_split,rm_fcnc}). We
review the relevant formalism in Appendix \ref{kk_appendix}.  A key
feature of our current study is that, by design, our solution for the
fermion wave function centers given in Table
\ref{lepton_centers_table} and Table \ref{quark_centers_table}, and
shown in Fig. (\ref{fermion_centers_figure}), yields nearly diagonal
charged lepton and down quark mass matrices, greatly suppressing FCNC
KK couplings for the charged leptons and charge $Q=-1/3$
quarks. Still, there are FCNC effects in the neutrino and up-quark
sector, as discussed in Appendix \ref{kk_appendix}.  Although there
are FCNC contributions from higher KK modes to decays such as
$D^0 \to \pi^+\pi^-$ and $D^0 \to 2\pi^0$, they are strongly suppressed,
relative to the SM contribution in amplitudes, by the
factor in Eq. (\ref{prefactor_NSI_eqn}) and hence are
negligible. As mentioned above, we have also estimated FCNC
contributions to $D^0 - \bar D^0$ mixing and find that it is tolerably
small, taking account of the fact that the dominant contributions to
$D^0 - \bar D^0$ mixing actually arise from long-distance
contributions \cite{pdg}.  Here we will focus on the neutrino sector
and demonstrate that these effects are sufficiently small for our
models with either the $G_{SM}$ or $G_{LRS}$ gauge symmetries to be in
accord with experimental constraints.

As discussed above with regard to the constraint from limits on baryon
number violation, in the split-fermion model with $G_{LRS}$ gauge
symmetry, below the scale of $v_R \sim 10^3$ TeV (recall
Eqs. (\ref{vr_min}) and (\ref{mnnb_vr})), the $G_{LRS}$ symmetry is
broken to $G_{SM}$. Hence, using usual low-energy field theory
methods, one analyzes the physics in terms of the fields of the SM
model.  This analysis applies to the split-fermion models with both a
$G_{SM}$ gauge symmetry and a $G_{LRS}$ gauge in the ultraviolet.


\subsection{Neutrino Non-Standard Interactions Mediated by KK Modes}
\label{NSI_section}

In this subsection we analyze the effects of the higher KK modes of
the $W$ and $Z$ bosons in
producing FCNC effects in the neutrino sector, commonly referred to as
neutrino non-standard interactions (NSI). (Some recent
papers on neutrino NSIs with further references to the literature
include \cite{farzan_NSI}-\cite{cgg_NSI}.) 
In a low-energy effective field theory approach,
non-standard interactions between neutrinos and matter beyond the SM
can be represented by the following neutral-current (NC) and
charged-current (CC) effective four-fermion operators
\begin{align}
  {\cal L}_{\rm NC}^{(\rm NSI)} & = -4 \frac{G_F}{\sqrt{2}} \,
  \sum_{X=L,R}\varepsilon_{a b}^{(f; X)}
   [\bar \nu_a \gamma^\lambda P_L \nu_b] [\bar f \gamma_\lambda P_X f]
      \nonumber \\
  {\cal L}_{\rm CC}^{(\rm NSI)} & = -4 \frac{G_F}{\sqrt{2}}\,
  \sum_{X=L,R}\varepsilon_{a b}^{(f f'; X)}
   [\bar \nu_a \gamma^\lambda P_L \ell_b] [\bar f' \gamma_\lambda P_X f) \ .
\label{NSI_Lagrangian_eqn}
\end{align}
Here $f \in (u, d, e)$, $P_{L, R} = (1 \mp \gamma_5)/2$
are the usual chiral projection operators, and $a,b$
denote the generational indices.  These new couplings modify the
neutrino propagation in matter \cite{MSW} and also alter the
production and detection in various neutrino experiments. Analyses of
data from from these 
experiments have yielded stringent bounds on the coupling strengths of
the new interactions, $\varepsilon_{a b}^{(f f'; X)}$, and
$\varepsilon_{a b}^{(f; X)}$ \cite{farzan_NSI,cgg_NSI}.

The $W$-boson KK modes contribute to the charged-current NSI parameter
$\varepsilon_{a b}^{(ud; L)}$. Using eqs. (\ref{Wkk_coupling_special},
\ref{LWkk_eqn}), we find
\beqs
\varepsilon_{a b}^{(ud; L)} &=& \bigg( \frac{m_W}{2 \pi \Lambda_L} \bigg )^2
V^*_{11} U^*_{ba} {\cal S}_W(\eta_{L_{b, L}},\eta_{Q_{1, L}}) \ , \cr\cr
&& 
\label{Wkk_NSI_eqn}
\eeqs
where, as above $V$ is the CKM quark mixing matrix, $U$ is the PMNS lepton
mixing matrix, and we have collected the terms that depends on the fermion
locations in the extra-dimensions and defined these as
\beqs
&& {\cal S}_W(\eta_{L_{b, L}},\eta_{Q_{1, L}}) \equiv
\sum_{m \in {\mathbb Z}_{\neq 0}^2}
\frac{e^{-\frac{\pi^2}{(\mu L)^2} \| m\|^2}}
      { \|m\|^2} \times \cr\cr
      &\times&
      \cos \bigg [\frac{2 \pi}{\mu L} m \cdot (\eta_{L_{b, L}}-\eta_{Q_{1, L}})
        \bigg ] \ .
\label{S_W_NSI_eqn}
\eeqs
The numerical values of the sum ${\cal S}_W(\eta_{L_{b,
    L}},\eta_{Q_{1, L}})$ are listed in Table \ref{S_W_NSI_table}.  As
is evident from this table, because of the oscillating cosine
functions and the damping by the exponential factors, the partial sums
converge rapidly. (To show this, we display these values to five
significant figures in this table; in the subsequent tables, ${\cal
  S}_W$ and ${\cal S}_Z$ values are usually listed to four significant
figures.)  Furthermore, the dependence on the locations of the fermion
wave function centers is embodied in a factor of order unity and is
not very sensitive to these locations. Therefore, the magnitudes of
the NSI interaction parameters are predominantly determined by the
prefactor in Eq. (\ref{Wkk_NSI_eqn}), which does not depend on the
details of the fermion wave function centers, but, instead, only on
the scale $\Lambda_L$. Numerically,
\beq
\bigg( \frac{m_W}{2 \pi \Lambda_L} \bigg )^2 = 1.64 \times 10^{-8} \ .
\label{prefactor_NSI_eqn}
\eeq
%


\begin{table}
  \caption{\footnotesize{Demonstration of convergence of the partial sums
      for ${\cal S}_W(\eta_{L_{b, L}},\eta_{Q_{1, L}})$ defined in
      Eq. (\ref{S_W_NSI_eqn}) as a function of generation index
      $b$ wavecenter. With the fermion wave function centers in
      Fig. (\ref{fermion_centers_figure}), the partial sum of
      contributions from $m=(m_1,m_2)$ up to
      $\|m \| = \sqrt{m_1^2+m_2^2} = \|m_0\|$ is displayed for each
      $\eta_{L_{b, L}}$.}}
\begin{center}
\begin{tabular}{|c|c|c|} \hline\hline
$b$ & $\|m_0\|/\sqrt{2}$ & ${\cal S}_W(\eta_{L_{b, L}},\eta_{Q_{1, L}})$ \\ \hline
$1$ & $3$ & $0.056419$ \\
$1$ & $30$ & $0.10331$ \\
$1$ & $300$ & $0.10331$ \\ \hline
$2$ & $3$ & $0.526714$ \\
$2$ & $30$ & $0.53884$ \\
$2$ & $300$ & $0.53884$ \\ \hline
$3$ & $3$ & $0.40892$ \\
$3$ & $30$ & $0.43437$ \\
$3$ & $300$ & $0.43437$ \\
\hline\hline
\end{tabular}
\end{center}
\label{S_W_NSI_table}
\end{table}

Using this result, we can estimate the CC NSI interaction strengths produced by
the higher $W$ boson KK modes. These are displayed in Table
\ref{epsilons_NSI_table}. The magnitudes of these CC NSI
parameters are largely determined the factor in
eq. (\ref{prefactor_NSI_eqn}). These values are far below current
experimental upper bounds on the magnitudes of these parameters, which are
of $O(1)$ \cite{farzan_NSI,cgg_NSI}.


\begin{table}
  \caption{\footnotesize{ Value of the KK $W$ boson mediated
      (charged- current) NSI parameters $|\varepsilon_{a b}^{(ud; L)}|$. Here
      $a,b$ are generational indices. See text for further details.}}
\begin{center}
\begin{tabular}{|c|c|c|c|} \hline\hline
$b$ & $a$ & ${\cal S}_W(\eta_{L_{b, L}},\eta_{Q_{1, L}})$ & $|\varepsilon_{a b}^{(ud; L)}|$\\ \hline
$1$ & $1$ & $0.1033$ & $1.36 \times 10^{-9}$ \\
$1$ & $2$ & $0.1033$ & $0.90 \times 10^{-9}$ \\
$1$ & $3$ & $0.1033$ & $0.25 \times 10^{-9}$ \\ \hline
$2$ & $1$ & $0.5388$ & $2.33 \times 10^{-9}$ \\
$2$ & $2$ & $0.5388$ & $5.20 \times 10^{-9}$ \\
$2$ & $3$ & $0.5388$ & $6.43 \times 10^{-9}$ \\ \hline
$3$ & $1$ & $0.4344$ & $3.43 \times 10^{-9}$ \\
$3$ & $2$ & $0.4344$ & $4.02 \times 10^{-9}$ \\
$3$ & $3$ & $0.4344$ & $4.47 \times 10^{-9}$ \\
\hline\hline
\end{tabular}
\end{center}
\label{epsilons_NSI_table}
\end{table}


In a similar manner, we can evaluate the NC NSI parameters due to the higher KK
modes of the $Z$ boson. For illustrative purposes, let us write down the NC NSI
parameters for $f = e, d$: 
\beq
\varepsilon_{a b}^{(f; X)} = \bigg( \frac{m_W}{2 \pi \Lambda_L} \bigg )^2
\frac{T_Z^{(f_X)}}{\cos^2 \theta_W} \times {\cal S}_{Z, ab} (\eta_{f_X}) \ ,
\label{Zkk_NSI_eqn}
\eeq
where $X=L,R$,
\beq
T_Z^{(f_X)} = T^{(f_X)}_{3L} - Q_f \sin^2 \theta_W \ ,
\label{tz}
\eeq
$Q_f$ is
the electric charge of fermion $f$, and the term that depends on the wavecenter
locations is defined as
\begin{multline}
  {\cal S}_{Z, ab} (\eta_{f_X}) \equiv \sum_{m \in {\mathbb Z}^2_{\neq 0}}
  \frac{e^{-\frac{\pi^2}{(\mu L)^2} \| m \|^2}}{ \| m \|^2}
  \sum_{k=1}^3 U^*_{ka} \times  \\
  \times \cos \bigg [ \Big ( \frac{2 \pi}{\mu L} \Big )
    \Big \{ m \cdot (\eta_{L_{k , L}} - \eta_{f_X}) \Big \} \bigg ]U_{kb} \ .
\label{S_Z_NSI_eqn}
\end{multline}
This sum can be evaluated numerically, and we show the resultant
$\varepsilon_{ab}^{(f ; X)}$ in Table \ref{epsilons_NSI_CC_table}.  As
is evident from this table, the magnitudes for these NC NSI parameters
are essentially determined by the prefactor in
Eq. (\ref{prefactor_NSI_eqn}), which does not depend on the locations
of wave function centers, but only on $\Lambda_L$. Similar comments
apply for $\varepsilon_{ab}^{(u ; X)}$. Thus, the strengths of the
non-standard neutrino operators generated by the higher $Z$ and $W$ KK
modes are much smaller than current experimental upper bounds on the
magnitudes of these parameters, which are of $O(1)$
\cite{farzan_NSI,cgg_NSI}.  We comment on the NSI interactions
generated by local four-fermion operators below.


    \begin{table*}
  \caption{\footnotesize{Values of the KK $Z$ boson mediated neutral
      current NSI parameters $|\varepsilon_{a b}^{(f; X)}|$, for $f=e,
      d$, and $X = L, R$. Here $a,b$ are generational
      indices. The sum $|{\cal S}_{Z, ab}(\eta_{f_X})|$, defined
      in Eq. (\ref{S_Z_NSI_eqn}), is numerically evaluated and
      displayed for $f=e, d$, and $X=L, R$.}}
\begin{center}
\begin{ruledtabular}
\begin{tabular}{|c|c|c|c|c|c|c|c|c|c|} 
  $a$ & $b$  & $|{\cal S}_{Z, ab}(\eta_{e_L})|$ &  $|{\cal S}_{Z, ab}(\eta_{e_R})|$ &  $|{\cal S}_{Z, ab}(\eta_{d_L})|$ & $|{\cal S}_{Z, ab}(\eta_{d_R})|$ &
  $|\varepsilon_{a b}^{(e; L)}|$ & $|\varepsilon_{a b}^{(e; R)}|$ & $|\varepsilon_{a b}^{(d; L)}|$ & $|\varepsilon_{a b}^{(d; R)}|$ \\
  \hline
$1$& $1$ & $2.8128$ & $0.8291$ & $0.2167$ & $0.2920$ & $1.64 \times 10^{-8}$ & $3.89 \times 10^{-9}$ & $1.94 \times 10^{-9}$ & $4.57 \times 10^{-10}$ \\
$1$& $2$ & $2.4595$ & $1.2181$ & $0.1659$ & $0.09353$ & $1.44 \times 10^{-8}$ & $5.72 \times 10^{-9}$ & $1.49 \times 10^{-9}$ & $1.46 \times 10^{-10}$ \\
$1$& $3$ & $0.5970$ & $0.4419$ & $0.02269$ & $0.02563$ & $3.48 \times 10^{-9}$ & $2.07 \times 10^{-9}$ & $0.20 \times 10^{-9}$ & $0.40 \times 10^{-10}$ \\
$2$& $1$ & $2.4595$ & $1.2181$ & $0.1659$ & $0.09353$ & $1.44 \times 10^{-8}$ & $5.72 \times 10^{-9}$ & $1.49 \times 10^{-9}$ & $1.46 \times 10^{-10}$ \\
$2$& $2$ & $0.6836$ & $0.1401$ & $0.3743$ & $0.4023$ & $3.99 \times 10^{-9}$ & $ 0.66 \times 10^{-9}$ & $3.36 \times 10^{-9}$ & $6.30 \times 10^{-10}$ \\
$2$& $3$ & $0.5216$ & $0.06354$ & $0.07358$ & $0.10035$ & $3.04 \times 10^{-9}$ & $0.30 \times 10^{-9}$ & $0.66 \times 10^{-9}$ & $1.57 \times 10^{-10}$ \\
$3$& $1$ & $0.6290$ & $0.4419$ & $0.02269$ & $0.02563$ & $3.67 \times 10^{-9}$ & $2.07 \times 10^{-9}$ & $0.20 \times 10^{-9}$ & $0.40 \times 10^{-10}$ \\
$3$& $2$ & $0.5216$ & $0.06354$ & $0.07359$ & $0.10035$ & $3.04 \times 10^{-9}$ & $0.30 \times 10^{-9}$ & $0.66 \times 10^{-9}$ & $1.57 \times 10^{-10}$ \\
$3$& $3$ & $0.83705$ & $0.8404$ & $0.48555$ & $0.47799$ & $4.88 \times 10^{-9}$ & $3.95 \times 10^{-9}$ & $4.35 \times 10^{-9}$ & $7.47 \times 10^{-10}$ \\
\end{tabular}
\end{ruledtabular}
\end{center}
\label{epsilons_NSI_CC_table}
\end{table*}
%


%
\section{Some Further Phenomenology Involving Leptons}
\label{phenom_section}

\subsection{Weak Decays}

Weak decays that proceed at the tree level have amplitudes involving
coefficients $\propto G_F$ multipled by four-fermion operators.
These include pure leptonic, semileptonic, and nonleptonic weak
decays.  The amplitudes for the latter two types of decays include 
CKM quark mixing matrix elements, which we denote as a coefficient $c_q$,
where $c_q=V_{ud}$ for decays such as $\pi^+ \to \mu^+ \nu_\mu$ and
nuclear beta decay (abbreviated as $N \beta D$);
$c_q=V_{us}$ for $K^+ \to \mu^+\nu_\mu$, $K^+ \to \pi^0 \ell^+
\nu_\ell$, and $\Lambda \to p e \bar\nu_e$;
$c_q=V_{us}^*V_{ud}$ for $K^+ \to \pi^+\pi^0$; and so
forth for weak decays of heavy-quark hadrons. We may retain this
factor $c_q$ for pure leptonic decays such as $\mu \to \nu_\mu e
\bar\nu_e$ also by setting $c_q = 1$ for these decays.  The amplitudes
for (tree-level) weak decays can thus be written generically as
\beq
Amp = 4c_V \frac{G_F}{\sqrt{2}} \, 
[\bar \psi_{4,L} \gamma_\lambda \psi_{3,L}]
[\bar \psi_{2,L} \gamma^\lambda \psi_{1,L}] \ ,
\label{sm_weak_amp}
\eeq
where $\psi_j$, $j=1,...,4$ are the fermions involved in the decay.
The wealth of data on tree-level weak decays yields a number of
constraints on possible BSM effects.  For example, the agreement of the
measured rate for $\mu$ decay with the Standard Model prediction provides
one such constraint, since BSM effects from split fermions would spoil
this agreement, just as, e.g., massive neutrino emission via mixing would
\cite{shrock80,shrock81}. The ratios of branching ratios
$R^{(\pi)}_{e/\mu} \equiv BR(\pi^+ \to e^+ \nu_e)/BR(\pi^+ \to \mu^+ \nu_\mu)$,
$R^{(K)}_{e/\mu}$, $R^{(D_s)}_{e/\tau}$, and $R^{D}_{e/\tau}$, and the measured
branching ratios for $B^+ \to \mu^+ \nu_\mu$ and $B^+ \to \tau^+\nu_\tau$ with
Standard Model predictions provide another set of constraints
\cite{shrock80}-\cite{pft}. 

In the split-fermion models, there are additional contributions to
these amplitudes arising from the respective four-fermion operators
composed of fermion fields defined in the $d=4+n$ dimensional space.
The effective Lagrangian that describes these decays has the form
(\ref{leff_higherdim}) with these four-fermion operators and hence
$k=4$. In this ${\cal L}_{eff,4+n}$, a four-fermion operator
$O_{r,(4)}$ has a coefficient of the form (\ref{kappagen}), namely
$c_{r,(4)}=\bar\kappa_{r,(4)}/M^{2+n}$.  For these SM weak decays, the
relevant mass scale $M$ that describes the new contributions from the
presence of the higher dimensions is $M=\Lambda_L$.  From
Eq. (\ref{crgen}), it follows that after integration over the extra
dimensions, the new split-fermion model contribution (in
addition to the SM contribution) to the amplitude, in four-dimensional
spacetime, for a given decay involves operator products of
four-dimensional fermion fields with coefficients of the form
\beq
c_{r,(4)} = \frac{\bar\kappa_{r,(4)}}{\Lambda_L^2} \, \Big (
\frac{\mu}{\pi^{1/2} \Lambda_L} \Big )^n \, e^{-S_{r,(4)}} \ , 
\label{ledlf_weak_decay_contrib}
\eeq
where $e^{-S_{r,(k)}}$ was defined in Eq. (\ref{irgen}). 
The full amplitude for a tree-level weak decay is
thus $A_{SM} + A_{SF}$.  Since $|A_{SF}|/|A_{SM}| \ll 1$, the leading
effect on the observed rate is due to the interference term
${\rm Re}(A_{SM}A_{SF}^*)$ The ratio of the SFM to the SM
contribution to a given tree-level weak decay is then
\begin{widetext}
\beq
\frac{|A_{SF}|}{|A_{SM}|} \sim 
\frac{|\sum_r \bar\kappa_{r,(4)}\, e^{-S_{r,(4)}}|} {2 c_V} 
\Big ( \frac{v}{\Lambda_L}\Big )^2 
\Big (\frac{\mu}{\pi^{1/2} \Lambda_L} \Big )^n  \ ,
\label{weak_decay_amp_ratio}
\eeq
\end{widetext}
where we have used the SM relation $4(G_F/\sqrt{2}) = 2/v^2$ with
$v=246$ GeV.
In the split-fermion model with $n=2$ and the values $\Lambda_L$ and $\mu$
taken here (as in \cite{ms}) and $|\bar\kappa_{r,(4)}| \sim O(1)$, for
a leptonic or CKM-favored semileptonic or nonleptonic weak decay, this
ratio is generically
\beq
\frac{|A_{SF}|}{|A_{SM}|} \sim \frac{10^{-3}}{|c_V|} \, |\sum_r
\bar\kappa e^{-S_{r,(4)}}| \ ,
\label{weak_decay_amp_ratio_value}
\eeq
where the sum $\sum_r$ is over the four-fermion operators that
contribute to this decay.
The exponential factor $e^{-S_{r,(4)}}$ depends on the type of
decay. For example, with the assignments for
locations of wave function centers in Tables \ref{quark_centers_table}
and \ref{lepton_centers_table}, shown graphically in Fig.
\ref{fermion_centers_figure} \cite{sigfigs}, a factor contributing to
$\mu$ decay is
\beq
\mu \to \nu_\mu e \bar\nu_e: \quad
e^{-\|\eta_{L_{1,L}}-\eta_{L_{2,L}}\|^2} = 0.618 \times 10^{-31} \ .
\label{mudec_LLfactor}
\eeq
Exponential factors that occur for semileptonic weak decays are
extremely small because of the separation of quark and lepton wave
function centers required to suppress proton decay.  In general, we
find that the ratio (\ref{weak_decay_amp_ratio_value}) is negligibly
small for Standard-Model weak decays. Consequently, the split-fermion
models satisfy constraints from data on these weak decays.


\subsection{Neutrino Reactions}

We next discuss neutrino reactions.  We focus on the reactions $\nu_e
e \to \nu_e e$ and $\bar\nu_e e \to \bar\nu_e e$, since these involve
lepton fields located at the same point in the extra dimensions and
hence could exhibit especially large non-SM effects.  As is well
known, in the SM, these involve both charged-current and
neutral-current contributions. For example, the amplitude for $\nu_e e
\to \nu_e e$ is
\begin{widetext}
\beq
A_{\nu_e e,SM} = 4 \frac{G_F}{\sqrt{2}} \,
[\bar\nu_{eL}\gamma_\lambda \nu_{eL} ]\Big [
  \Big ( \frac{1}{2}+\sin^2\theta_W \Big ) \,
  [\bar e_L \gamma^\lambda e_L] + \sin^2\theta_W \,
  [\bar e_R \gamma^\lambda e_R] \ \Big ] \ ,
\label{nue_e_amplitude}
\eeq
\end{widetext}
where $\sin^2\theta_W \simeq 0.23$.  Since $\Lambda_L \gg v$, it
follows that the operators in the effective Lagrangian in $4+n$
dimensions must be invariant under the ${\rm SU}(2)_L \otimes {\rm
  U}(1)_Y$ electroweak SM gauge symmetry.  The lowest-dimension
operators are four-fermion operators. Of particular importance are the
operators
\beqs
O^{(\nu_e e)}_{LLLL}(x,y)
&=& \kappa^{(\nu_e e)}_{LLLL} [\bar L_{1,L}(x,y)\gamma_\lambda  L_{1,L}(x,y)]
\cr\cr
&\times& [\bar L_{1,L}(x,y)\gamma^\lambda L_{1,L}(x,y)] + h.c., \cr\cr
&&
\label{oll}
\eeqs
where here the Lorentz index $\lambda$ runs over all $4+n$ values
and we write $\kappa^{(\nu_e e)}_{LLLL,(4)} \equiv
\kappa^{(\nu_e e)}_{LLLL}$. From 
the $k=4$ special case of Eq. (\ref{kappagen}), we have
$\kappa^{(\nu_e e)}_{LLLL}=
\bar\kappa^{(\nu_e e)}_{LLLL}/M^{2+n}$, and, as before, the relevant mass
in the higher-dimensional theory is $\Lambda_L$, so
$\kappa^{(\nu_e e)}_{LLLL} =
\bar\kappa^{(\nu_e e)}_{LLLL}/\Lambda_L^{2+n}$, where
$\bar\kappa^{(\nu_e e)}_{LLLL}$ is dimensionless.
This operator gives the dominant correction to the SM amplitude for
the $\nu_e e \to \nu_e e$ and $\bar\nu_e e \to \bar\nu_e e$ reactions
because the lepton fields are located at the same point in the
extra-dimensional space, so the integration of the four-fermion
operator products over the extra dimensions does not involve any
exponential suppression factor.  In contrast, the integration of the
operator $[\bar L_{1,L}(x,y)\gamma_\lambda L_{1,L}(x,y)]
[\ell_{1,R}(x,y)\gamma^\lambda \ell_{1,R}(x,y)]$ over the extra
dimensions does does yield an exponential suppression factor; with our
assignments for wave function centers, this exponential factor is 
$e^{-\|\eta_{L_{1,L}}-\eta_{\ell_{1,R}}\|^2} = 0.78 \times 10^{-11}$.

Now we estimate the correction in the amplitudes for the $\nu_e e$ and
$\bar \nu_e e$ reactions due to these new  contributions.  Performing
the integration over the operator product (\ref{oll}) over the higher
dimensions, we obtain the operator in four-dimensions
\begin{widetext}
\beq
O_{LLLL}(x) = \frac{\bar\kappa^{(\nu_e e)}_{LLLL}}{\Lambda_L^2}
\Big ( \frac{\mu}{\pi^{1/2} \Lambda_L} \Big )^n \,
[\bar L_{1,L}(x)\gamma_\lambda  L_{1,L}(x)]
         [\bar L_{1,L}(x)\gamma^\lambda L_{1,L}(x)] \ . 
\label{oll_4d}
\eeq
\end{widetext}
The amplitude for the $\nu_e e \to \nu_e e$ reaction can be written as
$Amp = A_{\nu e,SM} + A_{\nu e, SF}$ and
similarly for $\bar\nu_e e \to \bar \nu_e e$.
As before, the leading
correction arises from the interference term. The relative importance of
this is given by the ratio $|A_{\nu_e e,SF}|/|A_{\nu_ e,SM}|$. We find
\beq
\frac{|A_{\nu_e e,SF}|}{|A_{\nu_e e,SM}|} \simeq
|\bar\kappa^{(\nu_e e)}_{LLLL}| \, \Big (\frac{v}{\Lambda_L} \Big )^2 \,
\Big ( \frac{\mu}{\pi^{1/2}\Lambda_L} \Big )^n \ . 
\label{ampratio_nu_e}
\eeq
With $n=2$, $\bar\kappa^{(\nu_e e)}_{LLLL} \sim O(1)$, and our values of
$\mu$ and $\Lambda_L$, we obtain
\beq
\frac{|A_{\nu_e e,SF}|}{|A_{\nu_e e,SM}|} \simeq 10^{-3} \ , 
  \label{ampratio_nu_e_value}
\eeq
and similarly for $|A_{\bar\nu_e e,SF}|/|A_{\bar\nu_e e,SM}|$.
This is sufficiently small to be in accord with data on these
neutrino reactions.  There are also contributions to the ratio
(\ref{ampratio_nu_e_value}) from the
neutrino NSI terms generated by higher KK modes of the $W$ and $Z$,
but these are much smaller than the contribution that we have
calculated in Eq. (\ref{ampratio_nu_e_value}) because they enter
with a factor of $10^{-8}$ suppression from Eq. (\ref{prefactor_NSI_eqn}). 


\subsection{Charged Lepton Flavor-Violating Decays
  $\mu \to e \gamma$ and $\tau \to \ell \gamma$}
\label{meg_subsection}

Here we discuss charged lepton flavor violation (CLFV).  A
particularly stringent constraint is the upper limit on the decay $\mu
\to e \gamma$, namely \cite{pdg,cpv}
\beq
BR(\mu \to e \gamma) < 4.2 \times 10^{-13}  .
\label{megbr_limit}
\eeq
This and other experimental limits are given at the 90 \% confidence
level (90 \% CL).  Recall that the rate for regular $\mu$ decay,$\mu
\to \nu_\mu e \bar\nu_e$ is, to very good accuracy, given by
$\Gamma_\mu = G_F^2 m_\mu^5/(192\pi^3)$. The contribution to the
decay $\mu \to e \gamma$ (abbreviated $\mu e \gamma$) from diagrams in
the Standard Model, as extended to include massive neutrinos, is
smaller than this upper limit by many orders of magnitude
\cite{p77,lee_shrock77} and is thus negligible.  Given the lower limit
on $v_R \gsim 10^3$ TeV, and hence on $m_{W_R}$, in an LRS split-fermion model
from the non-observation of $n-\bar n$ oscillations \cite{nnblrs}, it
is also the case that diagrams with $W_R$ exchange make a
negligigible contribution to $\mu \to e \gamma$ \cite{rs82}.  In a
low-energy effective field theory applicable below the EWSB scale, the
terms in the effective Lagrangian that are responsible for the decay
$\mu \to e \gamma$ involve the operators
\beq
\{ \ [\bar e_L \sigma_{\lambda \rho} \mu_R]F_{em}^{\lambda\rho}, \quad 
[\bar e_R \sigma_{\lambda \rho} \mu_L]F_{em}^{\lambda\rho} \ \} \ ,
\label{meg_ops0}
\eeq
where $\sigma_{\lambda \rho} = (i/2)[\gamma_{\lambda},\gamma_\rho]$ is
the antisymmetric Dirac tensor and $F_{em}^{\lambda\rho}$ is the
electromagnetic field strength tensor. These lepton bilinears connect
left-handed and right-handed components of the lepton fields and hence
violate both the ${\rm SU}(2)_L \otimes {\rm U}(1)_Y$ SM electroweak
gauge symmetry and the ${\rm SU}(2)_L \otimes {\rm SU}(2)_R$ part of
the LRS gauge symmetry.

We begin our analysis with the split-fermion model with $G_{SM}$ gauge
and fermion content and will then consider the corresponding SF model
with $G_{LRS}$.  The effective field theory relevant for
the SM LEDLF theory in the energy interval $v < E < \Lambda_L$ (i.e.,
$250 \ {\rm GeV} < E < 100 \ {\rm TeV}$) the effective Lagrangian for
this decay must be invariant under $G_{SM}$ and hence must involve the
Higgs field, $\phi$.  This effective Lagrangian for $\mu \to e \gamma$
is
\beqs
 && {\cal L}_{eff,\mu e \gamma,4+n} = \Big [
  \kappa^{(\mu e \gamma) \, \prime}_1 [\bar L_{1,L}
    \sigma_{\lambda \rho} \ell_{2,R}] \phi
  \cr\cr
  &+&
  \kappa^{(\mu e \gamma) \, \prime}_2
        [\bar \ell_{1,R}\sigma_{\lambda \rho} L_{2,L}   ]
  \tilde \phi \Big ] \, F^{\lambda \rho}_B + h.c. \ , 
\label{meg_ops}
\eeqs
where here $F^{\lambda\rho}_B$ is the U(1)$_Y$ field strength tensor and,
as before, $\tilde \phi = i\tau_2 \phi^*$. 
We have discussed above how the Higgs and gauge fields are taken to
have flat profiles in the extra dimensions. Hence, as in our earlier
operator analyses of operators involving Higgs fields in
\cite{nnblrs}, although a boson field in $d=4+n$ dimensions has
Maxwellian mass dimension $1+(n/2)$, in the integration of
the boson field over the $n$ extra dimensions, the normalization
constant for the $d$-dimensional boson field just cancels the
additional powers of $1/\Lambda_L$ that appear in coefficients.
Hence, it suffices to consider just the fermionic part of the
operators in the integration over the extra dimensions.  After this
integration, the operators (\ref{meg_ops0}) result from the vacuum
expectation value, $v/\sqrt{2}$, of the
$\phi$ field in Eq. (\ref{meg_ops}). Hence, in the effective theory
below this EWSB scale, the operators (\ref{meg_ops}) involve a factor
of $v/\sqrt{2}$.  Furthermore, since the decay is absent unless
$m_\mu$ is nonzero (with $m_\mu > m_e$), the operators involve, as
prefactors, not just $v/\sqrt{2}$, but also the requisite Yukawa couplings
that yield $m_\mu$.  Because there is an emission of a photon in the
$\mu \to e \gamma$ decay, the amplitude also contains a factor of the
electromagnetic gauge coupling, $e$.  To make the factors of $e$ and
$m_\mu$ explicit, we write
\beq
\kappa^{(\mu e \gamma) \ '}_j = e \, m_\mu \kappa^{(\mu e \gamma)}_j,
\quad j=1,2 \ .
\label{kappa_meg_rel}
\eeq
Starting with the operators in $d=4+n$ dimensions, and using the
property that $b_2=1$, the integration of the fermion bilinear 
$[\bar L_{1,L}(x,y) \sigma_{\lambda \rho} \ell_{2,R}(x,y)]$ over the
$y$ coordinates yields the operator in four spacetime dimension 
\beqs
&& [\bar L_{1,L}(x) \sigma_{\lambda \rho} \ell_{2,R}(x)] \,
e^{-(1/2)\|\eta_{L_{1,L}}-\eta_{\ell_{2,R}}\|^2} \cr\cr
&=& [\bar L_{1,L}(x) \sigma_{\lambda \rho} \ell_{2,R}(x)] \times
(1.63 \times 10^{-12}) \ ,
\label{L1_ell2_integral}
\eeqs
where we have used the value of $\|\eta_{L_{1,L}}-\eta_{\ell_{2,R}}\|$
listed in Table \ref{lepton_distances_table}. Similarly, the integration of
the operator $[\bar \ell_{1,R}(x,y) \sigma_{\lambda \rho}
  L_{2,L}(x,y)]$ over the $y$ coordinates yields the operator
\beqs
&& [\bar \ell_{1,R}(x) \sigma_{\lambda \rho} L_{2,L}(x)]
e^{-(1/2)\|\eta_{\ell_{1,R}}-\eta_{L_{2,L}}\|^2} \cr\cr
&=& [\bar \ell_{1,R}(x) \sigma_{\lambda \rho} L_{2,L}(x)] \times 
(0.695 \times 10^{-25}) \ ,
\label{ell1_L2_integral}
\eeqs
where we have used the value of $\|\eta_{\ell_{1,R}}-\eta_{L_{2,L}}\|$
listed in Table \ref{lepton_distances_table}. 
Reverting to general notation, the resultant effective Lagrangian for
$\mu \to e \gamma$ in $d=4$ dimensions is (suppressing the $x$
arguments)
\begin{widetext}
\beqs
&& {\cal L}_{eff,\mu e \gamma,4D} = \frac{e \, m_\mu}{\Lambda_L^2} \, \Big
    [ \,
      \bar\kappa^{(\mu e \gamma)}_1 \,
                [\bar L_{1,L} \sigma_{\lambda \rho} \ell_{2,R}] \,
       e^{-(1/2)\|\eta_{L_{1,L}}-\eta_{\ell_{2,R}}\|^2} +
       \bar\kappa^{(\mu e \gamma)}_2 \,
                 [\bar \ell_{1,R} \sigma_{\lambda \rho}L_{2,L}] \, 
   e^{-(1/2)\|\eta_{\ell_{1,R}}-\eta_{L_{2,L}}\|^2} \ \Big ] \,
   F^{\lambda \rho} + h.c. \cr\cr
   &&
\label{meg_ops_4d}
\eeqs
\end{widetext}
Since the Maxwellian (mass) dimension of the operators
$\bar L_{1,L} \sigma_{\lambda \rho} \ell_{2,R}] \, F^{\lambda\rho}$ and
  $[\bar \ell_{1,R} \sigma_{\lambda \rho}L_{2,L}] \, F^{\lambda\rho}$ in
  four-dimensional spacetime is 5, their coefficients in
  ${\cal L}_{eff,\mu e \gamma,4D}$ have 
dimension $-1$ and since the operators have
$m_\mu$ as a prefactor, this means that $\kappa^{(\mu e \gamma)}_j$, $j=1,2$
have dimension $-2$. In Eq. (\ref{meg_ops_4d}) we have conservatively
taken the normalization mass to be $\Lambda_L$, writing
\beq
\kappa^{(\mu e \gamma)}_j =
\frac{\bar\kappa^{(\mu e \gamma)}_j}{\Lambda_L^2} \ , \ j=1,2 \ ,
\label{kappa_bar_meg}
\eeq
where the $\bar\kappa^{(\mu e \gamma)}_j$ are dimensionless, by construction.
Combining these results
with the general formulas (specifically, Eqs. (2.63) and (2.65)) in Ref.
\cite{lee_shrock77}, we calculate the branching ratio
\begin{widetext}
\beqs
BR(\mu \to e \gamma) &=& \frac{192 \pi^3 \alpha_{em}}{(G_F \Lambda_L^2)^2} \,
\Big [|\bar\kappa^{(\mu e \gamma)}_1|^2 \,
  e^{-\|\eta_{L_{1,L}}-\eta_{\ell_{2,R}}\|^2}
  +     |\bar\kappa^{(\mu e \gamma)}_2|^2 \,
  e^{-\|\eta_{\ell_{1,R}}-\eta_{L_{2,L}}\|^2}
  \ \Big ] \cr\cr
&=& (0.908 \times 10^{-32}) \Big [ \, |\bar\kappa^{(\mu e \gamma)}_1|^2 +
   (1.81 \times 10^{-27}) |\bar\kappa^{(\mu e \gamma)}_2|^2 \ \Big ]\ .
\label{meg_br}
\eeqs
\end{widetext}
With $|\bar\kappa^{(\mu e \gamma)}_j| \sim O(1)$ for $j=1,2$, this is
considerably smaller than the experimental upper limit on $BR(\mu \to
e \gamma)$. 

In a similar manner, we calculate the branching ratios for the decays
$\tau \to e \gamma$ and $\tau \to \mu \gamma$ in the $G_{SM}$
split-fermion model with the locations of the lepton wave function
centers given above.  Since both $m_e^2/m_\tau^2 \ll 1$ and
$m_\mu^2/m_\tau^2 \ll 1$, the rates for each of the two leptonic decay
modes of the $\tau$ are given, to very good accuracy, by
\beqs
\Gamma_{\tau \to \nu_\tau \ell \bar\nu_\ell}
&=& \frac{G_F^2 m_\tau^5}{192 \pi^3} \quad {\rm for} \ \ell=e, \mu \ .
\cr\cr
&& 
\label{gamma_tau_leptonic}
\eeqs
The corresponding measured branching ratios are \cite{pdg}
\beq
BR(\tau \to \nu_\tau e \bar\nu_e) = (17.82 \pm 0.04) \ \% \equiv B_{\tau-e}
\label{br_tau_e}
\eeq
and
\beq
BR(\tau \to \nu_\tau \mu \bar\nu_\mu) = (17.39 \pm 0.04) \ \% \ \equiv B_{\tau-\mu} .
\label{br_tau_mu}
\eeq
Analogously to Eq. (\ref{meg_ops_4d}, we calculate the effective
Lagrangian for $\tau \to \ell \gamma$ with $\ell=e$ or
$\ell=\mu$ (symbolized as $\tau \ell \gamma$) in $d=4$ dimensions to be 
\begin{widetext}
\beqs
&& {\cal L}_{eff,\tau\ell\gamma,4D} = \frac{e \, m_\tau}{\Lambda_L^2} \, \Big
    [ \,
      \bar\kappa^{(\tau\ell\gamma)}_1 \,
                [\bar L_{a,L} \sigma_{\lambda \rho} \ell_{3,R}] \,
       e^{-(1/2)\|\eta_{L_{a,L}}-\eta_{\ell_{3,R}}\|^2} +
       \bar\kappa^{(\tau\ell\gamma)}_2 \,
                [\bar \ell_{a,R} \sigma_{\lambda \rho}L_{3,L}] \, 
   e^{-(1/2)\|\eta_{\ell_{a,R}}-\eta_{L_{3,L}}\|^2} \ \Big ] \,
   F^{\lambda \rho} + h.c., \cr\cr
   &&
\label{tau_ell_gamma_ops_4d}
\eeqs
\end{widetext}
where $a=1,2$ corresponds to $\ell=e,\mu$. Here, analgously to Eq.
(\ref{kappa_meg_rel}), we set
\beq
\kappa^{(\tau \ell \gamma) \ '}_j = e \, m_\tau \kappa^{(\tau \ell \gamma)}_j,
\quad j=1,2 \ .
\label{kappa_tel_rel}
\eeq
Substituting the values of the distances
$\|\eta_{   L_{a,L}}-\eta_{\ell_{3,R}}\|$ and
$\|\eta_{\ell_{a,R}}-\eta_{   L_{3,L}}\|$ with $a=1,2$ from Table
\ref{lepton_distances_table} and again using Eqs. (2.63) and (2.65)) in
Ref. \cite{lee_shrock77}, we calculate the following branching ratios
in this $G_{SM}$ split-fermion model:
\begin{widetext}
\beqs
BR(\tau \to e \gamma) &=& \frac{\Gamma_{\tau \to e \gamma}}{\Gamma_\tau} =
\frac{B_{\tau-e} \Gamma_{\tau \to e \gamma}}
{\Gamma_{\tau \to \nu_\tau e \bar\nu_e}} \cr\cr
&=& \frac{192 \pi^3 \alpha_{em}B_{\tau-e}}{(G_F \Lambda_L^2)^2} 
\Big [|\bar\kappa^{(\tau e \gamma)}_1|^2 
  e^{-\|\eta_{L_{1,L}}-\eta_{\ell_{3,R}}\|^2}
  +     |\bar\kappa^{(\tau e \gamma)}_2|^2 
  e^{-\|\eta_{\ell_{1,R}}-\eta_{L_{3,L}}\|^2}
  \ \Big ] \cr\cr
&=& (2.93 \times 10^{-60}) \Big [ \,
  (3.08 \times 10^{-2}) |\bar\kappa^{(\tau e \gamma)}_1|^2 +
  |\bar\kappa^{(\tau e \gamma)}_2|^2 \ \Big ] \ .
\label{tau_e_gamma_br}
\eeqs
\beqs
BR(\tau \to \mu \gamma) &=& \frac{\Gamma_{\tau \to \mu \gamma}}{\Gamma_\tau} =
\frac{B_{\tau-\mu} \Gamma_{\tau \to \mu \gamma}}
{\Gamma_{\tau \to \nu_\tau \mu \bar\nu_\mu}} \cr\cr
&=& \frac{192 \pi^3 \alpha_{em}B_{\tau-\mu}}{(G_F \Lambda_L^2)^2} 
\Big [|\bar\kappa^{(\tau \mu \gamma)}_1|^2 
  e^{-\|\eta_{L_{2,L}}-\eta_{\ell_{3,R}}\|^2}
  +     |\bar\kappa^{(\tau \mu \gamma)}_2|^2 
  e^{-\|\eta_{\ell_{2,R}}-\eta_{L_{3,L}}\|^2}
  \ \Big ] \cr\cr
&=& (2.30 \times 10^{-20}) \Big [ \,
  |\bar\kappa^{(\tau e \gamma)}_1|^2 +
  (1.52 \times 10^{-4})|\bar\kappa^{(\tau e \gamma)}_2|^2 \ \Big ]  \ .
\label{tau_mu_gamma_br}
\eeqs
\end{widetext}
As is the case with the predictions of $BR(\tau \to \ell \gamma)$ in
the Standard Model extended to include massive neutrinos, these
$G_{SM}$ split-fermion model predictions are many orders of magnitude
below the respective experimental upper limits \cite{pdg,cpv}
\beq
BR(\tau \to e \gamma) < 3.3 \times 10^{-8}
\label{tau_e_gamma_exp}
\eeq
and
\beq
BR(\tau \to \mu \gamma) < 4.4 \times 10^{-8} \ .
\label{tau_mu_gamma_exp}
\eeq
%


\subsection{CLFV Decays $\ell_a \to \ell_b \ell_c \bar\ell_c$}
\label{CLFV_section}

Here we analyze the CLFV decays $\ell_a \to \ell_b \ell_c
\bar\ell_c$, where $a$, $b$, and $c$ are generation indices.  In the
Standard Model extended to include massive neutrinos, the rates for
these decays were calculated in \cite{lee_shrock77,petcov77b}. The
decay is very strongly suppressed by a cancellation between different
contributions, and the resultant branching ratio is many orders of
magnitude smaller than the current limit \cite{pdg,cpv}
\beq
BR(\mu \to e e \bar e) < 1.0 \times 10^{-12} \ .
\label{br_mu3e_limit}
\eeq
An analogous comment applies to the corresponding leptonic CLFV decays
of the $\tau$ lepton, for which experimental searches have obtained
the upper limits \cite{pdg,cpv} 
\beq
BR(\tau \to e e \bar e) < 2.7 \times 10^{-8} 
\label{br_tau3e_limit}
\eeq
\beq
BR(\tau\to \mu \mu \bar \mu) < 2.1 \times 10^{-8}
\label{br_tau3mu_limit}
\eeq
\beq
BR(\tau \to e \mu \bar \mu) < 1.7 \times 10^{-8} 
\label{br_tauemm_limit}
\eeq
\beq
BR(\tau \to \mu e \bar e) < 1.5 \times 10^{-8} \ .
\label{br_taumee_limit}
\eeq

In the $G_{SM}$ split-fermion model these decays can arise in several
ways. We begin with an analysis of contributions from higher KK modes
of SM gauge bosons. Recall that the locations of the lepton wave
function centers in the higher dimensions, as listed in Table
\ref{lepton_centers_table} and displayed in
Fig. \ref{fermion_centers_figure}, produce a nearly diagonal charged
lepton mass matrix.  Hence, the photon and $Z$ boson KK mode
couplings to the charged leptons are flavor-diagonal up to very small
correctioms, as can be seen in Eqs. (\ref{AKK_coupling_special}) and
(\ref{Zkk_coupling_diag_special}). Therefore, they do not contribute
significantly to the decays $\ell_a \to \ell_b \ell_c \bar\ell_c$.
We next consider the contributions of the higher KK modes of the
Higgs boson.  From Eq. (\ref{HKK_couplings_special}), we see that
the non-diagonal couplings of the Higgs boson to the charged leptons
are heavily suppressed. This is a consequence of the fact that the Higgs
interaction connects opposite-chirality components of fermion fields,
and, due to the large
separation among $L_{a, L}$ and $\ell_{b, R}$ for $a \ne b$, the
flavor-violating Higgs KK mode couplings are also suppressed. We will
estimate the contribution of the higher KK modes of the Higgs to the branching
ratios for $\ell_a \to \ell_b \ell_c \bar \ell_c$ here.

For illustrative purposes, let us consider the contribution of Higgs
boson KK modes to the branching ratios of CLFV decays $\ell_a \to 
\ell_b \ell_b \bar \ell_b$, where $a, b$ are generational
indices. Using eq. (\ref{HKK_couplings_special}), and assuming ${\cal          
  O}(1)$ higher-dimensional Yukawa couplings, the branching ratio for
$\ell_a \to \ell_b \ell_b \bar \ell_b$ mediated by the Higgs KK modes
is
\begin{multline}
  BR(\ell_a \to \ell_b \ell_b \bar \ell_b) \simeq
  BR(\ell_a \to \nu_a \ell_b \bar \nu_b)   \
  \frac{1}{2^{10}  \pi^4  (G_F \Lambda_L^2)^2}
  \\ \times \bigg( |{\cal S}^{(L)}_{H, ab}|^2 +
  |{\cal S}^{(R)}_{H, ab}|^2 \bigg) \ ,
\label{BR_CLFV_KK_eqn}
\end{multline}
where we have defined the term that depends on fermion wavecenter locations as
\begin{multline}
  {\cal S}^{(L)}_{H, ab} \equiv \sum_{m \in {\mathbb Z}^2_{\neq 0}}
  \frac{e^{-\frac{\pi^2}{(\mu L)^2} \| m \|^2}}{ \| m \|^2}
  \cos \bigg [  \frac{\pi}{\mu L}
  \{ m \cdot (\eta_{\ell_{a, L}}-\eta_{\ell_{b, L}}) \} \bigg ] \\
  \times \exp \bigg[ -\frac{1}{2} \Big (
    \|\eta_{\ell_{a, L}}-\eta_{\ell_{b, R}}\|^2 +
    \|\eta_{\ell_{b, L}}-\eta_{\ell_{b, R}}\|^2 \Big ) \bigg] \ .
  \label{SH_kk_eqn}
\end{multline}
Similarly, the expression for ${\cal S}^{(R)}_{H, ab}$ is obtained from
Eq. (\ref{SH_kk_eqn}) via the replacement $L \to R$.
From Eq. (\ref{lepton_mass_matrices}) it follows that the factor
$e^{-(1/2)\|\eta_{\ell_{b, L}}-\eta_{\ell_{b, R}}\|^2}$
is proportional to the mass of the lepton                   
$\ell_b$. This is a result of the fact that it arises from the Higgs KK
modes. The resulting numerical branching ratios are as follows:
\begin{align}
     BR(\mu \to e e \bar e) &\simeq 2.6 \times 10^{-50} \\
     BR(\tau \to e e \bar e) &\simeq 4.5 \times 10^{-78} \\
     BR(\tau \to \mu \mu \bar \mu) &\simeq 5.3 \times 10^{-35} \ .
\label{BR_CLFV_KK_results}
\end{align}
Evidently, these contributions from the KK modes are extremely small,
many orders of magnitude below experimental limits. Similarly we have
analyzed other CLFV $\ell_a \to \ell_b \ell_c \bar \ell_c$ processes,
and we find that their branching ratios are also far below experimental
bounds because of the exponential suppression in the non-diagonal Higgs KK
mode couplings.

A second way that $\ell_a \to \ell_b \ell_c \bar \ell_c$ decays can occur
is via local four-lepton operator products not directly involving KK modes
of gauge or Higgs fields.  We write these as 
\beq
{\cal L}^{(\ell_a \to \ell_b \ell_c \bar\ell_c)}_{4\ell}(x) =
\sum_r c^{(\ell_a \to \ell_b \ell_c \bar\ell_c)}_r
{\cal O}^{(\ell_a \to \ell_b \ell_c \bar\ell_c)}_r(x) \ .
\label{leff_ell3ell}
\eeq
These operators are local at the level of the low-energy
effective theory in four-dimensional spacetime but arise from four-fold
products of lepton fields in the higher-dimensions with wave function
centers located at different points in the higher-dimensional space.
These are given by the effective Lagrangian in the $4+n$ dimensional
space
\beq
{\cal L}^{(\ell_a \to \ell_b \ell_c \bar\ell_c)}_{4\ell,4+n}(x,y) =
\sum_r \kappa^{(\ell_a \to \ell_b \ell_c \bar\ell_c)}_r
O^{(\ell_a \to \ell_b \ell_c \bar\ell_c)}_r(x,y) \ .
\label{leff_ell3ell_higherdim}
\eeq
As before, one obtains the operators and their coefficients in the 4D
Lagrangian (\ref{leff_ell3ell}) by integration of the Lagrangian
(\ref{leff_ell3ell_higherdim}) over the higher dimensions.

A third way in which the decay $\ell_a \to \ell_b \ell_c \bar \ell_c$
can arise is via a combination of an operator involving $\ell_a$,
$\ell_b$ and some set of virtual SM fields producing the $\ell_c
\bar\ell_c$ pair.  For example, an initial $\ell_a$ can make a
transition to $\ell_b$ and a photon, as mediated by the operators in
${\cal L}^{(\ell_a \to \ell_b\gamma)}_{eff}$ in
Eq. (\ref{meg_ops_4d}), but instead of the photon being onshell, it is
virtual and materializes into the $\ell_c \bar\ell_c$ in the final
state.  The amplitude for this contribution to $\ell_a \to \ell_b
\ell_c \bar\ell_c$ does not involve a four-lepton local operator of
the form (\ref{leff_ell3ell}), but instead the operator in
Eq. ({\ref{meg_ops_4d}) combined with the virtual photon propagator
  connected to the $\ell_c \bar \ell_c$ bilinear.  A similar
  contribution arises from diagrams in which the virtual photon is
  replaced by a $Z$ boson.  These are analogous to the diagrams shown
  in Figs. 2(a) and 2(b) of Ref. \cite{lee_shrock77}.  A third type of
  contribution arises from a box diagram involving virtual $W^+$ and
  $W^-$ vector bosons and two internal neutrino lines, analogous to
  Fig. 2(g) in Ref. \cite{lee_shrock77}.

  We first calculate the contribution from the four-lepton operators
  in Eq. (\ref{leff_ell3ell}) in the $G_{SM}$ split-fermion model.  As in
  \cite{bvd,nnblrs}, we classify these according to the resultant
  integrals that they yield upon integration over the extra
  dimensions. We find six classes of integrals. Since the effective
  mass scale governing the decay is required to be large compared with
  the electroweak symmetry breaking scale, it follows that the
  operators ${\cal O}^{(\ell_a \to \ell_b \ell_c \bar\ell_c)}_r$ must
  be invariant with respect to the SM gauge group, $G_{SM}$. Six such
  operators are listed below (with Roman indices being SU(2)$_L$
  indices here and the subscripts such as $LLLL$ indicating the
  chirality of the four lepton fields):
  \beqs
  {\cal O}^{(\ell_a \to \ell_b \ell_c \bar\ell_c)}_{LLLL} &=& 
  [\bar L_{b,L,i} \gamma_\lambda L^i_{a,L}]
  [\bar L_{c,L,j}\gamma^\lambda L^j_{c,L}] \cr\cr
  &=&\bigg [ [\bar\nu_{b,L}\gamma_\lambda \nu_{a,L}] +
     [\bar\ell_{b,L}\gamma_\lambda \ell_{a,L}] \Big ] \times \cr\cr
& \times & \Big [[\bar\nu_{c,L}\gamma^\lambda \nu_{c,L}] +
    [\bar\ell_{c,L}\gamma^\lambda \ell_{c,L}] \bigg ] \cr\cr
  &&
\label{O_ell3ell_LLLL}
\eeqs
\beqs
  {\cal O}^{(\ell_a \to \ell_b \ell_c \bar\ell_c)}_{LLRR} &=& 
  [\bar L_{b,L,i} \gamma_\lambda L^i_{a,L}]
  [\bar \ell_{c,R}\gamma^\lambda \ell_{c,R}] \cr\cr
  &=&\Big [ [\bar\nu_{b,L}\gamma_\lambda \nu_{a,L}] +
     [\bar\ell_{b,L}\gamma_\lambda \ell_{a,L}] \Big ] 
  [\bar\ell_{c,R}\gamma^\lambda \ell_{c,R}] \cr\cr
  &&
\label{Oell3ell_LLRR}
\eeqs     
\beqs
  {\cal O}^{(\ell_a \to \ell_b \ell_c \bar\ell_c)}_{RRLL} &=& 
  [\bar \ell_{b,R} \gamma_\lambda \ell_{a,R}]
  [\bar L_{c,L,i}\gamma^\lambda L^i_{c,L}] \cr\cr
  &=&    [\bar\ell_{b,R}\gamma_\lambda \ell_{a,R}]
  \Big [ [\bar \nu_{c,L}\gamma^\lambda \nu_{c,L}] +
    [\bar\ell_{c,L}\gamma^\lambda \ell_{c,L}] \Big ] \cr\cr
  &&
\label{Oell3ell_RRLL}
\eeqs     
\beq
  {\cal O}^{(\ell_a \to \ell_b \ell_c \bar\ell_c)}_{RRRR} = 
  [\bar \ell_{b,R} \gamma_\lambda \ell_{a,R}]
  [\bar \ell_{c,R} \gamma^\lambda \ell_{c,R}]
\label{Oell3ell_RRRR}
\eeq     
\beqs
    {\cal O}^{(\ell_a \to \ell_b \ell_c \bar\ell_c)}_{LRRL} &=&
    [\bar L_{b,L,i} \ell_{a,R}][\bar\ell_{c,R} L^i_{c,L}] \cr\cr
    &=& [\bar \nu_{b,L} \ell_{a,R}][\bar\ell_{c,R} \nu_{c,L}] + 
    [\bar \ell_{b,L} \ell_{a,R}][\bar\ell_{c,R} \ell_{c,L}] \cr\cr
    &&
    \label{Oell3ell_LRRL}
    \eeqs
\beqs
    {\cal O}^{(\ell_a \to \ell_b \ell_c \bar\ell_c)}_{RLLR} &=&
    [\bar \ell_{b,R} L^i_{a,L}][\bar L_{c,L,i} \ell_{c,R}] \cr\cr
    &=& [\bar \ell_{b,R} \nu_{a,L}][\bar\nu_{c,L} \ell_{c,R}] + 
    [\bar \ell_{b,R} \ell_{a,L}][\bar\ell_{c,L} \ell_{c,R}] \ . \cr\cr
    &&
\label{Oell3ell_RLLR}
\eeqs
(Here we show all terms arising from these $G_{SM}$-invariant operators,
but only the ones with all charged leptons are relevant for our analysis in
this section.)

Integrating these four-lepton operator products over the extra dimensions
and using the integration formula (\ref{intform}), we obtain the following
results: 

\begin{widetext}
\beq
I^{(\ell_a \to \ell_b \ell_c \bar\ell_c)}_{LLLL} = b_4 \,
\exp \bigg [ -\frac{1}{4}\Big \{
  \|\eta_{L_{a,L}}-\eta_{L_{b,L}}\|^2 +
2 \|\eta_{L_{a,L}}-\eta_{L_{c,L}}\|^2 +
2 \|\eta_{L_{b,L}}-\eta_{L_{c,L}}\|^2 \Big \} \bigg ] 
\label{int_op_ell3ell_LLLL}
\eeq
\beq
I^{(\ell_a \to \ell_b \ell_c \bar\ell_c)}_{LLRR} = b_4 \,
\exp \bigg [ -\frac{1}{4}\Big \{
 \|\eta_{L_{a,L}}-\eta_{L_{b,L}}\|^2 +
2\|\eta_{L_{a,L}}-\eta_{\ell_{c,R}}\|^2 +
2\|\eta_{L_{b,L}}-\eta_{\ell_{c,R}} \|^2 \Big \} \bigg ] \quad 
\label{int_op_ell3ell_LLRR}
\eeq

\beq
I^{(\ell_a \to \ell_b \ell_c \bar\ell_c)}_{RRLL} = b_4 \,
\exp \bigg [ -\frac{1}{4}\Big \{
 \|\eta_{\ell_{a,R}}-\eta_{\ell_{b,R}}\|^2 +
2\|\eta_{\ell_{a,R}}-\eta_{L_{c,L}}\|^2 +
2\|\eta_{\ell_{b,R}}-\eta_{L_{c,L}}\|^2 \Big \} \bigg ]
\label{int_op_ell3ell_RRLL}
\eeq
\beq
I^{(\ell_a \to \ell_b \ell_c \bar\ell_c)}_{RRRR} = b_4 \,
\exp \bigg [ -\frac{1}{4}\Big \{
 \|\eta_{\ell_{a,R}}-\eta_{\ell_{b,R}}\|^2 +
2\|\eta_{\ell_{a,R}}-\eta_{\ell_{c,R}}\|^2 +
2\|\eta_{\ell_{b,R}}-\eta_{\ell_{c,R}}\|^2 \Big \} \bigg ]
\label{int_op_ell3ell_RRRR}
\eeq
\beqs
I^{(\ell_a \to \ell_b \ell_c \bar\ell_c)}_{LRRL} &=& b_4 \,
\exp \bigg [ -\frac{1}{4}\Big \{
 \|\eta_{L_{b,L}}-\eta_{\ell_{a,R}}\|^2 +
 \|\eta_{L_{b,L}}-\eta_{\ell_{c,R}}\|^2 +
 \|\eta_{L_{b,L}}-\eta_{L_{c,L}}\|^2 \cr\cr
 &+&
 \|\eta_{\ell_{a,R}}-\eta_{\ell_{c,R}}\|^2 +
 \|\eta_{\ell_{a,R}}-\eta_{L_{c,L}}\|^2 +
 \|\eta_{\ell_{c,R}}-\eta_{L_{c,L}}\|^2 \Big \} \bigg ]
\label{int_op_ell3ell_LRRL}
\eeqs
\beqs
I^{(\ell_a \to \ell_b \ell_c \bar\ell_c)}_{RLLR} &=& b_4 \,
\exp \bigg [ -\frac{1}{4}\Big \{
 \|\eta_{\ell_{b,R}}-\eta_{L_{a,L}}\|^2 +
 \|\eta_{\ell_{b,R}}-\eta_{L_{c,L}}\|^2 +
 \|\eta_{\ell_{b,R}}-\eta_{\ell_{c,R}}\|^2 \cr\cr
 &+&
 \|\eta_{L_{a,L}}-\eta_{L_{c,L}}\|^2 +
 \|\eta_{L_{a,L}}-\eta_{\ell_{c,R}}\|^2 +
 \|\eta_{L_{c,L}}-\eta_{\ell_{c,R}}\|^2 \Big \} \bigg ] \ .
\label{int_op_ell3ell_RLLR}
\eeqs
\end{widetext}
%


Following the notation of Eq. (\ref{irgen}), we list the values of the
integrals of these operators over the extra dimensions in Table
\ref{ell3ell_integrals_table} for the various $\ell_a \to \ell_b
\ell_c \bar\ell_c$ decays.  


\begin{table*}
  \caption{\footnotesize{Integrals $I^{(\ell_a \to \ell_b \ell_c \bar\ell_c)}$
      for $\ell_a \to \ell_b \ell_c \bar\ell_c$ decays.}}
  \begin{center}
\begin{ruledtabular}
  \begin{tabular}{|c|c|c|c|c|c|c|c|c|c|}
  decay & $a$ & $b$ & $c$ & $I_{LLLL}$ & $I_{LLRR}$ & $I_{RRLL}$ &
  $I_{RRRR}$ & $I_{LRRL}$ & $I_{RLLR}$ \\ \hline
  $\mu \to e e \bar e$  & 2 & 1 & 1 &
  $3.92 \times 10^{-24}$  &
  $3.06 \times 10^{-39}$  &
  $0.709 \times 10^{-31}$ &
  $3.74 \times 10^{-42}$  &
  $0.709 \times 10^{-31}$ &
  $3.06 \times 10^{-39}$  \\ \hline
  $\tau \to e e \bar e$ & 3 & 1 & 1 &
  $3.21 \times 10^{-29}$   &
  $0.618 \times 10^{-40}$  &
  $3.23 \times 10^{-45}$   &
  $0.846 \times 10^{-39}$  &
  $3.23 \times 10^{-45}$   &
  $0.618 \times 10^{-40}$  \\ \hline
  $\tau \to \mu \mu \bar \mu$ & 3 & 2 & 2 &
  $0.0709$  &
  $1.88\times 10^{-11}$   &
  $2.29 \times 10^{-17}$  &
  $2.40 \times 10^{-25}$  &
  $2.29 \times 10^{-17}$  &
  $1.88 \times 10^{-11}$  \\ 
\end{tabular}
\end{ruledtabular}
\end{center}
\label{ell3ell_integrals_table}
\end{table*}

In presenting a result for the contribution of these four-lepton
operators to the decay $\ell_a \to \ell_b \ell_c \bar\ell_c$ in the
case where $\ell_a = \tau$, it is convenient to normalize
relative to one of the allowed leptonic decays of the $\tau$, using
the identity
\begin{widetext}
\beq
BR(\ell_a \to \ell_b \ell_c \bar\ell_c) =
\frac{\Gamma_{\ell_a \to \ell_b \ell_c \bar\ell_c}}
     {\Gamma_{\ell_a}}  = 
     \frac{BR(\ell_a \to \nu_a \ell_b \bar\nu_b)
       \Gamma_{\ell_a \to \ell_b \ell_c \bar\ell_c}}
     {\Gamma_{\ell_a \to \nu_a \ell_b \bar\nu_b}} \ ,
\label{ell3ell_identity}
\eeq
where $BR(\mu \to \nu_\mu e \bar\nu_e)=1$, and the values of $BR(\tau
\to \nu_\tau \ell \bar\nu_\ell)$ were given in Eqs. (\ref{br_tau_e})
and (\ref{br_tau_mu}) for $\ell=e,\mu$. The contribution of these
local four-lepton ($4\ell$)operators to the branching ratio for the
decay $\ell_a \to \ell_b \ell_c \bar\ell_c$ is
\beq
BR(\ell_a \to \ell_b \ell_c \bar\ell_c)_{4\ell} =
\xi_{abc}BR(\ell_a \to \nu_a \ell_b \bar\nu_b)\bigg (
\frac{v^2}{2\Lambda_L^2)^2} \bigg )^2 
    \Big | \sum_r \bar\kappa^{(\ell_a \to \ell_b \ell_c \bar\ell_c)}_r
      e^{-S^{(\ell_a \to \ell_b \ell_c \bar\ell_c)}_r} \Big |^2
  \bigg ( \frac{\mu^2}{\pi \Lambda_L^2} \bigg )^2 \ .
\label{br_ell3ell_4lepton_contrib}
\eeq
\end{widetext}
Here, $\xi_{abc}$ is a factor that takes account of the presence of
the direct minus the crossed diagram and the factor of 1/2 in the rate
in the case where two of the fermions in the final state are identical;
this will not be important for our conclusions here.

For $\mu \to e e \bar e$ in this $G_{SM}$ split-fermion model with
$n=2$ and assuming the dimensionless coefficients $\bar\kappa^{(\mu
  \to e e \bar e)}_r \sim O(1)$, the operator ${\cal O}^{(\mu \to e e
  \bar e)}_{LLLL}$ gives the dominant contribution, and we find
\beq
BR(\mu \to e e \bar e)_{4\ell} \simeq 10^{-53} \ .
\label{br_mu3e_4ell}
\eeq
We conclude that, with our set of lepton wave function centers,
contributions of the second kind dominate over these contributions of
four-lepton operators to the decay $\mu \to e e \bar e$.  A rough
estimate of the second type of contributions can be obtained by
focusing on the process $\ell_a \to \ell_b + \gamma$
mediated by ${\cal L}_{\mu e \gamma}$ that was studied
in Section \ref{meg_subsection}, but with the modification here that
the photon is virtual instead of real, and produces the $e^+e^-$ pair
in the final state.  We obtain the approximate estimate
\beq
\Gamma_{\mu \to e e \bar e} \simeq (4\pi\alpha_{em})
\bigg[\frac{\bar R_3^{(e e \bar e)}}{R_2^{(e \gamma)}}\bigg]
\Gamma_{\mu \to e \gamma} \ ,
\label{mu3e_mug_relation}
\eeq
where $R_2^{(e \gamma)}$ and $\bar R_3^{(e e \bar e)}$ are the
two-body and dimensionless 3-body phase space respectively. Since
$m_e^2 \ll m_\mu^2$, these phase space factors reduce to $R_2^{(e
  \gamma)} =1/(2^3 \pi)$ and $\bar R_3^{(e e \bar e)} = 1/(2^8
\pi^3)$. Denoting the four-momentum carried by the virtual photon as
$q$, we note that the $1/q^2$ factor in the amplitude from the photon
propagator is cancelled by momenta of order $m_\mu^2$ in the
calculation of the rate.  From Eq. (\ref{mu3e_mug_relation}), using
the result (\ref{meg_br}), we thus obtain the following estimate for
the contribution to $\mu \to e e \bar e$ decay in this $G_{SM}$
split-fermion model:
\beqs
&& BR({\mu \to e e \bar e}) \simeq \bigg( \frac{\alpha_{em}}{8 \pi} \bigg)
  BR({\mu \to e \gamma}) \cr\cr
  &=& (2.6 \times 10^{-36})  \Big [ \, |\bar\kappa^{(\mu e \gamma)}_1|^2
    + (1.8 \times 10^{-27})|\bar\kappa^{(\mu e \gamma)}_2|^2 \ \Big ] \ .
  \cr\cr
  &&
\label{mu3e_mug_branching}
\eeqs
This is many orders of magnitude below the current
experimental upper limit, (\ref{br_mu3e_limit}), on the branching
ratio for this decay.

In a similar manner, we can analyze the CLFV processes $\tau \to
\ell_b \ell_c \bar \ell_c$, where $\ell_b$ and $\ell_c$ can be $e$ or
$\mu$.  We focus on the decay $\tau \to \mu \mu \bar \mu$, because,
for our choice of lepton wave function centers, the integral $I^{(\tau
  \to \mu \mu \bar\mu)}_{LLLL}$ is relatively unsuppressed.  This is
due to the fact that the dimensionless distance $\|\eta_{L_{3, L}} -
\eta_{L_{2, L}}\| = 1.878$ is relatively small compared with other
distances entering into the relevant integrals.  Therefore, the
operator ${\cal O}^{(\tau \to \mu \mu \bar\mu)}_{LLLL}$ provides the
dominant contribution to the decay $\tau \to \mu \mu \bar \mu$.  The
corresponding effective Lagrangian in the four-dimensional low-energy
field heory is
\begin{widetext}
\beq
    {\cal L}_{eff, 4D}^{(\tau \to \mu \mu \bar \mu)}(x)  =
    \frac{\bar \kappa^{(\tau \to \mu \mu \bar \mu)}}{\Lambda_L^2}
    \bigg(\frac{\mu^2}{\pi \Lambda_L^2} \bigg)
    \exp \bigg[-\frac{3}{4} \|\eta_{L_{2,L}}- \eta_{L_{3,L}}\|^2 \bigg]
         [\bar L_{2,L}(x)\gamma_\lambda L_{3,L}(x)]
         [\bar L_{2,L}(x)\gamma^\lambda L_{2,L}(x)] + h.c. 
        \label{leff_tauto3mu}
\eeq
Using Eq. (\ref{ell3ell_identity}), we find that in the 
$G_{SM}$ split-fermion model with $n=2$,
\beq
BR(\tau \to \mu \mu \bar \mu) \simeq 
2BR(\tau \to \nu_\tau \mu \bar \nu_\mu)
|\bar \kappa^{(\tau \to \mu \mu \bar \mu)}_{LLLL}|^2 \,
 \bigg(\frac{v^2}{2 \Lambda_L^2}\bigg)^2
        \bigg(\frac{\mu^2}{\pi \Lambda_L^2} \bigg)^2
        e^{-\frac{3}{2} \|\eta_{L_{2, L}}- \eta_{L_{3, L}}\|^2 } \ . 
\label{BR_tauto3mu_eqn}
\eeq
\end{widetext}
Substituting the value of $BR(\tau \to \nu_\tau \mu \bar\nu_\mu)$ from
Eq. (\ref{br_tau_mu}) and the value of
$\|\eta_{L_{2,L}}-\eta_{L_{3,L}}\|$ from Table \ref{lepton_distances_table},
we obtain the resulting estimate
\beq
BR(\tau \to \mu \mu \bar \mu) \simeq 10^{-9} \,
|\bar \kappa^{(\tau \to \mu \mu \bar \mu)}_{LLLL}|^2 \ .
\label{BR_tauto3mu_value}
\eeq
For $|\bar\kappa^{(\tau \to \mu \mu \bar \mu)}_{LLLL}| \lsim O(1)$,
this is in accord with the current experimental upper bound $BR(\tau
\to \mu \mu \bar \mu) < 2.1 \times 10^{-8}$ given in
Eq. (\ref{br_tau3mu_limit}). As discussed in Appendix
\ref{lepton_centers_appendix}, although the nine distance constraints
in Table \ref{L_nu_distances_table} do not restrict $\|\eta_{L_{3, L}}
- \eta_{L_{2, L}}\|$, the selection criteria in this appendix, in
conjunction with the symmetries observed there for a class of
solutions essentially fix this distance. This, in turn, produces the
branching ratio (\ref{BR_tauto3mu_value}), which, for $|\bar
\kappa^{(\tau \to \mu \mu \bar \mu)}_{LLLL}| \simeq 1$, is
approximately a factor of 20 smaller than the current experimental
limit on $BR(\tau \to \mu \mu \bar\mu)$ in
Eq. (\ref{br_tau3mu_limit}).  Using similar methods, we find that the
branching ratios for the other CLFV $\tau$ decays $\tau \to e e \bar
e$, $\tau \to e \mu \bar \mu$, and $\tau \to \mu e \bar e$ are many
orders of magnitude below the respective experimental upper bounds.


\subsection{Transition Magnetic Moments of Majorana Neutrinos}

The diagonal magnetic and electric dipole moments vanish for a Majorana
(i.e., self-conjugate) neutrino, but the transition magnetic and electric
dipole moments are nonzero.  These are given by the following terms in
the matrix element $\langle \nu_b | J_{em,\lambda} | \nu_a \rangle$: 
\beq
 [\bar\nu_b \sigma_{\lambda\rho} \{ (F_2^V)_{\nu,ba} +
      (F_2^A)_{\nu,ba}\gamma_5 \}) \nu_a] \, q^\rho \ , 
\label{dipolemoments}
\eeq
where $q$ is the four-momentum of the photon.  The transition magnetic
and electric dipole moments of a Majorana neutrino $\nu_a$ to $\nu_b$
(with $a \ne b$) in the SM (extended to include neutrino masses) have
the respective magnitudes \cite{rs82,svem,pw,nieves}
\begin{widetext}
\beq
|\mu_{\nu,ba}| = \frac{3eG_F (m_{\nu_a} + m_{\nu_b})}{16\pi^2 \sqrt{2}} \,
\bigg | \sum_{k=1}^3 {\rm Im}(U^*_{kb}U_{ka})
\Big ( \frac{m_{\ell_a}^2}{m_W^2} \Big ) \bigg |
\label{munu_majorana}
\eeq
and
\beq
|d_{\nu,ba}| = \frac{3eG_F |m_{\nu_a} - m_{\nu_b}|}{16\pi^2 \sqrt{2}} \,
\bigg | \sum_{k=1}^3 {\rm Re}(U^*_{kb}U_{ka})
\Big ( \frac{m_{\ell_a}^2}{m_W^2}  \Big ) \bigg | \ .
\label{dnu_majorana}
\eeq
\end{widetext}
where $m_{\ell_a}$ denotes the mass of the charged lepton $\ell_a$. 
In contrast, for example, the diagonal magnetic moment of a Dirac neutrino is
\cite{fs}
\beqs
\mu_\nu &=& \frac{3eG_F m_\nu}{8\pi^2 \sqrt{2}} \cr\cr
&=& \Big ( \frac{3G_F m_\nu m_e}{4\pi^2 \sqrt{2}}\Big ) \, \mu_B \cr\cr
&=& (3.20 \times 10^{-19}) \Big ( \frac{m_\nu}{1 \ {\rm eV}}\Big ) \, \mu_B \ .
\label{magmom}
\eeqs
The most stringent upper limit on a diagonal Dirac or transition
magnetic or electric moment of a neutrino arises from astrophysics,
specifically stellar cooling rates, and is $\sim 10^{-12} \mu_B$
\cite{pdg,gs_emrev,bk_emrev}.

We proceed to calculate contributions to the transition magnetic moment
of a Majorana neutrino in the split fermion model with $n=2$.
the operator contributing to transition magnetic moment in the
6-dimensional theory is
\beq
O_{mm;ba;4+n} = \frac{g'}{\Lambda_L^3} A_f^2 A_{bos.}^2 A_F [L^T_{b,L} C 
  \sigma_{\lambda\rho}L_{a,L}]\, \phi^2 F_B^{\lambda \rho} \ ,
\label{o_magmom_higherdim}
\eeq
where $g'=e/\sin\theta_W$ and $F_B^{\lambda\rho}$ are the weak
hypercharge U(1)$_Y$ gauge coupling and field strength tensor, and the
dimensionful normalization constants $A_f$, $A_{bos.}$, and $A_F$ were
given in Eqs. (\ref{af}), (\ref{ab}), and (\ref{afmunu}).  We have
also incorporated the mass dimension of the gauge coupling $g'$
in the prefactor.  The integral of the fermion bilinear over the extra
dimensions yields the factor
\beq
I_{mm;ba} = e^{-(1/2)\|\eta_{L_{b,L}}-\eta_{L_{a,L}}\|^2} \ .
\label{munu_Lb_nuaR}
\eeq
With our solution for the locations of the wave function centers of the
leptons in the extra dimensions, these have the values
$I_{mm;12}=2.49 \times 10^{-16}$, $I_{mm;13}=1.01 \times 10^{-19}$, and
$I_{mm;23}=0.171$. The fact that $I_{mm;23}$ is much larger than
$I_{mm;12}$ and $I_{mm;13}$ is a consequence of the property that the
distance $\|\eta_{L_{2,L}}-\eta_{L_{3,L}}\|$ is considerably smaller than
the distances $\|\eta_{L_{1,L}}-\eta_{L_{2,L}}\|$ and 
$\|\eta_{L_{1,L}}-\eta_{L_{3,L}}\|$, combined with the exponential sensitivity
of $I_{mm;ba}$ to the squares of these distances.) 

After this integration, the operator reduces to 
\beq
\frac{e \,(v/\sqrt{2})^2}{\Lambda^3_L \, \sin\theta_W} \,
I_{mm;ba} \, [L^T_{b,L}(x)\sigma_{\lambda\rho}L_{a,L}(x)]  \ .
\eeq
The resultant transition magnetic moments of a Majorana
neutrino,$\mu_{\nu;ba}$, are of order
\beqs
\mu_{\nu;ba} &  \simeq & \frac{e v^2}{2 \Lambda^{3}_{L} \, \sin \theta_W} \,
I_{mm;ba} \cr\cr
& \simeq & \Big ( \frac{(2m_e) v^2}{\Lambda^3_L} \, I_{mm;ba} \Big ) \mu_B
\ .
\eeqs
The largest of these is $\mu_{\nu;23} \simeq 10^{-14} \ \mu_B$, while
$\mu_{\nu;12}$ and $\mu_{\nu;13}$ are much smaller.  A similar analysis
applies for the transition electric dipole moments of the neutrinos. 
These transition magnetic and electric dipole moments are all well below
the astrophysical upper bound of $\sim 10^{-12} \mu_B$ on the (magnitude)
of diagonal or transition magnetic or electric neutrino dipole moments
\cite{pdg,gs_emrev,bk_emrev}.


\subsection{Neutrinoless Double Beta Decay and $|\Delta L|=2$ Hadron Decays}

Here we analyze predictions for neutrinoless double beta decay and
$|\Delta L|=2$ hadron decays in split-fermion models.  Neutrinoless double beta
decay is a $\Delta L=2$ process and its occurence would indicate that neutrinos
are self-conjugate, Majorana fermions. As discussed above, the Majorana
nature of the neutrino is natural in both (a) the SM extended to include
SM-singlet right-handed neutrinos, since the right-handed neutrino mass terms
are $|\Delta L|=2$ operators, and (b) in the LRS model, since the vacuum
expectation value of the $\Delta_R$ Higgs breaks $B-L$ by two units and,
among other things, yields $|\Delta L|=2$ right-handed neutrino bilinears
\cite{mm80,lrs81,rmv,rmm,schechter_valle82}.
Searches for neutrinoless double beta decay
have been performed for many decades and have set quite stringent lower
bounds on the half-lives of various decays of this type involving parent
nuclei such as ${}^{76}$Ge and ${}^{136}$Xe; some recent reviews are
\cite{aee}-\cite{rodejohann}. With calculations of nuclear matrix
elements, these lower limits can be transformed into upper limits for the
effective Majorana mass quantity
$m_{\beta\beta} = |\sum_{j=1}^3 U_{ej}^2 m_{\nu_j}|$; at present the
non-observation of neutrinoless double beta decays yields upper limits of
$m_{\beta\beta} \lsim 0.3$ eV.  At a nucleon level, neutrinoless double
beta decay is the process $nn \to ppee$, and at the quark level,
$dd \to uuee$; in both cases, the coefficient $c_{\beta\beta}$ of the
corresponding six-fermion operator in an effective Lagrangian has 
Maxwellian mass dimension $-5$ (in four-dimensional spacetime).
Thus, an equivalent way of expressing the experimental limits from the
non-observation of neutrinoless beta decay is as an upper limit on this
coefficient.  Current data give the upper limit
  \cite{pdg},\cite{vergados_ejiri}-\cite{rodejohann}.
\beq
|c_{\beta\beta}| \lsim 10^{-19} \ {\rm GeV}^{-5} \ .
\label{cbblimit}
\eeq

There are many operators arising from physics beyond the Standard
Model that can contribute to neutrinoless double beta decay
\cite{mm80,lrs81,vergados_ejiri,engel,rodejohann,vicenzo}.
We first consider
one of the lowest-dimension operators invariant under the SM gauge
group, namely the six-fermion operator in the $d=4+n$ space,
\beq
    O_{\beta\beta} =
    \kappa_{\beta\beta} \, [\bar{d}_R\gamma^\lambda u_R]
    [\bar{d}_R\gamma^\lambda u_R][e^T_R C e_R] + h.c.,
    \label{obb}
\eeq
where here the Lorentz index runs over all $4+n$ dimensions and we set
$\kappa_{\beta\beta,(6)} \equiv \kappa_{\beta\beta}$.
The $\eta$-dependent part of the six-fermion operator product in (\ref{obb})
is
\beq
A^6 \, e^{ -2\|\eta-  \eta_{u_{1,R}}\|^2
          -2\|\eta-   \eta_{d_{1,R}}\|^2
          -2\|\eta-\eta_{\ell_{1,R}} \|^2} \ .
\label{bb_vfun}
\eeq
Carrying out the
integration over the extra dimensions, we obtain the corresponding
${\cal O}_{\beta\beta}$ in $d=4$ with coefficient
\beq
c_{\beta\beta,SF} =
\frac{\bar\kappa_{\beta\beta}}{\Lambda_L^5} \,
\Big ( \frac{2\mu^2}{ 3^{1/2}\pi \Lambda^2} \Big )^n \,
e^{-S_{\beta\beta,(6)}} \ , 
\label{obb4D}
\eeq
where, from an application of Eq. (\ref{intform}), 
\begin{widetext}
\beq
e^{-S_{\beta\beta,(6)}} = \exp \bigg [ -\frac{2}{3}\Big \{
\|\eta_{u_{1,R}}-   \eta_{d_{1,R}} \|^2 +
\|\eta_{u_{1,R}}-\eta_{\ell_{1,R}} \|^2 +
\|\eta_{d_{1,R}}-\eta_{\ell_{1,R}} \|^2 \Big \} \bigg ] \ .
\label{sr_opbb}
\eeq
\end{widetext}
With our choice of locations of wave function centers, we calculate
that $\|\eta_{u_{1,R}}- \eta_{d_{1,R}}\|=4.72$,
$\|\eta_{u_{1,R}}-\eta_{\ell_{1,R}}\|=18.22$, and
$\|\eta_{d_{1,R}}-\eta_{\ell_{1,R}} \|=17.08$. Substituting these
values into Eq. (\ref{sr_opbb}) yields an extremely small value,
$e^{-S_{\beta\beta,(6)}}$, since $S_{\beta\beta,(6)}=430.70$.  When
one encounters such a small number, one naturally inquires how stable
it is to perturbations in the distances of the wave function centers
of the fermion fields.  The property that this quantity
$e^{-S_{\beta\beta,(6)}}$ is extremely small remains true under small
perturbations of the positions of the relevant wave function centers
and hence the relevant distances entering into $S_{\beta\beta,(6)}$.
With $|\bar\kappa_{\beta\beta}| \sim O(1)$ and $n=2$, using our values
of $\mu$ and $\Lambda_L$, the prefactor multiplying
$e^{-S_{\beta\beta,(6)}}$ in Eq. (\ref{obb4D}) is
$\simeq 1 \times 10^{-20}$, yielding an even smaller value for the coefficient
$c_{\beta\beta}$, which is many order of magnitude smaller than the
current upper limit (\ref{cbblimit}). Thus, the contribution from the
split-fermion models to neutrinoless double beta decay are negligibly
small.  For operators with higher dimensions, the contributions are
even smaller, and therefore we do not discuss them. Similar comments
apply to other $|\Delta L|$=2 processes such as $K^+ \to
\pi^-\mu^+\mu^+$ \cite{littenberg_shrock,na62}.  As noted above, in
the LRS split-fermion model, below the scale of $v_R \sim 10^3$ TeV
the $G_{LRS}$ symmetry is broken to $G_{SM}$. Hence, using usual
low-energy field theory methods, one analyzes the physics in terms of
the fields of the SM model.  Therefore, our discussion above applies
to the split-fermion model with a $G_{SM}$ gauge symmetry and also a
$G_{LRS}$ gauge in the ultraviolet.

  Returning to the 4D LRS LRS theory, for completeness, we add some
  further remarks.  In this theory, one contribution to neutrinoless
  double beta decay arises from a graph in which two $d$ quarks make
  transitions to $u$ quarks via vertices connecting to two virtual
  $W^-_L$ vector bosons, which then connect via an internal light
  neutrino line with emission of the two electrons.  There are also
  additional contributions, which have been analyzed in a number of
  papers; some early studies were
  \cite{mm80,lrs81,rmv,rmm,schechter_valle82,picciotto,hpr,bbm} and some recent
  ones are \cite{goswami}-\cite{sppm}. These contributions include
  (i) a graph in which two $d$ quarks make transitions to $u$ quarks
  via vertices connecting to two $W_R^-$ vector bosons, which then
  connect to an internal heavy $\nu_R$ neutrino line, with emission of
  the $2e^-$; (ii) the corresponding graph in which the two $W_R^-$
  lines meet at a single vertex, producing a virtual $\Delta_R^{--}$
  in the $s$ channel, which then materializes to the $2e^-$ pair;
  (iii) the corresponding graph in which the $W_R^-$ lines are
  replaced by $W_L^-$ lines and the $s$-channel $\Delta_R^{--}$ by a
  $\Delta_L^{--}$; and (iv) other corresponding graphs with the
  $W_R^-$ or $W_L^-$ lines replced by $\phi_1^-$ or $\phi_2^-$ lines
  meeting to produce $s$-channel $\Delta_L^{--}$ or $\Delta_R^{--}$
  Higgs. In the present LRS extra-dimensional model, because $v_R$ and
  hence $m_{W_R}$ are quite large, $\sim 10^3$ TeV, the graphs
  involving $W_R$ and/or $\Delta_R$ internal propagators make
  negligibly small contributions. These additional contributions are
  also in accord with the bound (\ref{cbblimit}). For example, the
  $W_L W_L \Delta_L$ vertex in the graph (iii) can be suppressed by
  a large $\Delta_L$ mass and, moreover, contains a factor of
  $v_L$, which is required to be $\ll v$ in order not to upset the
  experimentally observed property that
  $m_{W_L}^2/(m_Z^2\cos^2\theta_W)$ is very close to 1. 


\section{Baryogenesis and Dark Matter in the Models}
\label{baryogensis_dm_section} 

We now discuss ways to incorporate baryogenesis and dark matter in the
models. We argue that all the ingredients for baryogenesis are already
present in the models discussed above, whereas to understand dark matter
of the universe, one needs a very minimal extension.

\subsection{Baryogenesis}

Baryogenesis requires that the three Sakharov conditions are
satisfied: (i) baryon number violation, (ii) C and CP violation, and
(iii) dynamical evolution that is out of thermal equilibrium
\cite{sakharov}.  One of the mechanisms that can account for
baryogenesis is to generate the baryon asymmetry via a first step
involving leptogenesis~\cite{fukugita}.  In our models, this
leptogenesis mechanism can be used to explain baryogenesis.  The basic
mechanism of leptogenessis~\cite{fukugita} requires the presence of
right-handed neutrinos producing a seesaw mechanism for small neutrino
masses, together with CP-violating Dirac neutrino Yukawa coupling coupling
that leads to the Dirac mass for the neutrinos in the seesaw mechanism.  Both of
these ingredients are present in the models, as is evident from Eq.
(5.25). Moreover, since the right-handed neutrinos (RHNs) are in the
multi-TeV range, the Dirac Yukawa couplings are too small to yield a
sufficient amount of baryon asymmetry if the RHNs are hierarchical in mass. 
However it is well known that  if the RHNs are quasi-degenerate, 
 the mechanism can be resonantly enhanced~\cite{pilaftsis, pu}.  In our models, as
Eq. (\ref{mr}) shows, we have chosen a quasi-degenerate right-handed
neutrino spectrum to fit neutrino masses. We have not computed the
magnitude of the baryon asymmetry generated by our models but it is
known that with resonant leptogenesis mechanism, the mass and width of
the RHNs generically provide sufficient enhancement to give the right
order of magnitude for the baryon asymmetry.


\subsection{Dark Matter}

There is compelling cosmological evidence for dark matter (DM), which
makes up about 85 \% of the matter in the universe.  An intriguing
possibility is that dark matter is comprised of one or more particles,
and there has been, and continues to be, an intense experimental
effort to detect dark matter particles predicted by various models.
(Some recent reviews with references to the extensive literature are
\cite{dmreview,lindner,dm_pdg}.)  Many possible dark matter candidates have
been proposed and studied, ranging in mass from $\sim 10^{-22}$ eV
\cite{fdm} to primordial black holes. While the cold dark matter (CDM)
paradigm has received much attention, scenarios with warm dark matter
have also been studied (e.g., \cite{ak}).  Here we shall suggest a CDM
scenario in the context of the split-fermion models. In thermal dark
matter models, to account for the observed value of the dark matter,
$\Omega_{DM} =0.265(7)$ \cite{pdg,dmnotation}, the (co)annihilation
cross section $\sigma_{\rm DM \ ann.}$ times velocity $v$ (in the
center-of-mass of the (co)annihilating DM particles) should satisfy
\beq
\langle \sigma_{\rm DM \ ann.} v \rangle \simeq (2-3) \times 10^{-26} \
        {\rm cm}^3 \, {\rm s}^{-1} \ , 
\label{sigmav_dma}
\eeq
i.e., $\langle \sigma_{\rm DM \ ann.} (v/c) \rangle \sim 10^{-36}$
cm$^2$ \cite{dmreview,dm_pdg,steigman}. In the thermal dark matter
scenario, the freeze-out of the $\chi$ DM, which thus determines the
relic DM density, occurs as the temperature $T$ in the early universe
decreases below a value given by $T/m_\chi \simeq 0.05$.  Since
most of the DM $\chi$ particles are nonrelativistic at this point,
the corresponding $v/c$ value is $\sim 0.3$.

In order to account for dark matter, we will extend our minimal
split-fermion models with the addition of a dark matter particle which
is a chiral fermion, denoted $\chi$, that, like the other fermions,
has a wave function that is strongly localized with a Gaussian profile
in the $n=2$ extra dimensions. (Here we follow a common convention of
using the symbol $\chi$ for a dark matter fermion; the reader should
not confuse this with the $\chi$ in Eq. (\ref{psiform}).) There is a
nonzero overlap between the wave function of this $\chi$ field and the
SM fermions if there are gauge-invariant operators connecting a single
$\chi$ and an appropriate number of SM fermions, as we discuss below.
In this case, these operators will lead to the dark matter being an
unstable particle. However for a $\chi$ that transforms according to a
sufficiently high-dimensional representation of the gauge group and
for sufficiently large separation distances between $\chi$ and SM
fermions in the extra dimensions, the resultant couplings in the
low-energy effective Lagrangian in four spacetime dimensions will be
highly suppressed, so the DM fermion can be considered effectively
stable. Thus the first thing we note is that the dark matter in the SM
split-fermion model is necessarily a decaying dark matter, whose
decay rate depends on the representation of $\chi$ under the SM
gauge group, $G_{SM}$.  In order for $\chi$ to be sufficiently weakly
interacting, it must be a singlet under color SU(3)$_c$, so in the SM
split fermion model, we are referring to the representation of the
weak isospin group, SU(2)$_L$.  For instance, if $\chi$ belongs to
very high-dimensional representation of SU(2)$_L$, there can be
high-dimensional operators connecting the DM to SM fermions.

The situation can be very different if the gauge group is $G_{LRS}$,
since it is known that certain kinds of fermions in the LRS case do
not have any operator connecting a single DM field to SM fermions. The
DM fermion can therefore be absolutely stable dark
matter~\cite{heeck}. We briefly elaborate on these ideas below.


\subsubsection{Dark matter with SM gauge group}

As noted above, the simplest dark matter particle in the case of the
SM gauge group is a chiral fermion, which we can denote as $\chi$,
with a Majorana mass term of the form $\chi^T C \chi + h.c.$,
belonging to a higher-dimension (color-singlet) representation of the
SU(2)$_L$ SM weak isospin group~\cite{cs}.  Clearly, such a mass
term is allowed only for certain representations, namely those which
have zero weak hypercharge, $Y_\chi=0$. The $Y_\chi=0$ property is
also necessary to avoid a tree-level coupling of the $\chi$ with the
$Z$.  Let us denote the value of weak isospin of the SU(2)$_L$
representation containing $\chi$ as $(T_L)_\chi$.  This value must be
an integer, since the relation $Q_{em}=T_{3L} + (Y/2)$ implies that if
$\chi$ were in an SU(2)$_L$ representation with a half-integer value
of $(T_L)_\chi$, then no component in the corresponding weak
isomultiplet would be electrically neutral, as required for dark
matter. Since $(T_L)_\chi$ is an integer, $\chi$ actually transforms
as a representation of SO(3) and does not produce any triangle gauge
anomaly or global anomaly in the SU(2)$_L$ sector of the Standard
Model. The values $(T_L)_\chi=0$ and $(T_L)_\chi=1$ will be excluded
below.  The fact that $T_\chi$ is nonzero means that the full
SU(2)$_L$ weak isomultiplet will contain electrically charged
components. However, gauge interactions naturally raise the masses of
the charged components of this weak isomultiplet, so that $\chi$ is
the lightest member of this multiplet \cite{cs}.  In the
split-fermion models with $n=2$ and thus $d=6$ spacetime dimensions, a
chiral fermion is a 4-component fermion (denoted by $\psi_{+}$),
which, in a domain-wall background, plays the role of a two component
Weyl fermion in four spacetime dimensions.  After the
extra-dimensional wave function overlap effect is taken into account,
this will induce an effective operator in the 4D low-energy effective
theory, which can let the $\chi$ field decay to SM fields.  It follows
that a DM fermion transforming according to a smaller representation
would have a shorter lifetime compared to the required lifetime
$\tau_\chi \gsim 10^{25}$ sec. \cite{queiroz}
and hence could not be dark matter. For
example, if $\chi$ were to have $(T_L)_\chi=0$ and would thus be 
an SM-singlet, then there would be an effective operator
\beq
\sum_{a=1}^3 c^{(\chi L \phi)}_a \epsilon_{ij} \,
[\chi_L^T C L^i_{a,L}]\phi^j + h.c.
\label{nop1}
\eeq
(where $\phi$ is the SM Higgs doublet, $i, \ j$ are SU(2)$_L$ group
indices, and $\epsilon_{ij}$ is the antisymmetric SU(2) tensor) that
would enable $\chi$ to mix with known leptons with a mass mixing
$\propto v$ (where $v$ is the SM Higgs VEV) and would hence make it
decay very fast. Furthermore, the weak isovector value $(T_L)_\chi=1$
is also forbidden \cite{cs}, since there would then be an
operator
\beq
\sum_{a=1}^3 c^{(\chi L \phi \prime)}_a
\, (\epsilon_{ki}\epsilon_{mj}+\epsilon_{kj}\epsilon_{mi}) \, 
[\chi_L^{km \ T}  C L^i_{a,L}]\phi^j + h.c.
\label{nop2}
\eeq
Here, one uses the property that the isovector representation of
SU(2) is equivalent to the symmetric rank-2 tensor representation to
write $\chi^{km}$ as a two-index symmetric tensor. 
So one needs to choose $\chi$ to be in a
higher-dimensional SU(2)$_L$ representation. The study in Ref. \cite{cs}
concluded that the value $(T_L)_\chi=2$ is allowed for a fermion
(and $(T_L)=3$ would be allowed for a scalar DM particle, which we do not
consider here.)

For a DM $\chi$, belonging to the $(1,{\bf N})_0$ representation
under ${\rm SU}(3)_c \otimes {\rm SU}(2)_L \otimes {\rm U}(1)_Y$,
where $N=2(T_L)_\chi+1$ (and the subscript denotes the weak hypercharge, $Y$),
an effective operator in $4+n=6$ dimensions will be 
\beq
    {\cal O}_{DM} =  \sum_{a=1}^3 \sum_r \kappa^{(\chi,L_a)}_r \,
[\chi_L^T C L_{a,L}] \, \phi \, {\cal O}_{SM,r} \ ,
\label{odm}
\eeq
where the operator(s) ${\cal O}_{SM,r}$ consist of SM fermions whose
effective $G_{SM}$ representation is such that it makes the full
${\cal O}_{DM}$ operator an SM singlet.  (In Eq. (\ref{odm}) we have
left the SU(2)$_L$ indices implicit on the fields, with it being
understood that they are contracted to make an SU(2)$_L$ singlet.  The
corresponding effective operator in 4D, after the Higgs VEV is
substituted, and after the integration over the two extra dimensions
is performed, has a prefactor that depends on the the number and types
of fields in ${\cal O}_{DM}$ and an exponential factor $e^{-S_{{\cal
      O}_{SM,r}}}$ that depends on the separation distances between
their wave function centers.  In particular, with $k$ fermions
comprising (part or all of) ${\cal O}_{SM,r}$, and hence $k_{{\cal
    O}_{DM}} = k+2$, this prefactor is given by the $n=2$ special case
of $c_{r,(k_{{\cal O}_{DM}})}$ in Eq. (\ref{crgen}) with the parameter
$k$ in that equation set equal to $k_{{\cal O}_{DM}}$ and the mass
scale $M=\Lambda_L$.  This prefactor contains an exponential prefactor
$e^{-S_{{\cal O}_{DM},r}}$ that depends on the separation distances
between the fields comprising ${\cal O}_{DM}$.  Thus, just as with
proton decay operators, one can suppress this operator very
strongly. For the case $(T_L)_\chi=2$ and $m_\chi \sim O(10)$ TeV, the
relevant SM-invariant interaction in the 6-dimensional space is of the
form
\beq
{\cal O}_{DM} =  \sum_{a=1}^3 \kappa^{({\cal O}_a)} \,
[\chi_L^T C L_{a,L}] \, \phi \, (\phi^\dagger \phi) \ ,
\label{odm2}
\eeq
where the weak isospinors $L_{a,L}$ and $\phi$ are combined to make a
$T_L=1$ state; the $\phi^\dagger \phi$ product is also in a $T_L=1$
state; these are combined to make $T_L=2$; and this is contracted with
$\chi$ to form an SU(2)$_L$ singlet.  In this case, the exponential
factor $e^{-S_{{\cal O}_{SM}}} =
\exp[-(1/2)\|\eta_{L_{a,L}}-\eta_\chi\|^2]$.  With
$|\bar\kappa^{({\cal O}_a)}| \sim O(1)$, and a separation distance
$\|\eta_{L_{a,L}}-\eta_\chi\| \gsim 10$, we calculate that the $\chi$
lifetime $\tau_\chi$ satisfies the requisite condition of being
greater than $10^{25}$ sec. This separation distance can be arranged
in the model with the solutions for the quark and lepton wavefunction
centers that we have obtained.  It should be recalled that in this
type of model, because the $\chi$ is an SU(2)$_L$ nonsinglet, it has
tree-level couplings with $W$ and $Z$. 


\subsubsection{Dark Matter with $G_{LRS}$ Gauge Group}

The situation in the left-right split-fermion model is, however, very
different. As has been pointed out in \cite{heeck}, in this case there
are representations to which a dark matter fermion can belong which do
not have any effective operator that consists of a single DM field
together with SM fields coupled in a gauge-invariant way. As a result,
the dark matter can be stable and can only annihilate to give the
relic density. Examples of some $G_{LRS}$ representations for which
this situation holds are $\{(1,1,1)_0, \ (1,1,3)_0, \ (1,3,1)_0,
\ (1,1,5)_0, \ (1,5,1)_0,...\}$, where we use the same notation for a
representation of $G_{LRS}$ as in Eqs. (\ref{lrs_quarks}) and
(\ref{lrs_leptons}).  Note that in this case a $\chi$ which is a
singlet under $G_{LRS}$ is allowed.  Here the stability of dark matter
is guaranteed by the remnant $(Z_2)_{B-L}$ symmetry in the model.
 It can be the conventional WIMP
DM and  its relic density is determined by their annihilation to SM
fields via $W_{L,R}$ exchange~\cite{heeck2,hooper}.

The detailed phenomenology of such dark matter in the split-fermion
model with $G_{LRS}$ gauge symmetry is beyond the scope of this paper;
however, we would like to make some general comments about it. The
first point is that, at the tree level, the various members of the DM
multiplet are degenerate in mass, and their masses are split only by
radiative corrections due to the exhange of $W_L,R$
bosons~\cite{heeck2,hooper}.  Furthermore, the relic DM density in the
LRS model depends on the representation of $\chi$ under both the
SU(2)$_L$ and SU(2)$_R$ gauge groups. In particular, if $\chi$ is a
nonsinglet under SU(2)$_R$, then the relic density depends on the
$W_R$ and $Z'$ masses. The heavier the $W_R$ and $Z'$, the higher the
relic density. Since in our model, the $W_R$ and $Z'$ are already in
the $10^3$ TeV range (recall Eqs. (\ref{vr_min}) and (\ref{mnnb_vr})),
the cross section for the reaction $\bar\chi\chi \to f \bar f$ from
$Z'$ (where $f$ is an SM fermion) is quite suppressed. If the $\chi$
is a nonsinglet under SU(2)$_L$, then a calculation and result similar
to those obtained in \cite{cs} would apply.  The new point about such
models is that, depending on the location of $\chi$ in the extr two
dimensions, there will be an additional contribution to the DM
(co)annihilation in the early universe.  The dominant operator will be
of four-fermion type, with a bilinear $\chi$ term multiplying
bilinears composed of SM fermions.  After integration over the extra
dimensions, the resultant 4D operators will have prefactors of the
form
\beq
\Big ( \frac{\mu^2}{\pi \Lambda_L^4} \Big ) e^{-S_{\chi f}} \ . 
\label{dmstrength}
\eeq
The generic size of this new contribution is of order $10^{-3}G_F$ and can
be adjusted by suitably choosing the location of $\chi$ in the extra
dimensions to give the right relic density. For example, if the
fermion is very close to the SM fermions, the exponential can be close
to unity and the DM annihilation cross section will be of order
\beq
\sigma_{\chi\chi \ {\rm ann.}}
\sim \frac{10^{-6}\, G_F^2 \, m_\chi^2}{4\pi} \ .
\label{sigma_dm_annihilation}
\eeq
this gives the desired relic density for $m_\chi \sim 30$ TeV. The
detailed phenomenological implication of such dark matter in the LRS
version of this split-fermion model are beyond the scope of this
paper, but merit further investigation.


\section{Conclusions}
\label{conclusion_section} 

In this paper we have studied several properties of models with large
extra dimensions, in which quarks and leptons have localized wave
functions in the extra dimensions.  We have focused on the case of
$n=2$ extra dimensions and have considered models with two types of
gauge groups: (i) the Standard-Model gauge group, and (ii) the
left-right symmetric (LRS) group. In particular, we have investigated
how well these split-fermion models can account for neutrino masses
and mixing.  With an extension to include a gauged U(1)$_{B-L}$
symmetry, the SM version of the split-fermion model can be in accord
with current data.  As compared with the SM version, the LRS version
of the split fermion model has the advantage that it can account for
data on neutrino masses and mixing without the need for any extension,
provided that the vacuum expectation value of the $\Delta_L$ Higgs is
$\sim O(1)$ eV. The LRS solution involves a seesaw mechanism arising
from a naturally large vacuum expectation value of the $\Delta_R$
Higgs.  As part of our work, we have also calculated a new solution
for quark wave functions. In order to suppress flavor-changing neutral
current effects due to higher KK modes of gauge and Higgs fields
sufficiently in the split-fermion models, we have chosen locations for
the wave function centers of $Q=-1/3$ quark fields and charged leptons
so as to render the corresponding mass matrices diagonal, up to small
corrections.  We have shown that, within the context of this approach,
the LRS and augmented SM split fermion models are in accord with
experimental constraints, including those from limits on
non-Standard-Model contributions to weak decays and neutrino
reactions, FCNC processes, neutrino electromagnetic properties, and
neutrinoless double beta decays.  We have also discussed baryogenesis
and dark matter in the context of the models with each types of gauge
symmetry group and suggested extensions of these models with a
candidate fermion that could comprise dark matter.


\begin{acknowledgments}

This research was supported in part by the US National Science Foundation
Grants PHY-1914631 (R.N.M.) and NSF-PHY-1915093 (S.G. and R.S.).

\end{acknowledgments}


\begin{appendix}

  \section{Generalities on Quark and Lepton Mixing}
  \label{mixing_appendix}

We recall the procedure for diagonalizing the various fermion mass
matrices.  For this purpose, let us denote the chiral components of
the weak eigenstates as $\xi_{f,L}$ and $\xi_{f,R}$, for $Q=2/3$ and
$Q=-1/3$ quarks, and charged leptons, denoted generically as
$f=u,d,\ell$, where each of these is a three-dimensional vector with
generation indices $a=1,2,3$, which are henceforth implicit in the
notation.  A generic mass term is then
\beq
\bar \xi_{f,L} M^{(f)} \xi_R + h.c.
\label{general_mass_term}
\eeq
Each $M^{(f)}$ can be diagonalized by a bi-unitary transformation
For $f=u$ or $f=d$, we write
\beq
U^{(f) \dagger}_L M^{(f)} U^{(f)}_R = M^{(f)}_{diag.} \ .
\label{umu}
\eeq
To do this, one constructs the hermitian products $M^{(f)}M^{(f) \dagger}$
and $M^{(f) \dagger}M^{(f)}$, which can be diagonalized according to
\beq
U^{(f) \dagger}_L M^{(f)}M^{(f) \dagger} U^{(f)}_L = M_{diag.}^2
\label{ul_m_uladjoint}
\eeq
and
\beq
U^{(f) \dagger}_R M^{(f) \dagger}M^{(f)} U^{(f)}_R = M_{diag.}^2 \ .
\label{ur_m_uradjoint}
\eeq
(See, e.g., Eqs. (2.12)-(2.15) in \cite{lee_shrock77} with requisite
changes in notation.)
The corresponding transformations relating the weak eigestates $\xi^{(f)}$
and mass eigenstates, denoted $\psi^{(f)}$ (each a three-dimensional vector)
for $f=u,d$ are
\beq
\psi^{(f)}_L = U^{(f) \dagger}_L \xi^{(f)}_L \ , \quad
\psi^{(f)}_R = U^{(f) \dagger}_R \xi^{(f)}_R \ .
\label{psi_xi_rel}
\eeq
The weak charged current involving quarks, in terms of weak eigenstates, is
\beq
J_\lambda = \bar\xi^{(u)}_L \gamma_\lambda \xi^{(d)}_L 
= \bar\psi^{(u)}_L \gamma_\lambda V \psi^{(d)}_L \ , 
\label{quark_charged_current}
\eeq
where the CKM quark mixing matrix is 
\beq
V = U^{(u) \dagger}_L U^{(d)}_L \ .
\label{vckm}
\eeq
One step in the construction of a split-fermion model is to work
backwards from the known CKM matrix $V$ to determine $M^{(u)}$ and
$M^{(d)}$ and, from these, a set of quark wave function centers that
yield these mass matrices. It should be recalled that these mass
matrices are not unique; in view of the relation (\ref{vckm}), a
different set of mass matrices $M^{(u)}$ and $M^{(d)}$ and hence
different $U^{(u)\prime}_L$ and $U^{(d)\prime}_L$ (and different
$U^{(u)\prime}_R$ and $U^{(d)\prime}_R$, yielding the same mass
eigenvalues) satisfying $U^{(u)\prime \dagger}_L U^{(d)\prime}_L =
U^{(u) \dagger}_L U^{(d)}_L$ would yield the same CKM quark mixing
matrix $V$. Indeed, many forms have been studied for quark mass
matrices (e.g. \cite{qmix}). 
A standard convention for parametrizing the CKM matrix $V$ is 
in terms of three rotation angles, $\theta_{12}$,
$\theta_{13}$, $\theta_{23}$, and a phase, $\delta$ is 
\begin{widetext}
\beqs
          V &=& R_{23}(\theta_{23}) K^* R_{13}(\theta_{13}) K R_{12}(\theta_{12})  \cr\cr
            &=& \left(\begin{array}{ccc}
     c_{12}c_{13}                                      & s_{12}c_{13}                                      & s_{13}e^{-i\delta} \\
    -s_{12} c_{23} - c_{12} s_{23} s_{13} e^{i\delta}  & c_{12} c_{23} - s_{12} s_{23} s_{13} e^{i\delta}  & s_{23} c_{13} \\
     s_{12} s_{23} - c_{12} c_{23} s_{13} e^{i\delta}  &-c_{12} s_{23} - s_{12} c_{23} s_{13} e^{i\delta}  & c_{23} c_{13}  \end{array}\right) \ ,
     \label{ugen}
     \eeqs
     \end{widetext}
where $R_{jk}(\theta_{jk})$ is the matrix for a rotation by an angle
$\theta_{jk}$ in the $j,k$ subspace and $K$ is the phase matrix
$K={\rm diag}(1,1,e^{-i\delta})$. (A different convention was used in
early papers \cite{km,cab}.)

The diagonalization of the charged lepton mass matrix and the neutrino
mass matrix is discussed in the text (see Eqs. (\ref{umulep}),
(\ref{unu_calc}), and the resultant lepton mixing matrix $U$ is given by
Eq. (\ref{uleptonic}).  The same convention as Eq. (\ref{ugen}) is
used for the lepton mixing matrix $U$, with respective leptonic mixing
angles $\theta_{ij}$ and phase $\delta$. If neutrinos are Majorana
fermions, then the transformation (\ref{nunuprel}) also involves a
Majorana phase matrix, which may be written as $K_{Maj} = {\rm
  diag}(1,e^{i\alpha_2},e^{i\alpha_3})$ \cite{alpha12_maj}. One often
writes
\beq
U =  \left(\begin{array}{ccc}
          U_{e1}     & U_{e2}     & U_{e3} \\
          U_{\mu 1}  & U_{\mu 2}  & U_{\mu 3} \\
          U_{\tau 1} & U_{\tau 2} & U_{\tau 3} \end{array}\right) \, K_{Maj}
 \ . 
\label{nunuprel}
\eeq
However, these Majorana phases do not affect the
fit to neutrino oscillation data.

Concerning the seesaw mechanism, we recall the algebraic origin of the
problem that one encounters with the SM version of the split-fermion
model.  To show this, it will suffice to illustrate the problem in a
simplified case of one-generation, where the matrix in
Eq. (\ref{numass_sm}) is a $2 \times 2$ matrix. We write this as
\beq
M =  \left(\begin{array}{cc}
          m_L  & m_D \\
          m_D  & m_R  \end{array}\right) 
 \ . 
\label{simpleseesaw}
\eeq
The neutrino mass eigenvalues of this matrix are given by
\beqs
m_\pm &=& \frac{1}{2}\Big [ m_R + m_L \pm \sqrt{(m_R-m_L)^2+4m_D^2} \ \Big ]
\ . \cr\cr
&&
\label{eigenvalues}
\eeqs
With $m_R \gg m_D$, these eigenvalues have the expansions
\beq
m_+ = m_R +\frac{m_D^2}{m_R} + ...
\label{mplus}
\eeq
and
\beq
m_- = m_L - \frac{m_D^2}{m_R} + ...
\label{mminus}
\eeq
where $...$ indicate higher-order terms.  For the small eigenvalues to
be of the seesaw form, it is necessary that $m_L < m_D^2/m_R$.  


\section{Couplings of KK Modes of Gauge Bosons with Fermions}
\label{kk_appendix}

Here we review the couplings of KK modes of gauge bosons with fermions
in split fermion models
\cite{dpq2000,kaplan_tait,ng_split1,ng_split2,grossman_perez,abel,hewett_split,rm_fcnc}
and show how the diagonality property of the charged lepton and
$Q=-1/3$ mass matrices greatly reduces FCNC effects.  We first recall
that the diagonalization of a general charged lepton mass matrix is
carried out with the bi-unitary transformation $U^{(\ell) \dagger}_L
M^{(\ell)} U^{(\ell)}_R = M^{(\ell)}_{diag.}$, as in
Eq. (\ref{umulep}).  Since $M^{(\ell)}$ is diagonal here, we have
$U^{(\ell)}_L = U^{(\ell)}_R = {\mathbb I}$. A corresponding comment
applies to the bi-unitary transformations that diagonalize the
$Q=-1/3$ quark mass matrix, so $U^{(d)}_L = U^{(d)}_R = {\mathbb I}$.
We often display formulas for the general case of $n$ dimensions,
although we focus on the case $n=2$ in this work. As in the text, $m$
is an $n$-dimensional integer-valued vector, $m \in {\mathbb Z}^n$
with components $m=(m_1,...,m_n)$, (Euclidean) norm $\| m
\|=(\sum_{i=1}^n m_i^2)^{1/2}$ and scalar products such as $m \cdot
\eta = \sum_{i=1}^n m_i \eta_I$.  Because the fermions are localized
on a scale $\sigma \ll L$, the integration, over the extra dimensions,
of an operator product involving the $m$'th KK mode of a generic
neutral gauge field $V_\lambda$ and the $y$-dependent part of a chiral
fermion bilinear, $V^{(m)}_\lambda[\bar \chi_f(y)_L \gamma^\lambda
\chi_f(y)_L]$ or $V^{(m)}_\lambda[\bar \chi_f(y)_R \gamma^\lambda
\chi_f(y)_R]$ (where possible non-Abelian group indices are suppressed)
essentially picks out the value of the gauge field at the location of
a chiral fermion $f_L$:
\beqs
  C^{(m)}_{\eta_f} &=& |A|^2  \int_{-L/2}^{L/2} d^{n}y \,
  e^{\frac{2 \pi i}{L} (m \cdot y)} e^{-2\mu^2 \| y-y_f \|^2} \cr\cr
  &\simeq& \exp \Big ( \frac{2 \pi  i}{\mu L} (m \cdot \eta_{f}) \Big )
  \, \exp\Big ( -\frac{\pi^2 }{2(\mu L)^2} \| m \|^2\Big ) \ , \cr\cr
  &&
\label{intkk}
\eeqs
where here $f$ refers to $f_L$ or $f_R$ and the generation indices on
$f$ are implicit. In Eq. (\ref{intkk}), $A$ is the fermion field
normalization constant defined in Eq. (\ref{af}), and the factor of
$1/(L^{n/2})$ in Eq. (\ref{v_kk}) is cancelled by the $L^{n/2}$
dependence of the coupling. In accordance with our effective field
theory approach, the KK modes with $\| m \|$ so large as to probe
distances much smaller than $\sigma$ are excluded, and hence 
$\exp[-\pi^2 \|  m \|^2/(2(\mu L)^2)] \simeq 1$.

With the help of Eq. (\ref{intkk}), let us write down the effective
interaction Lagrangians for the SM gauge bosons and their
corresponding KK modes with the fermion zero-modes. As has been
mentioned in the text, since $\mu \gg \Lambda_L$, the higher-order
fermion modes effectively decouple from the theory. The coupling of
the photon $A_{\lambda}$ with the (zero-mode) fermion $f$ is given by
the following Lagrangian:
\beqs
    {\cal L}^{(Aff)}_{\rm eff} &=& e \, q_f \, [\bar \psi_{f_{a}} \,
    \Big (  A_{\lambda}^{( 0 )} \delta_{ab} +
      \sum_{ m \in {\mathbb Z}_{\neq  0}^n}
          K_{A; f, ab}^{(m)} A_{\lambda}^{(m)} \Big ) \,
          \gamma^\lambda \, \psi_{f_{b}}] \cr\cr
    &&
\label{LAffkk_eqn}
\eeqs
where $q_f$ is the electric charge of the fermion $f$, and $a, b$ are
generational indices. $A_{\lambda}^{( 0 )}$ (where the superscript
denotes the $n$-dimensional zero vector) is identified with the
SM photon, which couples in a flavor-diagonal manner with fermions.
The $m$'th mode of the photon with $m$ not equal to the zero vector,
denoted $A_{\lambda}^{( m)}$ with $m \ne 0$, has the 
following coupling with the fermions:
\begin{multline}
  K_{A; f, ab}^{(m)} = \sum_{k=1}^3 \bigg[ \big(U_{L}^{(f)}\big)^\ast_{ka} \,
    C_{\eta_{f_{k, L}}}^{(m)} (U_{L}^{(f)})_{kb} \, P_L + \\ 
    + \big(U_{R}^{(f)}\big)^\ast_{ka} \,
    C_{\eta_{f_{k, R}}}^{(m)} (U_{R}^{(f)})_{kb} \, P_R \bigg] \ ,
\label{Akk_coupling_eqn}
\end{multline}
where $P_{L,R} \equiv (1 \mp \gamma_5)/2$ are the usual chiral 
projection operators. Similarly, suppressing the color indices, the
coupling of the gluons to the SM quarks $q = u, d$ is determined from
the following Lagrangian:
\beq
    {\cal L}^{(Gqq)}_{\rm eff} = g_s  [\bar \psi_{q_{a}}
    \Big ( {\vec G}_{\lambda}^{ 0)}\delta_{ab} +
      \sum_{ m \in {\mathbb Z}_{\neq  0}^n}
      K_{G; q, ab}^{(m)} {\vec G}_{\lambda}^{ (m)} \Big ) \,
      \cdot {\vec T} \gamma^\lambda \, \psi_{q_{b}}] \ ,
\label{LGqqkk_eqn}
\eeq
where $g_s$ is the strong coupling constant, and $\vec T$ denotes a generator of the algebra of color SU(3)$_c$. 
The KK gluons, ${\vec G}^{(m)}_\lambda$ have the following coupling to the SM quarks
\begin{multline}
  K_{G; q, ab}^{(m)} = \sum_{k=1}^3 \bigg[ \big(U_{L}^{(q)}\big)^\ast_{ka} \,
    C_{\eta_{q_{k, L}}}^{(m)} (U_{L}^{(q)})_{kb} \, P_L + \\ 
    + \big(U_{R}^{(q)}\big)^\ast_{ka} \,
    C_{\eta_{q_{k, R}}}^{(m)} (U_{R}^{(q)})_{kb} \, P_R \bigg] \ ,
 \label{Gkk_coupling_eqn}
\end{multline}
From eqs. (\ref{Akk_coupling_eqn}, \ref{Gkk_coupling_eqn}) it is
evident that if $U_{L, R}^{(\ell)}$, $U_{L, R}^{(d)}$ were different
from the identity matrix, then the the presence of $K_{G; d,
  ab}^{(m)}$, $K_{A; \ell, ab}^{(m)}$, and $K_{A; d, ab}^{(m)}$
would lead to FCNC terms for KK modes other than the zero-mode term
with $m=0$, the zero vector in ${\mathbb Z}^n$.  However, in our case,
since $U^{(\ell)}_{L, R} = U^{(d)}_{L, R} = {\mathbb I}$, these FCNC
terms are absent for the charged leptons and down-quark sector.  They
are, however, present in the neutrino and up quark sector, and we
discuss the resultant effects in the text.

The coupling of the KK modes of the $W$-boson to SM fermions is given by
\begin{multline}
  {\cal L}^{(W;\rm KK)}_{\rm eff} = \frac{g}{\sqrt{2}}  \sum_{ m \in {\mathbb Z}_{\neq  0}^n}
  \Big ( [\bar \nu_{a, L}  W_\lambda^{(m)} K_{W; L, ab}^{(m)}  \gamma^\lambda \ell_{b, L}] + \\
        +  [\bar u_{a, L}  W_\lambda^{(m)} K_{W; Q, ab}^{(m)}  \gamma^\lambda d_{b, L}] \Big ) + h.c. \ ,
\label{LWkk_eqn}
\end{multline}
where 
\begin{align}
        K_{W; L, ab}^{(m)} &= \sum_{k=1}^3 (U_{L}^{(\nu)})_{ka}^\ast \, C_{\eta_{L_{k, L}}}^{(m)} (U_{L}^{(\ell)})_{kb} \nonumber \\
        K_{W; Q, ab}^{(m)} &= \sum_{k=1}^3 (U_{L}^{(u)})_{ka}^\ast \,   C_{\eta_{Q_{k, L}}}^{(m)} (U_{L}^{(d)})_{kb}
\label{Wkk_coupling_eqn}
\end{align}
Similarly, the coupling of the KK modes of the $Z$ boson to the fermion
$f$ is given by the effective Lagrangian 
\begin{multline}
  {\cal L}^{(Zff)}_{\rm eff} = \sqrt{g^2+g'^2} \, \sum_{X=L,R}
  T_Z^{(f_X)} \, [\bar\psi_{f_{a,X}}
      \Big ( Z^{(0)}_\lambda \delta_{ab} + \\
      +
      \sum_{ m \in {\mathbb Z}_{\neq  0}^n} K_{Z; f_X, ab}^{(m)}
      Z^{(m)}_\lambda
        \Big ) \gamma^\lambda \psi_{f_{b,X}}] + h.c.
\label{Lzffkk}
\end{multline}
where $g$ and $g'$ denote the SM SU(2)$_L$ and U(1)$_Y$ gauge
couplings, $T_Z$ was defined in Eq. (\ref{tz}), and the matrix product
$K_{Z; f}^{(m)}$ has elements given by
\begin{multline}
    K_{Z; f, ab}^{(m)} = \sum_{k=1}^3 \bigg [ (U^{(f)}_L)^\ast_{ka}
    C^{(m)}_{\eta_{f_{k, L}}} (U^{(f)}_L)_{kb} P_L + \\ 
    + (U^{(f)}_R)^\ast_{ka}
    C^{(m)}_{\eta_{f_{k, R}}} (U^{(f)}_R)_{kb} P_R  \bigg ] \ , 
\label{Zkk_coupling_eqn}
\end{multline}
which is the same as $K_{A; f, ab}^{(m)}$.

In a similar manner, we can write down the coupling of the KK modes of
the Higgs boson with the SM fermions. Let us illustrate the coupling
of the KK modes of the Higgs boson to the leptons:
\begin{multline}
  {\cal L}_{\rm eff}^{(H \ell \ell)} =
  [\bar L_{a, L} \Big [
    \big ( m_{\ell_b} + \frac{g \, m_{\ell_b}}{2 \, m_W} H^{(0)} \big )
    \delta_{a b} + \\
    + \sum_{ m \in {\mathbb Z}_{\neq  0}^n} K_{H; \ell, ab}^{(m)}
    H^{(m)} \Big ] \ell_{b, R}] + h.c. \ , 
       \label{LHll_eqn}
\end{multline}
and $K_{H; \ell, ab}^{(m)}$ is given by
\begin{multline}
  K^{(m)}_{H; \ell, ab} = y^{(\ell)}_{ab}
  \sum_{k=1}^3 \sum_{q=1}^3 (U^{(\ell)}_L)^\ast_{ka} \
  e^{-\frac{1}{2} \|\eta_{L_{k, L}} - \eta_{\ell_{q, R}} \|^2} \times \\
  \times C^{(m)}_{\bar\eta_{kq}} (U^{(\ell)}_R)_{qb} \ ,
\label{HKK_coupling_eqn}
\end{multline}
where the notation $\bar\eta_{kq}$ is defined as 
\beq
\bar \eta_{kq} \equiv \frac{\eta_{L_{k, L}}+\eta_{\ell_{q, R}}}{2} \ , 
\label{etabar}
\eeq
and $y_{ab}^{(\ell)}$ is the 
higher-dimensional Yukawa coupling, taken to be of
${\cal O}(1)$. Here $H^{(\bm{0})}$ is the SM Higgs boson.

Having noted the general coupling formulas for $n$-extra dimensions
and arbitrary mixing matrices, let us specialize for the case
applicable to our current work, namely $n=2$, and
$U_{L,R}^{(\ell)}$ = $U_{L,R}^{(d)} = \mathbb{I}$, and hence
$U^{(\nu)} = U$, the lepton mixing matrix, and
$U_L^{(u)} = V^\dagger$, where $V$ is the CKM quark mixing matrix. In
this scenario, the couplings of the KK gauge bosons to the SM fermions
reduce to the following forms. We have denoted this special case with
a tilde over the coefficients. The photon KK couplings are
\begin{align}
  \tilde K^{(m)}_{A; \ell, ab} &= \delta_{ab}
  \bigg( C^{(m)}_{\eta_{L_{a, L}}} P_L +
         C^{(m)}_{\eta_{\ell_{a, R}}}P_R \bigg) \nonumber \\
         \tilde K^{(m)}_{A; d, ab} &= \delta_{ab}
         \bigg( C^{(m)}_{\eta_{Q_{a, L}}} P_L +
                C^{(m)}_{\eta_{d_{a, R}}} P_R \bigg) \nonumber \\
                \tilde K_{A; u, ab}^{(m)} &= \sum_{k=1}^3
         \bigg [ V_{ak} \, C_{\eta_{Q_{k, L}}}^{(m)} V^*_{bk} \, P_L +
                  \nonumber \\ 
                  & + \big(U_{R}^{(u)}\big)^\ast_{ka} \,
                C_{\eta_{f_{u, R}}}^{(m)} (U_{R}^{(u)})_{kb} \, P_R \bigg] \ .
\label{AKK_coupling_special}
\end{align}
From Eq. (\ref{Gkk_coupling_eqn}) we see that the gluon KK modes will
have the same coupling form as above, namely $K^{(m)}_{G;
  f, ab} = K^{(m)}_{A; f, ab}$, for $f = u, d$. For our
mixing matrices, Eq. (\ref{Wkk_coupling_eqn}) will reduce to
\begin{align}
  \tilde K_{W; L, ab}^{(m)} &=
  U^*_{ba} \, C_{\eta_{L_{b, L}}}^{(m)}  \nonumber \\
  \tilde K_{W; Q, ab}^{(m)} &= V_{ab} \, C_{\eta_{Q_{b, L}}}^{(m)} 
\label{Wkk_coupling_special}
\end{align}
Similarly, Eq. (\ref{Zkk_coupling_eqn}) reduces to Eqs.
(\ref{Zkk_coupling_diag_special}) and
(\ref{Zkk_coupling_non_diag_special}) below: 
\begin{align}
  \tilde K_{Z; \ell, ab}^{(m)} &=  \delta_{a b } \bigg(
  C_{\eta_{   L_{a, L}}}^{(m)} P_L +
  C_{\eta_{\ell_{a, R}}}^{(m)} P_R \bigg) \nonumber \\ 
  \tilde K_{Z; d, ab}^{(m)} &=  \delta_{a b } \bigg(
  C_{\eta_{Q_{a, L}}}^{(m)} P_L +
  C_{\eta_{d_{a, R}}}^{(m)} P_R \bigg) \ ,
\label{Zkk_coupling_diag_special}
\end{align}
and
\begin{align}
  \tilde K_{Z; \nu, ab}^{(m)} &= \sum_{k=1}^3 \bigg [
    U^*_{ka} \, C^{(m)}_{\eta_{L_{k, L}}} U_{kb} P_L \bigg ] \nonumber \\
  \tilde K_{Z; u, ab}^{(m)} &= \sum_{k=1}^3 \bigg [
    V_{ak} \,  C^{(m)}_{\eta_{Q_{k, L}}} V^*_{bk} P_L + \nonumber \\
    &+(U_{R}^{(u)})^\ast_{ka}
    C^{(m)}_{\eta_{u_{k, R}}} (U_{R}^{(u)})_{kb} P_R \bigg ] \ ,
\label{Zkk_coupling_non_diag_special}
\end{align}
The couplings of the $Z$ KK modes are diagonal in Eq.
(\ref{Zkk_coupling_diag_special}),  but are nondiagonal in Eq.
(\ref{Zkk_coupling_non_diag_special}).
Similarly, the Higgs coupling to the charged leptons is
\beq
\tilde K^{(m)}_{H; \ell, ab} = y^{(\ell)}_{ab} \
e^{-\frac{1}{2} \|\eta_{L_{a,L}} - \eta_{\ell_{b, R}}\|^2}
C^{(m)}_{\bar \eta_{ab}}  \ ,
\label{HKK_couplings_special}
\eeq
where $\bar \eta_{ab}$ was defined in Eq. (\ref{etabar}). 


\section{Calculation of Locations of Lepton Wave
  Function Centers in the Extra Dimensions}
\label{lepton_centers_appendix}

In this Appendix we describe the method that we use to determine the
wave function centers for the lepton fields $L_{a, L}$, $\nu_{b, R}$,
and $\ell_{c, R}$. As in the text, the subscripts $a, b, c$ are
generational indices. Although the Dirac neutrino mass matrix
$M^{(D)}$ in Eq. (\ref{md}) does not uniquely fix the lepton wave
function centers, the distances among different lepton wave functions
in the extra dimensions, as displayed in Table
\ref{lepton_distances_table}, enter cross sections and decay rates for
various physical processes. Therefore, it is important to stipulate a
set of necessary selection criteria for the wave function
centers, based on physical grounds. Our criteria are as
follows:
 \begin{itemize}

 \item C$_1$: The wave function centers for the leptons should reproduce
   the observed neutrino mixing matrix and charged lepton masses.

 \item C$_2$: The charged lepton mass matrix generated by the solution
   should be approximately diagonal, to justify the choice $U^{(\ell)}
   = {\mathbb I}$ in Section \ref{nu_sm_section}.

 \item C$_3$: The separations between lepton wave function centers for
   different generations should provide adequate suppression for the
   charged-lepton flavor-violating processes to be in accord with
   current experimental bounds. Together with criterion C$_2$, this condition
   requires adequate separation among the wave function centers of the
   $L_{a,L}$ and $L_{b,L}$ fields with $a \neq b$.

 \item C$_4$: The overall lepton wave function centers in the
   extra-dimensional space should be sufficiently separated from the
   quark wave function centers to yield adequate suppression of
   baryon-number-violating nucleon decays to be in accord with
   experimental bounds on these decays. 

\end{itemize}

To begin with, let us consider the set of $2n_{gen.}=6$ fields
\{$L_{a, L}$, $\nu_{b,R}$\}, where $n_{gen.}=3$ denotes the number of
fermion generations. If the set of solutions for wave function centers
\{$\eta^{(\ell)}_{L_{a, L}}$, $\eta^{(\ell)}_{\nu_{b,R}}$\} satisfies
the criteria that $L_{a, L}$, $L_{b, L}$ are sufficiently far apart
for $a \neq b$, then the positions for $\ell_{c, R}$ fields can easily
be chosen so as to produce an approximately diagonal charged lepton
mass matrix, thereby satisfying criterion C$_2$. Here, we follow the
notation in the text, where the superscript $\ell$ indicates that
these coordinates are relative and are to be translated by an
appropriate translation vector relative to the quark wave function
centers, as shown in figure \ref{fermion_centers_figure}.  Thus, let
us focus on determining the wave function centers for the fields
\{$L_{a, L}$, $\nu_{b,R}$\}.  For $n=2$ extra spatial-dimensions we
have $2n_{gen.}n-3=12-3=9$ parameters for the lepton wave function
centers of the above set of fields, where we have subtracted two
overall translational degrees of freedom and one rotational degrees of
freedom, since these do not affect the relative positions of the
lepton wave function centers. Thus, we have to satisfy the
$n_{gen.}^2=9$ distance constraints listed in table
\ref{L_nu_distances_table} using these nine parameters. Using the
three degrees of freedom (translational and rotational) mentioned
above, let us choose the origin and orientation of the axes of the
$\eta^{(\ell)}$ coordinate system  
such that we can parametrize two of the locations as follows:
\beq
 \eta^{(\ell)}_{L_{2,L}} = (d, 0) \quad\quad
 \eta^{(\ell)}_{L_{3,L}} = (-d, 0)
 \label{parametrization_eta_eqn1}
 \eeq
 We write the components of $\eta^{(\ell)}_{L_{1,L}}$ and
 $\eta_{\nu_{b,R}}$ with $1 \le b \le 3$ as
 \beqs
 &&\eta^{(\ell)}_{L_{1,L}} = (\eta_1^{(L_{1,L})}, \eta_2^{(L_{1,L})} )  \cr\cr
 &&\eta^{(\ell)}_{\nu_{b,R}} = (\eta_1^{(\nu_{b,R})}, \eta_2^{(\nu_{b,R})})
\label{parametrization_eta_eqn2}
\eeqs
using superscripts here to avoid overly cumbersome notation. 
Since the nine distance constraints in Table
\ref{L_nu_distances_table} are quadratic equations in terms of the
components of the wave function centers, they do not uniquely specify
these centers uniquely, but rather yield sets of solutions that are
related to each other by various reflections about the chosen axes. We
categorize the solutions into classes that have the same magnitude for
the nine parameters in Eqs. (\ref{parametrization_eta_eqn1}) and
(\ref{parametrization_eta_eqn2}).  As part of our analysis, we address
the following question: are the distances among different lepton
fields identical for different solutions in a class, or do the
reflections change the unconstrained distances? The unconstrained
distances are those that are not specified from the Dirac neutrino
mass matrix. These include the distances $\|\eta_{L_{a,L}}-
\eta_{L_{b, L}}\|$ and $\|\eta_{\nu_{a,R}}- \eta_{\nu_{b, R}}\|$ for
$1 \le a,b \le 3$. In other words, only the \{$L_{a, L}, \nu_{b, R}$\}
wave center distances $\|\eta_{L_{a,L}}-\eta_{\nu_{b,R}}\|$ in
Fig. \ref{fermion_centers_figure} are constrained from the neutrino
mixing matrix. So can we move these points in such a way that keeps
this constrained distances unaltered, but modifies the unconstrained
distances? Determining the answer to this question is crucial for our
analysis, since the distances $\|\eta_{L_{a,L}}- \eta_{L_{b, L}}\|$
appear in the decay rates for various charged-lepton flavor-violating
processes $\ell_a \to \ell_b \bar \ell_b \ell_c$, as discussed in
Section \ref{CLFV_section} in the text.  In answer to this question,
we will show that the various reflections that take us from one
solution to another form the Klein four-group ${\mathbb V}_4$
(Vierergruppe), which is defined by the group elements and
operations:
\beq
 {\mathbb V}_4: \ \langle {\mathbb I}, R, R_1 | R^2 = R_1^2 =
 (R \cdot R_1)^2 = {\mathbb I} \rangle \ ,
\label{fourgroup}
\eeq
where ${\mathbb I}$ is the identity element. (Note that 
${\mathbb V}_4$ is the smallest non-cyclic abelian group and is isomorphic to
${\mathbb Z}_2 \otimes {\mathbb Z}_2$ and the dihedral group of order
4, denoted ${\mathbb D}_4$.) This restricted symmetry for the reflections also
keeps the unconstrained distances unaltered. Thus for a class of
solution, all the distances among different lepton fields in the extra
dimensions are fixed.

We proceed as follows: from Table \ref{L_nu_distances_table} it is
evident that since the distances involved are less than $L/2= 15$, the
toroidal distance evaluation function for these points becomes
identical to the ordinary Euclidean distance function. Therefore, the
distance constraints in Table \ref{L_nu_distances_table}, in the
parametrization of Eqs. (\ref{parametrization_eta_eqn1}) and
(\ref{parametrization_eta_eqn2}) read:
\beqs
&&(\eta_1^{\nu_{b,R}} - \eta_1^{(L_{1,L})})^2 +
  (\eta_2^{\nu_{b,R}} - \eta_2^{(L_{1,L})})^2  =
     \|\eta_{L_{1,L}} - \eta_{\nu_{b, R}}\|^2 \cr\cr
&&    (\eta_1^{(\nu_{b,R})} - d)^2 + (\eta_2^{(\nu_{b,R})})^2 =
 \|\eta_{L_{2,L}} - \eta_{\nu_{b, R}}\|^2 \cr\cr
 &&
 (\eta_1^{(\nu_{b,R})} + d)^2 + (\eta_2^{(\nu_{b,R})})^2 =
\|\eta_{L_{3,L}} - \eta_{\nu_{b,R}}\|^2 \ ,
\label{nu_distance_eqn}
\eeqs
for $b=1, 2, 3$. We have already used the rotational and translational
degrees of freedom that keep the relative distances unaltered by
choosing the parametrization in
Eqs. (\ref{parametrization_eta_eqn1}) and
(\ref{parametrization_eta_eqn2}). Now we determine the set
of reflection operations that keeps Eq. (\ref{nu_distance_eqn})
invariant. These operations are as follows:
\beqs
&&{\mathbb I} : {\eta^{(\ell)}_f \to \eta^{(\ell)}_f} \cr\cr
&&   r : {\eta^{(\ell)}_f \to - \eta^{(\ell)}_f} \cr\cr
&& r_1 : \{\eta_2^{(\nu_{b,R})} \to -\eta_2^{(\nu_{b,R})},
\eta_2^{(L_{1, L})} \to -\eta_2^{(L_{1, L})} \} \cr\cr
&& r_2 = r \cdot r_1 :
\{ d \to -d, \eta_1^{(\nu_{b, R})} \to - \eta_1^{(\nu_{b, R})}, \cr\cr
&& \eta_1^{(L_{1, L})} \to -\eta_1^{(L_{1, L})} \} \ , 
\label{group_elements_eqn}
\eeqs
for $b =1, 2, 3$. Here $f$ represents elements of the set \{$L_{a, L},
\nu_{b, R}$\}, i.e., ${\eta^{(\ell)}_f \to -\eta^{(\ell)}_f}$ means
that all the wave function positions for \{$L_{a, L}, \nu_{b, R}$\} are
reflected in the origin of the $\eta^{(\ell)}$ coordinate system.  The
notation in Eq. (\ref{group_elements_eqn}) is such that those positions
omitted from an operation are left unaltered by that operation. For
example, $r_2$ leaves \{$\eta_2^{(\nu_{b, R})}, \eta_2^{(L_{1, L})}$\}
unaltered. The notation $\eta_2^{(\nu_{b, R})} \to -\eta_2^{(\nu_{b, R})}$
is used as a shorthand for representing $\{\eta_2^{(\nu_{1, R})} \to
-\eta_2^{(\nu_{1, R})}, \eta_2^{(\nu_{2, R})} \to -\eta_2^{(\nu_2, R)},
\eta_2^{(\nu_{3, R})} \to -\eta_2^{(\nu_{3, R})}\}$; thus $r_2, r_1$
denote only one operation each. This restricted nature of the
reflection is evident in Eq. (\ref{nu_distance_eqn}), where, for
example, $\eta_2^{(\nu_{b, R})}$ is linked with $\eta_2^{(L_{1, L})}$, for
all $b=1, 2, 3$; that is, if we reflect one of these points, the other
ones also have to be reflected. A similar comment applies for the set
\{$ d, \eta_1^{(\nu_{b, R})}, \eta_1^{(L_{1, L})}$\}. Therefore the set of
operations that generates other equivalent solutions for the \{$L_{a,
  L}, \nu_{b, R}$\} wave function centers from one solution in a class
is given by $G \equiv\{{\mathbb I}, r, r_1, r_2 = r \cdot r_1 \}$,
where $r^2 = r_1^2 = (r \cdot r_1)^2 = {\mathbb I}$. This shows that
these set of operations forms the group ${\mathbb V}_4$. We note that
each of the elements $g \in G$ leaves the unconstrained distances
$\|\eta_{L_{a, L}}- \eta_{L_{b, L}}\|$, $\|\eta_{\nu_{b, R}}-
\eta_{\nu_{c, R}}\|$ unchanged, thereby completing our proof that
solutions in a class generated by the action of $g \in G$ are
completely equivalent to each other in the sense that all the
distances among the lepton wave function centers are the same for the
class.

    Now we look for a class of solutions that satisfies the criteria C$_1$,
    C$_2$, C$_3$, and C$_4$. We find one class of solutions for the
    lepton wave function
    centers that satisfies these conditions and display the locations for
    these centers in Table \ref{lepton_centers_table}. The
    magnitudes for our chosen
parameters for this class are shown in Table \ref{parameters_table}.


\begin{table}
  \caption{\footnotesize{Set of values for the parameters that determine
      the locations for the wave function centers of $L_{a, L}, \nu_{b,
        R}$ fields. Different locations with these parameters
      are shown in Table \ref{different_solutions_table}.}}
\begin{center}
\begin{tabular}{|c|c|} \hline\hline
parameter & value  \\ \hline
$d$                       & $0.93907$   \\[3pt]
$\eta_1^{(L_{1, L})}$     & $4.15696$   \\[3pt]
$\eta_2^{(L_{1, L})}$     & $7.84265$     \\[3pt]
$\eta_1^{(\nu_{1, R})}$   & $0.32032$     \\[3pt]
$\eta_2^{(\nu_{1, R})}$   & $5.23999$     \\[3pt]
$\eta_1^{(\nu_{2, R})}$   & $0.02191$     \\[3pt]
$\eta_2^{(\nu_{2, R})}$   & $4.94447$     \\[2pt]
$\eta_1^{(\nu_{3, R})}$   & $0.07834$     \\[3pt]
$\eta_2^{(\nu_{3, R})}$   & $4.72885$     \\[2pt]
\hline\hline
\end{tabular}
\end{center}
\label{parameters_table}
\end{table}

\begin{table*}
  \caption{\footnotesize{Different solutions for the class defined by
      the values of the parameters in Table
      \ref{parameters_table}. The parameters mentioned in the
      parentheses are to be taken from Table
      \ref{parameters_table}. The operations $g \in G \equiv
      \{{\mathbb I}, r, r_1, r_2 \}$ take one solution to the other
      within the class. This ${\mathbb V}_4$ symmetry keeps all the
      distances among different lepton wave function centers unaltered.
      We use the first solution in the text, with the translation vector
      specified by Eq. (\ref{eta_etaell_relation}). }}
\begin{center}
\begin{tabular}{|c|c|c|c|c|c|c|} \hline\hline 
  $g \in G$ & $\eta^{(\ell)}_{L_{1,L}}$ & $\eta^{(\ell)}_{L_{2,L}}$ & $\eta^{(\ell)}_{L_{3,L}}$  &
  $\eta^{(\ell)}_{\nu_{1,R}}$ & $\eta^{(\ell)}_{\nu_{2,R}}$ & $\eta^{(\ell)}_{\nu_{3,R}}$  \\ \hline 
  ${\mathbb I}$ & $(+ \eta_1^{(L_{1,L})},+ \eta_2^{(L_{1,L})})$ &
  $(+ d,0)$ & $(-d,0)$  & 
  $(- \eta_1^{(\nu_{1,R})}, +\eta_2^{(\nu_{1,R})})$ &
  $(+ \eta_1^{(\nu_{2,R})}, +\eta_2^{(\nu_{2,R})})$ &
  $(+ \eta_1^{(\nu_{3,R})}, +\eta_2^{(\nu_{3,R})})$ \\[4pt] 
  $r$ & $(- \eta_1^{(L_{1,L})}, - \eta_2^{(L_{1,L})})$ & $(- d,0)$ & $(+d,0)$  & 
  $(+ \eta_1^{(\nu_{1,R})}, - \eta_2^{(\nu_{1,R}})$ &
  $(- \eta_1^{(\nu_{2,R})}, - \eta_2^{(\nu_{2,R}})$ &
  $(- \eta_1^{(\nu_{3,R})}, - \eta_2^{(\nu_{3,R})})$ \\[4pt] 
  $r_1$ & $(+ \eta_1^{(L_{1,L})}, - \eta_2^{(L_{1,L})})$ & $(+d,0)$ & $(-d,0)$  & 
  $(- \eta_1^{(\nu_{1,R})}, -\eta_2^{(\nu_{1,R})})$ &
  $(+ \eta_1^{(\nu_{2,R})}, -\eta_2^{(\nu_{2,R})})$ &
  $(+ \eta_1^{(\nu_{3,R})}, -\eta_2^{(\nu_{3,R})})$   \\[4pt]
  $r_2 = r_1 \cdot r$ & $(- \eta_1^{(L_{1,L})},+ \eta_2^{(L_{1,L})})$ & $(-d,0)$ & $(+d,0)$  & 
  $(+ \eta_1^{(\nu_{1,R})}, +\eta_2^{(\nu_{1,R})})$ &
  $(- \eta_1^{(\nu_{2,R})}, +\eta_2^{(\nu_{2,R})})$ &
  $(- \eta_1^{(\nu_{3,R})}, +\eta_2^{(\nu_{3,R})})$ \\[4pt]
  \hline\hline
\end{tabular}
\end{center}
\label{different_solutions_table}
\end{table*}

Table \ref{different_solutions_table} shows different equivalent
positions for the set \{$\eta^{(\ell)}_{L_{a, L}},
\eta^{(\ell)}_{\nu_{c,R}}$\} that forms the class defined by the
values in Table \ref{parameters_table}. For each of this solution
$\eta^{(\ell)}_{\ell_{b, R}}$ can be chosen such that they produce the
desired diagonal charged lepton mass matrix. The action of the group
elements in $G$ produces the other solutions, which are completely
equivalent to each other as the different distances among lepton
wave function centers that enters into the physical cross section and
decay rates are the same in a class of solution due to the ${\mathbb
  V}_4$ symmetry. As we are free to choose the overall translation of
these locations for the lepton fields, we do not show the table for
the quark-lepton separation matrix for these different solutions, but
mention that it is possible to place the leptons such that all the
constraints from ref. \cite{bvd} can be easily satisfied that provide
adequate suppression for the nucleon and dinucleon decays to leptonic
final states. We have chosen first row in Table
\ref{different_solutions_table} translated by $(5, 3)$ in the text for
the analysis. Our conclusions remain unchanged with the choice of any
of the other equivalent positions.

In a similar manner, we have evaluated a new set of solution for the
quark wavefunction centers that produces the CKM quark mixing matrix
$V$. Moreover, this has the desirable feature of greatly reducing the
flavor changing neutral currents of the higher KK modes of the gauge
bosons by generating a nearly diagonal $Q=-1/3$ quark mass matrix. In
other words, the wavefunction centers are such that $U_L^{(d)} \simeq
{\mathbb I}$ in Eq. (\ref{vckm}). Using Eqs. (\ref{ul_m_uladjoint},
\ref{ur_m_uradjoint}, \ref{vckm}) and following the notations in
appendix \ref{mixing_appendix}, we get $M^{(d)} = M_{\rm diag}^{(d)}$
and
    \beq
          M^{(u)} M^{(u) \dagger} = V^\dagger  (M_{\rm diag}^{(u)})^2 \  V \ .
    \label{Mu_vckm_eqn}
    \eeq
We choose $M^{(u)} = V^\dagger M_{\rm diag}^{(u)}$, where the mass
eigenvalues for the quarks in $M_{\rm diag}$ have been taken at the
same scale $m_t$ \cite{koide}. Eq. (\ref{mqa_distance_constraint})
gives the required wavefunction separation matrices in the higher
dimensions as shown in table \ref{Q_ud_distances_table}.
\begin{table}
  \caption{\footnotesize{Distances $\| \eta_{Q_{a,L}}-\eta_{u_{b,R}}
      \|$, and $\| \eta_{Q_{a,L}}-\eta_{d_{b,R}}
      \|$ determined from the CKM quark mixing matrix $V$.  As defined in the
      text, the numerical subscript on each fermion field is the
      generation index of the weak eigenstate, with $1 \le a,b \le 3$.}}
\begin{center}
\begin{tabular}{|c|c|c|c|} \hline\hline
$a$ & $b$ & $\| \eta_{Q_{a,L}}-\eta_{u_{b,R}} \|$ & $\| \eta_{Q_{a,L}}-\eta_{d_{b,R}} \|$ \\ \hline
1 & 1 & 4.752 & 4.597  \\
1 & 2 & 3.765 & far  \\
1 & 3 & 3.085 & far  \\ \hline
2 & 1 & 5.051 & far  \\
2 & 2 & 3.354 & 3.892   \\
2 & 3 & 2.537 & far   \\ \hline
3 & 1 & 5.817 & far   \\
3 & 2 & 4.190 & far   \\
3 & 3 & 0.920 & 2.868  \\
\hline\hline
\end{tabular}
\end{center}
\label{Q_ud_distances_table}
\end{table}
Following a similar approach as the lepton sector, we choose the
orientation of the axes and origin of the relative quark coordinate
system $\eta^{(q)}$ such that the fields $Q_{1,L}$ and $Q_{2, L}$ lie
on the $\eta_1^{(q)}$ axis and are equidistant from each
other. Therefore, we parametrize the coordinates as
    \begin{align}
        \eta_{Q_{1,L}}^{(q)} &\equiv (-d_q,0) \quad ; \quad \eta_{Q_{2,L}}^{(q)} \equiv  (d_q,0) \nonumber\\
        \eta_{u_{a,R}}^{(q)} &\equiv (\eta_1^{(u_{a,R})}, \eta_2^{(u_{a, R})}) \nonumber \\
         \eta_{d_{b,R}}^{(q)} &\equiv (\eta_1^{(d_{b,R})}, \eta_2^{(d_{b, R})}) \ ,
        \label{Qud_parametrizaton_eqn}
    \end{align}
where $a,b=1,2,3$ are the generational indices, and $d_q$ can have
both signs. To begin with, let us focus on the set
$\{\eta^{(q)}_{Q_{a, L}}, \eta^{(q)}_{u_{b, R}}\}$. Once we have a
solution for this set, $\eta_{d_{c, R}}$ can be chosen accordingly
that satisfy the constraints in Table
\ref{Q_ud_distances_table}. Hence, we look for the solutions of the
following equations where the right hand side values are taken from
Table \ref{Q_ud_distances_table}.
\begin{align}
& (\eta_1^{(u_{b,R})} - \eta_1^{(Q_{3,L})})^2 +
  (\eta_2^{(u_{b,R})} - \eta_2^{(Q_{3,L})})^2  =
     \|\eta_{Q_{3,L}} - \eta_{u_{b, R}}\|^2  \nonumber\\
& (\eta_1^{(u_{b,R})} - d_q)^2 + (\eta_2^{(u_{b,R})})^2 =
 \|\eta_{Q_{2,L}} - \eta_{u_{b, R}}\|^2  \nonumber \\
 & (\eta_1^{(u_{b,R})} + d_q)^2 + (\eta_2^{(u_{b,R})})^2 =
\|\eta_{Q_{3,L}} - \eta_{u_{b,R}}\|^2    \ ,
\label{ud_distance_eqn}
\end{align}
for $b=1,2,3$. Comparing with Eq. (\ref{nu_distance_eqn}), we again
identify the ${\mathbb V}_4$ symmetry that relates elements within one
class of solutions. Requiring that the quark wavecenters be
sufficiently spread out to suppress the effects of various BSM local
operators, we arrive at the solution listed in table
\ref{quark_centers_table}. Needless to say, similar to table
\ref{different_solutions_table}, there exist three other equivalent
sets of quark wavecenters that are related to each other by the
elements in the group ${\mathbb V}_4$.

In passing, we note that the results in refs. \cite{nnb02,nnblrs}
regarding $n-\bar n$ oscillations are not sensitively dependent on the
different solutions for the locations of the wave function centers in
the extra dimensions. This is because in the SM split-fermion model,
the corresponding amplitude for the dominant operator mediating $n -
\bar n$ oscillations depends only on the distance
$\|\eta_{Q_{1,L}}-\eta_{d_{1,R}}\|$, which, in turn, is determined by
the physical mass of the $d$-quark, $m_d$ \cite{nnb02}. Moreover, in
the LRS split-fermion model, the dominant operator contributing to $n
- \bar n$ oscillations involves quark fields at the same point in the
extra dimensions and hence does not yield any exponential suppression
factor \cite{nnblrs}.

\end{appendix}



\begin{thebibliography}{99}
\bibitem{lmugen}
  %
  We note that a generalization of the model could take the
  compactification lengths $L$ to be different for each of the $n$ extra
  dimensions.  Another type of generalization could take the localization
  lengths $\sigma$ to be different for the various quark and lepton fields.
  Here, we 
  will restrict the analysis to the simplest form of the model involving only
  one compactification length $L$ and only one localization width $\sigma$.
\bibitem{alpha12_maj}
%
If neutrinos are Majorana fermions, as is the generic case with the
  seesaw mechanism, then there are also two additional Majorana
  phases.  These additional phases enter in neutrinoless double beta
  decay but cancel out in neutrino oscillation experiments and hence
  are not determined by current neutrino oscillation data.
\bibitem{sigfigs}
%
  The locations of fermion wave function centers and distances between
  these are listed to a few significant figures in the tables, but are
  encoded to much higher precision in our computer programs. In all
  expressions involving exponentials of (sums of) squares of
  distances, we present results from our computer calculation using
  higher-precision inputs.
\bibitem{cpv}
%
  In the presence of leptonic CP violation, the rate for a specific decay
  mode of a particle is, in general, different from the conjugate decay,
  so, for example, $\Gamma(\mu^+ \to e^+ \gamma)$ would, in general, differ
  from $\Gamma(\mu^- \to e^- \gamma)$. Where experimental upper limits are
  different for a decay mode and its conjugate, we will use the more stringent
  limits.
\bibitem{dmnotation}
  Here $\Omega_{DM} = \rho_{DM}/\rho_{cr}$, where $\rho_{cr}=3H_0^2/(8\pi G_N)$
  is the critical closure mass density, with $H_0$ and $G_N$ being the present
  Hubble constant and the Newton gravitational constant.
\end{thebibliography}
\end{document}